\pdfoutput=1
\documentclass[12pt,hidelinks]{article}
\usepackage{hyperref}
\usepackage{amsmath}
\usepackage{amsthm}
\usepackage{xr}
\usepackage{graphicx}
\usepackage{enumerate}
\usepackage{natbib}
\usepackage{subfig}
\usepackage{caption}%subcaption
\usepackage{blkarray}
\usepackage{bbm}
\usepackage{stackrel}
\usepackage{booktabs}
\usepackage{multirow}
\usepackage{threeparttable}      % for compact notes
\usepackage{caption}             % adjust caption spacing
\usepackage{array}               % extra column control
\usepackage{amssymb,booktabs,bm,multirow,threeparttable,textcomp,soul,cancel,mathrsfs,amsfonts,xcolor,tikz,psfrag,epsf,cases,empheq,enumitem, float}
\usepackage{url} % not crucial - just used below for the URL 
\usepackage{bibunits} % for merge main and supp
\defaultbibliography{scwi}
\defaultbibliographystyle{apalike}
\usetikzlibrary{decorations.pathreplacing}

\newcommand{\blind}{1}

\theoremstyle{plain}
\newtheorem{assumption}{\sc Assumption}
\newtheorem{condition}{\sc Condition}

\newtheorem{proposition}{\sc Proposition}

\newtheorem{theorem}{\bf Theorem}

\newtheorem*{assumption*}{\assumptionnumber}
\providecommand{\assumptionnumber}{}

\providecommand{\assumptionnumber}{}


\def\it{{it}}

\def\ee{\varepsilon}

\def\t{\scriptscriptstyle\mathsf{T}}

\usepackage{easyReview} % for \comment, \alert, etc.

\begin{document}

\def\spacingset#1{\renewcommand{\baselinestretch}%
{#1}\small\normalsize} \spacingset{1}

%%%%%%%%%%%%%%%%%%%%%%%%%%%%%%%%%%%%%%%%%%%%%%%%%%%%%%%%%%%%%%%%%%%%%%%%%%%%%%

\if1\blind
{
\title{\bf 
A robust regression approach to synthetic control with interference
}
	\author{Peiyu He$^{1,\star}$, Yilin Li$^{2,\star}$, Xu Shi$^3$, and Wang Miao$^{2}$\\
 Center for Data Science, Peking University$^1$\\
 Department of Probability and Statistics, Peking University$^2$\\
 Department of Biostatistics, University of Michigan$^3$}
  \maketitle
} \fi

\if0\blind
{
  \bigskip
  \bigskip
  \bigskip

  \begin{center}
    {\bf \LARGE A robust regression approach to synthetic control with interference}
\end{center}
  \medskip
} \fi

\bigskip
\begin{abstract}
Synthetic control methods are widely used for policy evaluation, but most approaches rule out interference among units, compromising validity when these effects occur. 
Adopting an attributable-effects perspective, we consider direct and interference effects as post-intervention outcome contrasts between the realized intervention regime and the counterfactual no-intervention regime.
We develop a framework accommodating contaminated donor pools and unknown interference through two stages: factor-model adjustment for unobserved confounding, followed by robust regression in which average direct and interference effects appear as a sparse outlier component. With a fixed number of units and the majority of them unaffected by interference, high-breakdown-point robust regression yields consistent identification of valid controls and asymptotically normal inference. With many units, we allow for sparse large and dense weak interference, where robust M-estimation remains valid even with a short post-intervention period. Unlike approaches requiring prespecification of valid controls or parametric modeling of interference, our framework relies only on coarse sparsity information and enables formal inference on both direct and interference effects. Two applications illustrate the framework: an analysis of the US embassy relocation to Jerusalem reveals significant interference effects on Jordan conflicts, and an analysis of Beijing's air pollution policy uncovers spatial interference patterns consistent with prevailing winds.
 
	\end{abstract}
	\noindent

\noindent%
\vfill
\footnotetext{$\star$ for equal contribution.}
\newpage
\spacingset{1.9} % DON'T change the spacing!

\begin{bibunit}
\section{Introduction} 

Comparative case studies evaluate the effects of policy interventions by comparing outcomes across aggregate units such as organizations and geographic areas.
However, heterogeneity across units and unobserved confounding hinder valid comparisons.
The synthetic control method \citep{abadie2003economic,abadie2010synthetic} addresses these challenges by constructing a weighted average of a set of comparable units (the ``donor pool") that matches the treated unit’s pre-intervention outcomes and yields estimates of its post-intervention counterfactuals.
Policy effects are then obtained from post-intervention differences between the treated unit and its synthetic control.
Recent developments have expanded the method to augmented and penalized weighting \citep{ben2021augmented,abadie2021penalized}, model-based and matrix completion approaches \citep{agarwal2021pcr,athey2021matrix,xu2017generalized}, and improved inference procedures \citep{cattaneo2021prediction,chernozhukov2021exact}.
Despite these advances, most existing methods implicitly assume that the intervention on the treated unit does not affect control units—the no-interference assumption. 
This assumption is pervasive in causal inference as part of the stable unit treatment value assumption (SUTVA) \citep{rubin1980randomization}, but it can be untenable in comparative case studies.
For instance, the announcement of the relocation of the US embassy from Tel Aviv to Jerusalem may influence conflict dynamics throughout the Middle East, not only in Israel–Palestine.
Similarly, Beijing’s air pollution control policy is implemented in core urban areas, but may spread the PM$_{2.5}$ reduction effects to suburban areas through the meteorological system.

When interference is present, control units in the donor pool may be affected by the intervention, rendering the synthetic control constructed from them potentially biased for counterfactual estimation.
A common practical remedy is to exclude potentially interfered units, but this strategy has notable limitations.
Discarding such units may compromise efficiency \citep{cao2019estimation} or lead to poor matching and introduce bias \citep{distefano2024inclusive}, as these units often exhibit strong geographic proximity or similarity to the treated unit. 
Moreover, selecting non-interfered valid controls requires strong domain knowledge about the interference structure, which may be difficult to obtain. 
Developing methods that accommodate potential interference in comparative case studies is therefore of considerable interest.

While the broader interference literature largely adopts a design-based framework \citep{hudgens2008interference,forastiere2021identification,han2024population}, the synthetic control method treats the intervention as fixed.
Several recent attempts on synthetic control with interference have either imposed strong parametric models, such as linear structures \citep{cao2019estimation} and spatial autoregressive models \citep{sakaguchi2024identification} for interference effects, or required prior specification of potentially interfered units \citep{distefano2024inclusive,grossi2025direct}.
\cite{oRiordan2025spillover} propose detecting invalid controls via model-based prediction motivated by proximal causal inference, but their approach lacks theoretical guarantees and may be sensitive to model misspecification.

In this paper, we develop a novel framework for addressing interference in comparative case studies, building on latent factor models underlying synthetic control and allowing for sparse interference with minimal prior knowledge of its structure. 
Our main contribution is a robust regression perspective and a corresponding two-stage procedure for identifying and estimating direct and interference effects averaged over the post-intervention period.
In the first stage, we estimate time-invariant factor loadings from a pre-intervention panel under an untreated latent factor model; these loadings play a role analogous to synthetic control weights, controlling for unmeasured confounding.
In the second stage, identification would be straightforward if the interfered units were known, but this information is typically unavailable. 
We address this by recasting the problem as robust regression, in which average direct and interference effects enter as a sparse outlier component relative to an otherwise linear model, with interfered units corresponding to contaminated observations. 
This perspective  enables  formal inference on both direct and interference effects.

We study two complementary asymptotic regimes.
When the number of units is fixed, we leverage a long post-intervention period to average out idiosyncratic errors so that non-interfered units asymptotically lie on a common hyperplane.
Under a majority valid controls condition, assuming that more than half of the control units are unaffected without knowing which ones, the strict majority of data points align with this hyperplane.
We employ high-breakdown-point robust regression to exploit this geometry, locking onto the hyperplane supported by the majority while ignoring outliers, yielding consistent selection of valid controls and asymptotically normal estimators for average direct and interference effects.
When the number of units diverges, we leverage rich cross-sectional information and allow a small fraction of units to exhibit large interference effects while permitting dense but weak interference among remaining units.
Under this pattern, robust loss functions such as smoothed Huber render outlier influence asymptotically negligible through cross-sectional averaging, enabling recovery of the true regression coefficients even with a short post-intervention period.
Accordingly, we develop estimators that are consistent and asymptotically normal when both pre- and post-intervention periods diverge, and conformal permutation tests for inference when the post-intervention period is short.

Unlike existing methods that require prespecification of valid controls or parametric modeling of interference, our framework relies only on coarse information about the extent of interference, complementing the synthetic control literature by providing tools for settings with potentially contaminated donor pools and unknown interference structures.
The remainder of this paper proceeds as follows. Section~\ref{sec:basemodel} introduces the factor model underlying synthetic control and formalizes the causal framework with interference. 
Sections~\ref{sec: fixedN} and~\ref{sec: largeN} present the proposed methods when the number of units is fixed and when it diverges, respectively.
Section~\ref{sec:simulation} reports simulation results. Section~\ref{sec:application} presents two empirical applications: the US embassy relocation policy and Beijing's air pollution control policy.
Section~\ref{sec:discussion} concludes with limitations and discussions.

We adopt the standard $O(\cdot)$, $o(\cdot)$ operators for scalars and $O_p(\cdot), o_p(\cdot)$ for random variables, applied entry-wise to vectors and matrices.
Let $|\cdot|$ denote the cardinality of a set, and
$\mathbbm{1}(\cdot), \lfloor\cdot\rfloor$ denote the indicator and floor functions, respectively. 
For any matrix (or vector) $\bm{A}\in\mathbb{R}^{m\times k}$ and an index set $\mathcal{S}\subseteq \{1,\dots, m\}$, let $\bm{A}_{\mathcal{S}}$ denote the sub-matrix (or sub-vector) of $\bm{A}$ with rows indexed by $\mathcal{S}$, and define $\|\bm{A}\|_1 = \sum_{m,k}|A_{mk}|$ and $\|\bm{A}\|_0 = \sum_{m,k}\mathbbm{1}(A_{mk}\neq0)$.

\section{Factor model for synthetic control with interference}\label{sec:basemodel}

Suppose $N$ units, indexed by $1\leq i\leq N$, are  investigated over $T$ study times, indexed by $1\leq t\leq  T$. 
An intervention occurs at time $T_0$, yielding $T_0$ pre-intervention and $T-T_0$ post-intervention time points; let $Z_t = \mathbbm{1}(t>T_0)$ indicate the intervention.
At each time $t$, an outcome $Y_{it}$ is observed for each unit $i$.
Our analysis considers the following complementary asymptotic regimes, covering a wide range of panel data settings: (i) fixed $N$ with $T_0$, $T-T_0\to\infty$; and (ii) $N\to\infty$, $T_0\to\infty$ with either fixed $T-T_0$ or $T-T_0\to\infty$.
% ; and (iii) $N\to\infty$, $T_0\to\infty$, and fixed $T-T_0$. 

Suppose the   intervention targets $J$  units; 
let $Y_{it}(\underbrace{1, \dots, 1}_{J}, 0, \dots, 0)$ and $Y_{it}(0, \dots, 0)$ denote the potential outcomes of unit $i$ at time $t$ 
under intervention and no intervention, respectively.
Here, we allow the potential outcomes of untreated units to depend on the treatment status of treated units, i.e., interference may arise.
To simplify notation and emphasize that only two intervention regimes may be realized in comparative case studies,  we write potential outcomes as $Y_{it}(1), Y_{it}(0)$ for short, where ``1" and ``0" correspond to the length-$N$ intervention vector $(\underbrace{1, \dots, 1}_{J}, 0, \dots, 0)$ and $(0, \dots, 0)$, respectively.
We are interested in the post-intervention contrasts $Y_{it}(1) - Y_{it}(0)$, which capture the causal effects attributed to the realized intervention, referred to as (individual) attributable effects following \cite{rosenbaum2001effects,choi2024new}.
For treated units ($1 \leq i \leq J$), these effects are at least partially attributable to their own treatment, termed attributable direct effects; for untreated units ($J+1 \leq i \leq N$), we term them attributable interference effects.
While the proposed framework applies to this general setup, the remainder of the paper focuses on the classical synthetic control setting with a single treated unit $J=1$ to simplify presentation. 
In this case, $Y_{1t}(1)-Y_{1t}(0)$ represents the direct effect on the treated unit at time $t>T_0$ and $Y_{it}(1)-Y_{it}(0)$ for $i>1$ represents the interference effect \citep{hudgens2008interference, liu2014large}. 
We make the following two standard assumptions.

\begin{assumption}[Consistency]\label{asmp:consistency}
For any $i, t$, $Y_{it} = Z_tY_{it}(1)+(1-Z_t)Y_{it}(0)$.
\end{assumption}

\begin{assumption}[Linear factor model]\label{asmp:model1}
For any unit $i$ at time $t$, 
\begin{eqnarray*}
    Y_{it}(0)=\bm{\lambda}_i^{\t}\bm{f}_t+ \varepsilon_{it}, \quad Y_{it}(1)=Y_{it}(0) + \beta_{it}, 
\end{eqnarray*}
where $\bm{f}_t\in \mathbb{R}^r$ is a vector of unobserved common factors (assumed to be a fixed sequence); $\bm{\lambda}_i\in \mathbb{R}^r$ is a vector of unit-specific unknown fixed factor loadings; the number of factors $r$ is known a priori;
$\varepsilon_{it}$ is the unobserved random error with
$E(\varepsilon_{it}) =0$; and $\beta_{it} \in \mathbb{R}$ is the unknown unit- and time-varying effect, which is assumed to be fixed.
\end{assumption}
Assumption \ref{asmp:consistency} is the standard consistency assumption.
Assumption \ref{asmp:model1} posits a linear factor model for the potential outcomes, commonly used in the synthetic control literature \citep{abadie2010synthetic,xu2017generalized}. 
In the absence of intervention, the variation of potential outcome $Y_{it}(0)$ is driven by the time-varying latent factors $\bm{f}_t$ with time-invariant unit loadings $\bm{\lambda}_i$.
Loading stability over the study period is a crucial structural assumption and may be restrictive, especially when $T$ is large. 
Let $\bm{\Lambda}=(\bm{\lambda}_1,\ldots,\bm{\lambda}_N)^{\t}$ denote the pre-intervention loading matrix.
In the Supplementary Material, we discuss two relaxations of this assumption, which target distinct departures and can be accommodated simultaneously. 
The first allows systematic shifts that remain within the column space of $\bm{\Lambda}$: for $t>T_0$, $Y_{it}(0)=\bm{\lambda}_{\mathrm{post},i}^{\t}\bm{f}_t+\varepsilon_{it}$ with $\bm{\Lambda}_{\mathrm{post}} =(\bm{\lambda}_{\mathrm{post},1},\ldots,\bm{\lambda}_{\mathrm{post},N})^{\t} = \bm{\Lambda}\bm{H}$ for some fixed $\bm H\in\mathbb R^{r\times r}$. The second allows mean-zero perturbations of post-intervention loadings: for $t>T_0$, $Y_{it}(0)=(\bm{\lambda}_i+\bm{\xi}_{it})^{\t}\bm f_t+\varepsilon_{it}$ with $\{\bm{\xi}_{it}\}_{t>T_0}$ mean-zero and uncorrelated with $\{\bm{f}_t\}_{t>T_0}$.
The interactive term $\bm{\lambda}_i^{\t}\bm{f}_t$ is common in panel data models \citep{bai2009panel}, generalizing the two-way fixed effects in difference-in-differences models by setting $\bm{f}_t = (1, \xi_t)^{\t}$ and $\bm{\lambda}_i = (\gamma_i, 1)^{\t}$ for some unit and time fixed effects $\gamma_i$ and $\xi_t$.
The number of factors $r$ is assumed fixed and known, typically much smaller than $N$ and $T_0$ \citep{anderson1956factor,bai2012Statistical}; methods for determining $r$ are well established \citep{bai2002determining,owen2016bicross}, and one can assess robustness by varying $r$ in practice.
We treat $\bm{f}_t$ as fixed, so the mean-zero error $\varepsilon_{it}$ is exogenous to the latent factors from a conditional view.
We denote $\{{\beta_{1t} : t > T_0}\}$ and $\{{\beta_{it} : 2 \leq i \leq N,\ t > T_0}\}$ as the dynamic direct and interference effects, respectively. 
Following common practice in the literature   \cite[e.g.][]{xu2017generalized, ferman2021properties}, we treat the potential outcomes as random and the effects $\{\beta_{it} : 1 \leq i \leq N, T_0 < t \leq T\}$ as fixed unknown parameters.
Our primary  estimands are the direct and interference effects averaged over the post-intervention period, defined as: 
\begin{itemize}
    \item average direct effect: $\bar\beta_{1} = (T-T_0)^{-1}\sum_{t=T_0+1}^T \beta_{1t}$;
    \item average interference effects: $\bar\beta_{i} = (T-T_0)^{-1}\sum_{t=T_0+1}^T \beta_{it}, \quad 2\leq i \leq N.$
\end{itemize}
The average direct effect $\bar\beta_{1}$ coincides with the average treatment effect on the treated in the absence of interference, an estimand of central interest in the synthetic control literature \citep{li2020istatistical,masini2021counterfactual}.
% which is of common interest in the synthetic control literature \citep{li2020istatistical,masini2021counterfactual}.
The average interference effect $\bar{\beta}_i$ on unit $2\leq i \leq N$  measures the average impact of unit $1$'s treatment on unit $i$ over the post-intervention period. 
Let $\bar{\bm{\beta}} = (\bar{\beta}_1, \bar{\beta}_2,\cdots, \bar{\beta}_N)^{\t}$ collect these effects.

Under Assumptions \ref{asmp:consistency} and \ref{asmp:model1}, the observed outcome satisfies
\begin{equation}\label{eq:modelatit}
    Y_{it} = \beta_{it} Z_t + \bm{\lambda}_i^{\t}\bm{f}_t + \varepsilon_{it}.
\end{equation}
This can be viewed as a generalization of the commonly-used interactive fixed effects model in the panel data and synthetic control literature \citep{bai2009panel,xu2017generalized,shi2026theory} by allowing the heterogeneous effects $\beta_{it}$ to vary across both time and unit. 
The term $\bm{\lambda}^{\t}_i\bm{f}_t$ is correlated with the treatment status and can thereby be viewed as unobserved confounders between units.
The factor $\bm{f}_t$ can correlate with $Z_t$, accommodating potential distributional shifts in latent factors between the pre- and post-intervention periods.

However, the average effects $\bar{\bm{\beta}}$ are not identified without additional assumptions.
From the perspective of model complexity, for each $t>T_0$, the presence of interference induces extra $N-1$ unknown parameters $\beta_{it}, 2\leq i\leq N$, greatly increasing the complexity compared to a no-interference setting. 
% This complexity constitutes the essential difficulty in identifying the causal effects.
In the following, we introduce practical assumptions to facilitate the identification of causal effects in the presence of interference.

\section{Methodology for the fixed-$N$ setting}\label{sec: fixedN}
\subsection{Identification}
Model \eqref{eq:modelatit} can be written in matrix form
\begin{equation}\label{eq:model1at}
\begin{aligned}
    \underbrace{\bm{Y}_t}_{N\times 1} &= \underbrace{Z_t}_{1\times 1} \underbrace{\bm{\beta}_t}_{N\times 1} &+ \underbrace{\bm{\Lambda}}_{N\times r} \underbrace{\bm{f}_t}_{r \times 1} + \underbrace{\bm{\varepsilon}_t}_{N\times 1},
\end{aligned}
\end{equation}
where $\bm{Y}_t=(Y_{1t},\ldots, Y_{Nt})^{\t}$, 
$\bm{\beta}_t=(\beta_{1t},\ldots, \beta_{Nt})^{\t}$,  
$\bm{\ee}_t = (\ee_{1t}, \ldots, \ee_{Nt})^{\t}$, and $\bm{\Lambda} = (\bm{\lambda}_1, \ldots, \bm{\lambda}_N)^{\t}$. 
To achieve identification, we introduce the following assumptions on the factor model and the number of interfered units, respectively. 
Let $\bm{M}_{f} = T_0^{-1} \sum_{t=1}^{T_0} (\bm{f}_{t} - T_0^{-1} \sum_{t=1}^{T_0} \bm{f}_{t})( \bm{f}_{t} - T_0^{-1} \sum_{t=1}^{T_0} \bm{f}_{t})^{\t}$ be the empirical covariance of sequence $\{\bm{f}_t\}_{t\leq T_0}$.
\begin{assumption}\label{asmp:stationary}
(i) $T_0^{-1}\sum_{t=1}^{T_0}\bm{f}_t=\bm{\alpha}_0+o(1)$ and $(T-T_0)^{-1}\sum_{t=T_0+1}^{T}\bm{f}_t=\bm{\alpha}_1+o(1)$; (ii) $\operatorname{Var}(\bm{\varepsilon}_t) = \bm{\Sigma} = \operatorname{diag}(\sigma_1^2, \dots, \sigma_N^2)$ and $\bm{M}_f = \mathbf{I}_r$; (iii) $\bm{\Lambda}^{\t}\bm{\Sigma}^{-1}\bm{\Lambda}$ 
is a positive diagonal matrix with distinct diagonal entries arranged in decreasing order.
\end{assumption}

\begin{assumption}[Majority valid controls]\label{asmp:sparsity}
$|\mathcal{C}|\geq \lfloor N/2\rfloor+r$, where $\mathcal{C} = \{i: \bar{\beta}_i = 0\}$. 
\end{assumption}
Assumption 3(i) imposes stationarity of $\{\bm{f}_t\}$ 
within the pre- and post-intervention periods while allowing a mean shift 
$\bm{\alpha}_0 \neq \bm{\alpha}_1$.
This relaxes full-period stationarity and accommodates selection on 
time-varying unobservables, as emphasized by \citet{ferman2021imperfect}.
The difference $\bm{\alpha} = \bm{\alpha}_1 - \bm{\alpha}_0$ 
captures the mean shift in latent confounding factors and needs to be 
accounted for, but it is not part of the causal effect of the intervention 
on outcomes. Adjusting for $\bm{\alpha}$ is central to identification and 
estimation.
Conditions (ii)-(iii) are standard for factor analysis 
\citep{anderson1956factor} and identify $\bm{\Lambda}$ up to an orthogonal 
rotation.

Assumption \ref{asmp:sparsity} states that at least $\lfloor N/2 \rfloor +r$ units remain unaffected by the intervention in the post-intervention period. 
This implies that the majority of control units are not interfered with, without knowing the exact valid control set $\mathcal{C}$. 
From a parameter counting perspective, Assumption \ref{asmp:sparsity} reduces the number of unknowns by limiting the extent of interference.
The majority valid controls assumption shares congruent ideas as the majority rules in the invalid instrumental variables literature \citep{kang2016instrumental,guo2018confidence} and the confounding adjustment with multiple treatments or outcomes \citep{wang2017confounder,miao2023identifying}. 
Apart from limiting the extent of interference, the dynamic form of $\bm{\beta}_t$ can exhibit flexible patterns to accommodate common gradual or seasonal effects.

We briefly describe our identification strategy. 
Let $\bar{\bm{Y}}_{\rm pre}= T_0^{-1}\sum_{t = 1}^{T_0}\bm{Y}_{t}$ and $\bar{\bm{Y}}_{\rm post} = (T-T_0)^{-1}\sum_{t = T_0+1}^{T}\bm{Y}_{t}$ denote the pre-intervention and post-intervention outcome mean, respectively.
The temporal difference of outcomes $\bm{D} = \bar{\bm{Y}}_{\rm post} - \bar{\bm{Y}}_{\rm pre}$ can be written as 
\begin{equation}\label{eq:rl}
\bm D = \bar{\bm{\beta}} + \bm{\Lambda}\tilde{\bm{f}} + \tilde{\bm{\varepsilon}},
\end{equation}
where $\tilde{\bm{\varepsilon}} = (T-T_0)^{-1}\sum_{t=T_0+1}^T \bm{\varepsilon}_t -T_0^{-1}\sum_{t=1}^{T_0}\bm{\varepsilon}_t$ and $\tilde{\bm{f}} = (T-T_0)^{-1}\sum_{t=T_0+1}^T \bm{f}_t -T_0^{-1}\sum_{t=1}^{T_0}\bm{f}_t$. 
As $T_0$, $T-T_0\to\infty$, the error term $\tilde{\bm{\varepsilon}}$ vanishes under mild conditions and $\tilde{\bm{f}} = \bm{\alpha}+o(1)$ by Assumption \ref{asmp:stationary}.
Therefore, $\bm{D}$ identifies $\bar{\bm{\beta}} + \bm{\Lambda}\bm{\alpha}$: a mixture of the average causal effects $\bar{\bm{\beta}}$ and a bias term $\bm{\Lambda}\bm{\alpha}$ that captures temporal mean shift of the latent factor structures.

We exploit an exact-fit robust regression perspective to disentangle the two components. Viewing $\bm{D} = \bm{\Lambda}\bm{\alpha} + \bar{\bm{\beta}}$ as a noise-free linear model, we treat $\bm{D}$ as observations, $\bm{\Lambda}$ as the design matrix, $\bm{\alpha}$ as coefficients, and $\bar{\bm{\beta}}$ as an outlier component. This motivates a two-stage procedure: first, $\bm{\Lambda}$ is identified from a long pre-intervention panel via factor analysis \citep{anderson1956factor}, since outcomes follow a pure factor model $\bm{Y}_t = \bm{\Lambda} \bm{f}_t + \bm{\varepsilon}_t$ under Assumptions \ref{asmp:consistency} and \ref{asmp:model1}; second, $\bm{\alpha}$ is identified via robust regression of $\bm{D}$ on $\bm{\Lambda}$. 
The intuition is geometric: under Assumption \ref{asmp:sparsity}, a strict majority of units are non-interfered ($\bar{\beta}_i = 0$) and lie asymptotically on the hyperplane $\bm{D} = \bm{\Lambda}\bm{\alpha}$, while interfered units deviate as outliers. 
High-breakdown-point robust regression fits the hyperplane supported by this majority, effectively ignoring outliers to recover $\bm{\alpha}$, known as the exact fit property \citep{rousseeuw2005robust}. 
Finally, deviations from the hyperplane $\bm{D} - \bm{\Lambda}\bm{\alpha}$ identify $\bar{\bm{\beta}}$.

Crucially, our approach leverages a long post-intervention period to average out noise and exploit the exact fit robust regression.
By contrast, when $T-T_0$ is also fixed, \cite{ferman2021imperfect} show the synthetic control estimator is generally biased under imperfect pre-intervention fit even in the absence of interference.
Complementing their insight, our contribution is to establish that average direct and interference effects are identifiable when both $T_0$ and $T-T_0\to\infty$, provided the majority of units are unaffected by the intervention.

\subsection{Estimation and inference}\label{subsec: fixedNest}
Motivated by \eqref{eq:rl}, estimation procedures proceed as follows. Denote $T_* = \min\{T_0,T-T_0\}$.
\begin{itemize}
\item Step 1. Obtain $\hat{\bm{\Lambda}}$ as the estimated factor loading by factor analysis with pre-intervention data $\{\bm Y_t\}_{t\leq T_0}$. 
\item Step 2. Given $\bm D$ and $\hat{\bm{\Lambda}}$, solve the following robust linear regression   to obtain $\tilde{\bm{\alpha}}$,
\begin{equation}\label{eq:lts}
    \tilde{{\bm{\alpha}}} = \mathop{\arg\min_{{\bm{\alpha}}}} \sum_{i=1}^{\lfloor N/2\rfloor +1}  r_{(i)}\left({\bm{\alpha}}\right),
\end{equation}
where $r_{(i)}({\bm{\alpha}})$ is the $i$th smallest value among $\{r_i({\bm{\alpha}}) = (D_{i} - \bm{\hat\lambda}_i^{\t}{\bm{\alpha}})^2; 1\leq i\leq N\}$.

\item Step 3. Obtain $\hat{\mathcal{C}} = \{i: |D_{i} -\hat{\bm{\lambda}}^{\t}_i\tilde{{\bm{\alpha}}}| \leq \{2T_*^{-1}\log(NT_*)\}^{1/2} \hat{\phi}\}$ as the estimated set of non-interfered units, where $\hat{\phi}^2 = N^{-1}\operatorname{tr}(\hat{\bm{V}})$, and $\hat{\bm{V}} = T_*T_0^{-1}(T-T_0)^{-1}\{\sum_{t=1}^{T_0} (\bm{Y}_t- \bar{\bm{Y}}_{\rm pre})(\bm{Y}_t- \bar{\bm{Y}}_{\rm pre})^{\t} + \sum_{t=T_0+1}^{T} (\bm{Y}_t- \bar{\bm{Y}}_{\rm post})(\bm{Y}_t- \bar{\bm{Y}}_{\rm post})^{\t}\}.$

\item Step 4. 
Solve $\hat{\bm{\alpha}}_1 = \arg\min_{{\bm{a}}} \|\bar{\bm{Y}}_{{\rm post}, \hat{\mathcal{C}}}-\hat{\bm{\Lambda}}_{\hat{\mathcal{C}}}{\bm{a}}\|_2^2$ for $\bm{\alpha}_1$, and obtain the estimator  $\hat{\bm{\beta}} = \bar{\bm{Y}}_{{\rm post}} -\hat{\bm{\Lambda}}\hat{{\bm{\alpha}}}_1$ for $\bar{\bm{\beta}}$, where $\bar{\bm{Y}}_{{\rm post}, \hat{\mathcal{C}}}$ is the subvector of $\bar{\bm{Y}}_{{\rm post}}$ with indices in $ \hat{\mathcal{C}}$.
\end{itemize}

In Step 1,  we estimate $\bm{\Lambda}$ from pre-intervention data via classical factor analysis \citep{anderson1956factor}, exploiting the pure factor structure $\bm{Y}_t = \bm{\Lambda}\bm{f}_t + \bm{\varepsilon}_t, 1\leq t\leq T_0$ implied by \eqref{eq:model1at}.
In Step 2, we plug the factor-loading estimator $\hat{\bm{\Lambda}}$ into \eqref{eq:lts} and obtain $\tilde{\bm{\alpha}}$ via robust linear regression, specifically using the least trimmed squares (LTS) method.
The   estimator $\tilde{\bm{\alpha}}$ is anticipated to be $T_*^{1/2}$-consistent for $\bm{\alpha}$ but is generally not asymptotically normal, where $T_*$ reflects the additional growth requirements on $T-T_0$ beyond $T_0$.
Consequently, the plug-in estimator $\bm{D} - \hat{\bm{\Lambda}}\tilde{\bm{\alpha}}$  is $T_*^{1/2}$-consistent for $\bar{\bm{\beta}}$ but does not readily admit inference.
To promote asymptotic normality, Step 3 applies hard thresholding to the preliminary effect estimates $\bm{D} - \hat{\bm{\Lambda}}\tilde{\bm{\alpha}}$ to select non-interfered control units.
The threshold is set to $\{2T_*^{-1}\log(NT_*)\}^{1/2} \hat{\phi}$ as inspired by \cite{donoho1994Ideal}, see Supplementary Material for discussions. 
In Step 4,  based on the selected control units set $\hat{\mathcal{C}}$, we obtain the estimator $ \hat{\bm \alpha}_1$ of $\bm{\alpha}_1$ by solving $ \bar{\bm{Y}}_{{\rm post}, \hat{\mathcal{C}}}=\hat{\bm{\Lambda}}_{\hat{\mathcal{C}}}\hat{{\bm{\alpha}}}_1$ through least squares. 
The ultimate estimator of $\bar{\bm{\beta}}$ is $\hat{\bm{\beta}} = \bar{\bm{Y}}_{\rm post} -\hat{\bm{\Lambda}}\hat{{\bm{\alpha}}}_1$.

The theoretical results require regularity conditions C.1–C.4 in the Supplementary Material, which impose standard factor analysis conditions excluding dominating factors and mixing dependence constraints on the errors.
The following theorem establishes selection consistency and asymptotic normality for both the average direct and interference effects.

\begin{theorem}\label{thm:1}
Under Assumptions \ref{asmp:consistency}--\ref{asmp:sparsity} and regularity conditions C.1--C.4 in the Supplementary Material, as $T_* \to\infty$, we have $P(\hat{\mathcal{C}} = \mathcal{C})\to1$. If further $T_0/T\to\kappa\in[0,1]$, then $T_*^{1/2}(\hat{\bm{\beta}}-\bar{\bm{\beta}})\stackrel{d}{\to}\mathcal{N}(\bm{0},\bm{\Omega}(\kappa))$ for some nondegenerate covariance matrix $\bm{\Omega}(\kappa)$.
\end{theorem}
For inference under serial dependence, we employ the circular block bootstrap \citep{politis1992circular}; implementation details are provided in the Supplementary Material.
Since uncertainty from the data-driven selected set $\hat{\mathcal{C}}$ is difficult to quantify, we treat selection as an intermediate step to facilitate asymptotic approximation and conduct inference based on the resulting point estimates and bootstrap confidence intervals.
As a practical diagnostic, if the number of statistically significant estimated effects exceeds $N-\lfloor N/2\rfloor - r$, this suggests a potential violation of the majority valid controls assumption, in which case the proposed method may no longer be applicable.

Although no explicit weighting  is involved in the estimation procedure, 
it in fact encompasses a weighting scheme and  constructs a hypothetical  control in the same spirit as previous synthetic control methods.
Let $\hat{w}_j = \bm{\hat{\lambda}}_{1}^{\t}(\bm{\hat{\Lambda}}_{\hat{\mathcal{C}}}^{\t}\bm{\hat{\Lambda}}_{\hat{\mathcal{C}}})^{-1}\bm{\hat{\lambda}}_j$ denote  the estimated weights for $1\leq j \leq N$.
The    estimator of the average direct effect can be rewritten as 
\[\hat{\beta}_1 = \bar{ Y}_{\rm post,1} -  \sum_{j\in\hat{\mathcal{C}}}\hat{ w}_j \bar{Y}_{{\rm post},j},\]
which is a contrast between  the observed outcome mean and   the  counterfactual outcome mean estimated with the weighted average of the observed outcomes   among the selected control units.
Unlike previous  methods that select control units completely based on prior knowledge,
we adaptively select valid control units based on the assumption of majority valid controls and robust regression.

\begin{proposition}\label{prop:weight}
Under Assumptions \ref{asmp:consistency}--\ref{asmp:sparsity} and   regularity conditions C.1--C.4 in the Supplementary Material, as $T_0\to\infty$, we have $\bm{\lambda}_1 - \sum_{j\in\hat{\mathcal{C}}}\hat{w}_j\bm{\lambda}_j = O_p(T_0^{-1/2})$.
If further   $|\hat{w}_{j}|\leq C$ almost surely for all $j$ and some constant $C<\infty$, 
then $E\{\bar{Y}_{\mathrm{post}, 1}(0) - \sum_{j\in\hat{\mathcal{C}}}\hat{w}_j\bar{Y}_{\mathrm{post},j}\} = o(1)$.
\end{proposition}
Proposition \ref{prop:weight} shows that the estimated weights consistently reconstruct the treated unit's factor loading, yielding the synthetic control outcome mean asymptotically unbiased for the counterfactual mean $\bar{Y}_{\mathrm{post},1}(0)$. These results parallel the fundamental balancing role of synthetic control weights \citep{abadie2010synthetic,ferman2021imperfect}. 
Unlike the original synthetic control method, which enforces weights to be non-negative and sum to one, our approach relaxes these constraints to permit extrapolation, a common extension in the literature \citep{amjad2018robust,li2020istatistical,ben2021augmented}. Nevertheless, the constructed weights approximately satisfy the sum-to-one condition when a factor has constant loadings across units.
The boundedness condition $\max_j|\hat{w}_j|\le C$ is a standard regularity condition; similar conditions are used in \cite{bai2012Statistical,bai2016maximum}.

\section{Methodology for the large-$N$ setting}\label{sec: largeN}
\subsection{Identification}
The fixed-$N$ setup   requires long pre- and post-intervention periods.
As a rule of thumb from prior work \citep{maccallum2001sample}, also supported by our numerical experiments,  reliable pre-intervention factor analysis typically requires $T_0\geq5N$ for numerical stability, and a long post-intervention period is crucial for consistent selection of valid control units.
However, many   applications involve a moderate or large number of units with limited temporal observations. 
An asymptotic framework in which both $N$ and $T_0$ diverge can provide a more realistic approximation for such scenarios.
In this section, we extend our framework to a large-$N$ regime, 
considering a high-dimensional approximate factor model and robust M-estimation that accommodates a generalized sparse interference pattern.
We reuse the notation from the previous section to denote the same or directly analogous quantities in the model of Section \ref{sec:basemodel}.
The   key identification assumptions are parallel to Assumptions~\ref{asmp:stationary} and \ref{asmp:sparsity}.
Recall $\bm{M}_{f}$ is the empirical covariance of $\{\bm{f}_t\}_{t\leq T_0}$ and $T_* = \min\{T_0,T-T_0\}$.
\begin{assumption}[Approximate factor model]\label{asmp: AFM}
For some constant $C > 0$, the following holds for all $t=1,\ldots,T$:
(i) $|E(\varepsilon_{it}\varepsilon_{jt})|\leq \tau_{ij}$ and $\sum_{j = 1}^N \tau_{ij}\leq C$ for each $i$; moreover, $\bm{\Sigma} = \operatorname{diag}\{E(\bm{\varepsilon}_t\bm{\varepsilon}_t^{\t})\} = \operatorname{diag}(\sigma^2_1,\ldots,\sigma^2_N)$ satisfies $C^{-2}\leq \sigma^2_i\leq C^2$;
    (ii) $\|\bm{f}_{t}\|_2 \leq C$ and $\bm{M}_{f} = \bm{I}_r$;
    (iii) $\|\bm{\lambda}_{i}\|_2 \leq C$ for all $i$, and $
 N^{-1} \bm{\Lambda}^{\t} \bm{\Sigma}^{-1} \bm{\Lambda} \to\bm{Q}$ as $N\to\infty$ for some positive definite matrix $\bm{Q}$.

\end{assumption}

\begin{assumption}[Sparsity of large interference effects]\label{asmp: largeNsparse} There exists some set $\mathcal{S}\subset\{1,\ldots,N\}$ such that 
either (a) $|\mathcal{S}| = o(N)$ and $\|\bar{\bm{\beta}}_{\mathcal{S}^c}\|_{1} = o(N)$ as $N\to\infty$  or (b) $|\mathcal{S}| = o(NT_*^{-1/2})$ and $\|\bar{\bm{\beta}}_{\mathcal{S}^c}\|_1 = o(NT_*^{-1/2})$ as $N$, $T_*\to\infty$, where $\mathcal{S}^c = \{1,\ldots, N\}\backslash \mathcal{S}$.
\end{assumption}
Assumption~\ref{asmp: AFM} aligns with the high-dimensional approximate factor model \citep{bai2012Statistical,bai2016maximum} and ensures identifiability of both $\bm{\Lambda}$ and $\bm{\Sigma}$ under some further regularity conditions. 
Assumption~\ref{asmp: largeNsparse}(a) characterizes a sparse interference pattern: only a small proportion of units may exhibit large interference effects, captured by the set $\mathcal{S}$ with $|\mathcal{S}| = o(N)$, while interference effects on remaining control units can be dense but weak in the sense that $\|\bar{\bm{\beta}}_{\mathcal{S}^c}\|_{1} = o(N)$.
This condition nests two common sparsity patterns as special cases: (i) $\|\bar{\bm{\beta}}\|_0 = o(N)$ with $\mathcal{S} = \{i:\bar{\beta}_i \neq 0\}$, and (ii) $\|\bar{\bm{\beta}}\|_1 = o(N)$ with $\mathcal{S}=\emptyset$, demonstrating certain flexibility of permitted interference structures.
Assumption~\ref{asmp: largeNsparse}(b) imposes an analogous pattern with stronger rate conditions required for asymptotic normality.

The identification strategy follows the same spirit of combining factor analysis with robust regression.
Recall   $\bar{\bm{f}}_{\rm post} = (T-T_0)^{-1}\sum_{t=T_0+1}^{T}\bm{f}_t$, $\bar{\bm{\varepsilon}}_{\rm post} = (T-T_0)^{-1}\sum_{t=T_0+1}^{T}\bm{\varepsilon}_t$, and    $\bar{\bm{Y}}_{\rm post} = (T-T_0)^{-1}\sum_{t=T_0+1}^{T}\bm{Y}_t$. Write the decomposition
\begin{equation}\label{eq: rridenlargeN}
    \bar{\bm{Y}}_{\rm post} = \bar{\bm{\beta}}+\bm{\Lambda}\bar{\bm{f}}_{\rm post} + \bar{\bm{\varepsilon}}_{\rm post}.
\end{equation}
We interpret this as a robust regression model with response $\bar{\bm{Y}}_{\rm post}$, design matrix $\bm{\Lambda}$, and coefficient $\bar{\bm{f}}_{\rm post}$, where $\bar{\bm{\beta}}$ acts as a sparse outlier component.
Importantly, we allow $\bar{\bm{\varepsilon}}_{\rm post}$ to be nonvanishing, thereby accommodating fixed $T-T_0$.
This motivates a two-stage procedure paralleling the fixed-$N$ setting: first, $\bm{\Lambda}$ is identified via high-dimensional factor analysis \citep{bai2016maximum} using the pre-intervention panel with both $N$ and $T_0$ diverging; second, $\bar{\bm{f}}_{\rm post}$ is recovered by regressing $\bar{\bm{Y}}_{\rm post}$ on $\bm{\Lambda}$ via robust M-estimation.
Under Assumption~\ref{asmp: largeNsparse}, only a vanishing fraction of outliers exhibit large deviations, while the remainder adhere closely to the baseline model $\bar{\bm{Y}}_{\rm post} = \bm{\Lambda}\bar{\bm{f}}_{\rm post} + \bar{\bm{\varepsilon}}_{\rm post}$.
A robust loss function bounds the influence of these outliers, rendering their contamination asymptotically negligible as $N\to\infty$. 
Finally, $\bar{\bm{\beta}}$ is identified from the residual $\bar{\bm{Y}}_{\mathrm{post}} - \bm{\Lambda}\bar{\bm{f}}_{\rm post}$ up to the error term $\bar{\bm{\varepsilon}}_{\rm post}$.

The robust regression perspective underlying the two asymptotic regimes differs sharply.
The fixed-$N$ setting employs exact-fit regression to identify the hyperplane supported by the majority of valid controls, enabling precise outlier separation. 
In the large-$N$ setting, the regression is intrinsically noisy; identification does not separate outliers but instead relies on robust loss to dilute interference contamination through cross-sectional averaging as $N\to\infty$.
Accordingly, the fixed-$N$ setting demands a long post-intervention period and, per Assumption~\ref{asmp:sparsity}, that the majority of controls be valid; the large-$N$ setting remains valid with a short post-intervention period and, per Assumption~\ref{asmp: largeNsparse}, permits interference to be dense but weak across many control units. 
These differences illustrate a ``blessing of dimensionality". 
The contrast also mirrors factor model asymptotics: the factor structure is consistently estimable when $N$, $T_0\to\infty$ \citep{bai2003inferential,bai2012Statistical}, but latent factors are not consistently recoverable with fixed $N$ \citep{anderson1956factor,bai2003inferential}.

\subsection{Estimation}\label{subs: largeNest}
Motivated by \eqref{eq: rridenlargeN}, the estimation proceeds in three steps. Recall $T_* = \min\{T_0,T-T_0\}$.
\begin{itemize}
\item Step 1. Obtain $\hat{\bm{\Lambda}}$ and $\hat{\bm{\Sigma}} = \operatorname{diag}(\hat{\sigma}_1^2,\ldots, \hat{\sigma}_N^2)$ as the estimated factor loading and variances by factor analysis \citep{bai2016maximum} with pre-intervention data $\{\bm Y_t\}_{t\leq T_0}$. 
\item Step 2.  Given $\hat{\bm{\Lambda}}$ and $\hat{\bm{\Sigma}}$, 
obtain the estimator $\hat{\bm{f}}$ for $\bar{\bm{f}}_{\rm post}$ by solving the robust regression problem:
\begin{equation}\label{eq: huberregavg}
    \hat{\bm{f}} = \arg\min_{\bm{f}} \sum_{j=1}^N\rho\left(\frac{\bar{Y}_{\mathrm{post},j} - \hat{\bm{\lambda}}_j^{\t}\bm{f}}{\hat{\sigma}_j}\right),
\end{equation}
where $\rho(\cdot)$ is some prespecified robust loss function.
\item Step 3. Obtain the estimator $ \hat{\bm{\beta}} = \bar{\bm{Y}}_{\rm post}-\hat{\bm{\Lambda}}\hat{\bm{f}}$ for $\bar{\bm{\beta}}$.
\end{itemize}
In Step 1, we follow \cite{bai2016maximum} to apply an EM algorithm to maximize a quasi-likelihood function and obtain $\hat{\bm{\Lambda}}$ and $\hat{\bm{\Sigma}} = \operatorname{diag}(\hat{\sigma}_1^2,\ldots, \hat{\sigma}_N^2)$, implemented using the R package \texttt{cate}.
In Step 2, motivated by \cite{wang2017confounder}, 
we regress $\bar{\bm{Y}}_{\rm post}$ on $\hat{\bm{\Lambda}}$ under a robust loss $\rho(\cdot)$ to estimate the underlying model coefficient $\bar{\bm{f}}_{\rm post}$. 
Unlike the fixed-$N$ setting, $T-T_0$ may remain fixed here, so the target is the empirical rather than the population mean of factors.
The choice of robust loss is crucial for the validity of the proposed method; the following assumption specifies the requirements on $\rho$.
In practice, log-cosh loss $\rho(x) = \log\{(e^x + e^{-x})/2\}$ and smoothed Huber loss functions such as pseudo-Huber $\rho(x) = \delta^2(\{1 + (x/\delta)^2\}^{1/2} - 1 )$ satisfy Assumption \ref{asmp: largeNrrloss}. 
\begin{assumption}\label{asmp: largeNrrloss}
    The function $\rho : \mathbb{R} \to [0, \infty)$ satisfies $\rho(0)=0$, is nonincreasing on $(-\infty,0]$, and nondecreasing on $(0,\infty)$.
    In addition, $\rho$ is convex on $\mathbb{R}$ and the derivative $\varphi = \rho^{\prime}$ satisfies that $|\varphi(x)|\le C$, $|\varphi^{\prime}(x)|\le C$, and $|\varphi^{\prime\prime}(x)|\leq C$ for all $x\in\mathbb{R}$ and some constant $0<C<\infty$. 
\end{assumption}

For theoretical analysis, we impose standard technical conditions for factor analysis \citep{bai2016maximum} and high-dimensional time-series analysis \citep{fan2023bridging} (e.g., exponential tails and strong mixing), collected as Conditions D.1--D.6 in the Supplementary Material.

\begin{theorem}\label{thm: largeNasymptotics}
(i) Under Assumptions \ref{asmp:consistency}, \ref{asmp:model1}, \ref{asmp: AFM}, \ref{asmp: largeNsparse}(a), \ref{asmp: largeNrrloss}, and regularity conditions D.1--D.6 in the Supplementary Material, as $N$, $T_0\to\infty$ with $T_0^{1/2}N^{-1}=o(1)$ and $(\log N)^2/T_0=o(1)$, we have $\hat{\bm{f}}-\bar{\bm{f}}_{\rm post}=o_p(1)$ and $\hat{\bm{\beta}}-\bar{\bm{\beta}}=\bar{\bm{\varepsilon}}_{\rm post}+o_p(1)$. If further $T-T_0\to\infty$, then $\hat{\bm{\beta}}-\bar{\bm{\beta}}=o_p(1)$.\\
(ii) Under the same Assumptions and regularity conditions with \ref{asmp: largeNsparse}(b) in place of \ref{asmp: largeNsparse}(a), as $N$, $T_*\to\infty$ with $T_0^{1/2}N^{-1}=o(1)$, $(\log N)^2/T_0=o(1)$, and $T_0/T\to\kappa\in [0,1]$, assuming $\bar{\bm{f}}_{\rm post}=\bm{\alpha}_1+o(1)$ for some $\bm{\alpha}_1$, we have $\hat{\bm{f}}-\bm{\alpha}_1=o_p(1)$ and for each fixed $i$, $T_*^{1/2}(\hat{\beta}_i-\bar{\beta}_i)\stackrel{d}{\to} \mathcal{N}(0,\omega_i^2(\kappa))$ for some positive finite $\omega_i^2(\kappa)$, where $\bar{\bm{\beta}} = (\bar{\beta}_1,\ldots,\bar{\beta}_N)^{\t}$.
\end{theorem}

Theorem \ref{thm: largeNasymptotics}(i) shows that with many control units and a long pre-intervention period, robust regression consistently recovers true coefficients $\bar{\bm{f}}_{\rm post}$, regardless of whether $T-T_0$ diverges.
Consequently, the estimation error for the average effects $\bar{\bm{\beta}}$ decomposes into a negligible term and the inherent error term $\bar{\bm{\varepsilon}}_{\rm post}$. 
The latter reflects finite-sample uncertainty intrinsic to synthetic control designs, arising from small numbers of treated units and/or limited post-intervention periods \citep{abadie2010synthetic,cattaneo2021prediction}.
When $T-T_0\to\infty$, the error vanishes and $\hat{\bm \beta}$ becomes consistent.
Theorem \ref{thm: largeNasymptotics}(ii) further establishes asymptotic normality when both $T_0$ and $T-T_0$ diverge, under the stronger sparsity requirement in Assumption \ref{asmp: largeNsparse}(b) and a mild stationary condition on post-intervention factors.
The asymptotic variance $\omega_i^2(\kappa)$ is provided in the Supplementary Material.

The above estimation strategy extends the method of \cite{wang2017confounder}, originally developed for estimating treatment effects across multiple outcomes. 
We adapt their approach to address interference in the panel data setting and show that the robust regression estimator in \eqref{eq: huberregavg} consistently recovers the true coefficient even when $T-T_0$ is fixed and $\bar{\bm{Y}}_{\rm post}$ contains nonvanishing errors, whereas \cite{wang2017confounder} require both $T_0$ and $T-T_0$ to diverge.
Additionally, we establish consistency under a substantially more general sparsity condition, while \cite{wang2017confounder} obtain consistency only under global $L_1$ sparsity.

\subsection{Statistical inference for the post-intervention period}\label{subs: largeNinfer} 
When $T-T_0$ is sufficiently large, the asymptotic normality results in Theorem~\ref{thm: largeNasymptotics}(ii) enable Wald-type inference. 
However, the corresponding asymptotic variance depends on complex dependence structures and is hard to estimate directly. 
We therefore draw statistical inference using the circular block bootstrap; see the Supplementary Material for details.

However, when $T-T_0$ is small, Theorem~\ref{thm: largeNasymptotics}(i) implies that $\hat{\bm{\beta}}$ contains a non-negligible error, which hinders valid asymptotic inference. 
To address this challenge, we adapt the conformal permutation framework of \cite{chernozhukov2021exact} to our interference setting. 
For a given unit $j$, we first consider testing the sharp null hypothesis 
\[H_0:\beta_{jt}=\beta_{jt}^0 \text{ for all } t>T_0. \]
Denote the set of moving block permutations by $\Pi_{\rightarrow} = \{\pi_j: j = 1,\ldots,T-1\}$ with elements defined by the mapping $
\pi_j(i) = 1 + \{(i + j - 1) \bmod T\}$, for $i = 1,\ldots,T$; see
\cite{chernozhukov2021exact} for details. 
The inferential procedure proceeds as follows:
\begin{itemize}
\item Step 1. Obtain data $\bm{W} = (\bm{W}_1,\ldots, \bm{W}_{T})$, where $\bm{W}_t = \bm{Y}_t$ for $t\leq T_0$ and $\bm{W}_{t} = (Y_{1t},\ldots,Y_{j-1,t}, Y_{jt}-\beta^0_{jt},Y_{j+1,t}\ldots, Y_{Nt})^{\t}$ for $T_0 < t\leq T$.
\item Step 2. Obtain $\hat{\bm{\Lambda}}$ and $\hat{\bm{\Sigma}} = \operatorname{diag}(\hat{\sigma}_1^2,\ldots, \hat{\sigma}_N^2)$ by factor analysis \citep{bai2016maximum} with data $\{\bm{W}_t\}_{t\leq T_0}$.
Obtain $\hat{\bm{f}}_t$ by generalized least squares and robust regression:
\begin{equation}\label{eq: huberregdyn}
    \hat{\bm{f}}_{t} = \left\{\begin{aligned} &\left(\hat{\bm{\Lambda}}^{\t}\hat{\bm{\Sigma}}^{-1}\hat{\bm{\Lambda}}\right)^{-1}\hat{\bm{\Lambda}}^{\t}\hat{\bm{\Sigma}}^{-1}\bm{W}_t,\quad &\text{ for } 1\leq t\leq T_0, \\
    &\arg\min_{\bm{f}} \sum_{i=1}^N\rho\left(\frac{W_{it} - \hat{\bm{\lambda}}_i^{\t}\bm{f}}{\hat{\sigma}_i}\right) ,\quad & \text{ for } T_0 < t\leq T.
    \end{aligned}
    \right.
\end{equation}
\item Step 3. Construct estimators of the proxy for the counterfactual outcomes as $\hat{P}_{jt} = \hat{\bm{\lambda}}_j^{\t}\hat{\bm{f}}_t$ and obtain residuals $\hat{u}_{jt}$ = $W_{jt} - \hat{P}_{jt}$ for $1\leq t\leq T$.
\item Step 4. Obtain $p$-value $\hat{p} = 1- |\Pi_{\to}|^{-1}\sum_{\pi\in\Pi_{\to}}\mathbbm{1}\{S(\hat{\bm{u}}_{j,\pi})<S(\hat{\bm{u}}_j)\}$, where $S(\hat{\bm{u}}_j) = (T-T_0)^{-1/2}\sum_{t=T_0+1}^T|{\hat{u}_{jt}}|$, and similarly for $S(\hat{\bm{u}}_{j,\pi})$ with permuted residuals $\hat{\bm{u}}_{j,\pi}$. 
\end{itemize}

In Step 1, the constructed outcomes $\bm{W}$ under $H_0$ preserve the pre-intervention factor structure and the post-intervention sparsity pattern of interference effects.
Accordingly, in Step 2, under the conditions of Theorem \ref{thm: largeNasymptotics}(i), the factor model is consistently estimable from the pre-intervention data, with pre-intervention factors recovered by generalized least squares \citep{bai2016maximum}.
The post-intervention factors $\bm{f}_{t}$, $t>T_0$, are consistently estimated via robust regression as a corollary of Theorem~\ref{thm: largeNasymptotics}(i), provided each $\bm{\beta}_t$ follows the same sparsity pattern as $\bar{\bm{\beta}}$ in Assumption~\ref{asmp: largeNsparse}(a); see Section A.2 of the Supplementary Material for details.
These steps yield a consistent estimator $\hat{P}_{jt}$ of the mean-unbiased proxy $P_{jt} = \bm{\lambda}_j^{\t}\bm{f}_t$ for the counterfactual outcome, enabling valid conformal inference with finite-sample coverage guarantees as established in \cite{chernozhukov2021exact}.

For any post-intervention time point $t>T_0$, the procedure extends directly to pointwise testing of $H_0: \beta_{jt} = \beta^0_{jt}$ by constructing $\bm{W} = (\bm{W}_1,\ldots, \bm{W}_{T_0},\bm{W}_{t})$.
Inverting these tests yields pointwise confidence intervals for dynamic effects.
Average-effect hypothesis $H_0: (T-T_0)^{-1}\sum_{t=T_0+1}^T\beta_{jt} = \beta^0_j$ can be tested by partitioning the pre-intervention period into non-overlapping blocks and applying the same procedure; see the Supplementary Material of \cite{chernozhukov2021exact}.
These permutation-based inferences remain valid for arbitrary $T-T_0$, which complement asymptotic-normality-based inference and are particularly useful when $N$ is large but $T-T_0$ is small.

\section{Simulation}\label{sec:simulation}

\subsection{Simulation for the fixed-$N$ setting}\label{subsec:simufix}

We first evaluate the proposed method under the fixed-$N$ setting, comparing it to related methods and assessing robustness to misspecification of the number of factors $r$ and violation of Assumption \ref{asmp:sparsity}.
We generate panel data with 
$N=10$ units and $T_0  = T/2\in\{100, 200\}$.
Outcomes follow the data-generating process with autoregressive random errors and align with the model in Section \ref{sec:basemodel}:
$
\bm{f}_t = {\bm{\alpha}}_1Z_t + {\bm{\alpha}}_0(1-Z_t) + \bm{\zeta}_t,\  \bm{Y}_{t} =  \bm{\beta}_{t}Z_t + \bm{\Lambda} \bm{f}_t + \bm{\ee}_{t}, \ \bm{\ee_t} = 0.2\bm{\ee}_{t-1} + 0.1\bm{\ee}_{t-2} + \bm{\nu}_t, 
$
where $\bm{\zeta}_t \sim N(\bm{0}, \mathbf{I}_2), \bm{\nu}_t \thicksim  N(\bm{0}, \mathbf{I}_N), {\bm{\alpha}}_0 = (0,0)^{\t}, {\bm{\alpha}}_1 = (1,1)^{\t},$ and
{\footnotesize
\[
\bm{\Lambda} = 0.5\begin{pmatrix} 1.6&-0.6&1&1&1&-2&3&-3&1.5&-1.5\\0.6&1.6&1&-1&2&1&1&1&1&1 \end{pmatrix}^{\t}.\]}
The effects $\bm{\beta}_t$ are set to gradually increase and then become periodic:
$\beta_{1t} = \{(t-T_0)/12\}  \mathbbm{1}(T_0 < t \leq T_0 + 12)  + \{4 + \sin(\pi t/12)\} \mathbbm{1}(t>T_0+12)$, $\beta_{it} = 0.75\beta_{1t}\mathbbm{1}(t>T_0)$ for $1< i \leq N_0$, and $\beta_{it}=0$ for $i>N_0$.
Here $N_0$ is the number of units affected by intervention, varying from 1 (no interference) to 2, 3 (moderate interference), and 4 (violation of Assumption \ref{asmp:sparsity}).
Each setting was replicated 1000 times with the following methods.
\begin{itemize}
    \setlength{\itemsep}{-12pt}
    \item[SCI$_1$]: proposed method in Section \ref{subsec: fixedNest} with the correct number of factors $(r=2)$;
    \item[SCI$_2$]: proposed method in Section \ref{subsec: fixedNest} with misspecified number of factors $(r=3)$;
    \item[SDID]: the synthetic difference-in-differences estimator of \cite{arkhangelsky2021synthetic};
    \item[GSC]: the generalized synthetic control method of \cite{xu2017generalized};
    %\item[MD]: the empirical mean difference $\bm{D}$;
    \item[SC]: synthetic control method of  \cite{abadie2010synthetic};
    \item[DID]: the difference-in-differences estimator.
\end{itemize}

Additional simulation results under alternative data-generating processes are provided in the Supplementary Material.
Figure \ref{fig:interfere1} summarizes the bias and mean squared error (MSE) of different methods for estimating the average direct effect $\bar{\beta}_{1}$ under $T_0  = T/2\in\{100, 200\}$, respectively.
The correctly specified proposed method (SCI$_1$) exhibits minimal bias and the lowest MSE. 
When the number of factors is misspecified, the proposed method (SCI$_2$) has slightly larger MSE but remains less biased than competitors, except when $N_0=4$ where Assumption \ref{asmp:sparsity} is violated.
We therefore recommend assessing the robustness by varying the number of factors in practice.
Similar patterns hold for estimating average interference effects, and we present them in the Supplementary Material.
Figure \ref{fig:interfere1} also suggests that the competing methods have non-ignorable biases in the presence of interference, with both biases and MSE increasing as $N_0$ grows, underscoring the importance of correctly selecting valid control units when applying synthetic control methods that do not account for interference. 
Even for $N_0=1$, where the set of valid controls is correctly specified, our estimator achieves smaller bias and variability than competing methods.

\begin{figure}[h]
    \captionsetup[subfloat]{captionskip=-1pt}
    \subfloat[$T_0=100, T=200$]{
  		 	\includegraphics[width=0.9\textwidth, height=0.3\textwidth]{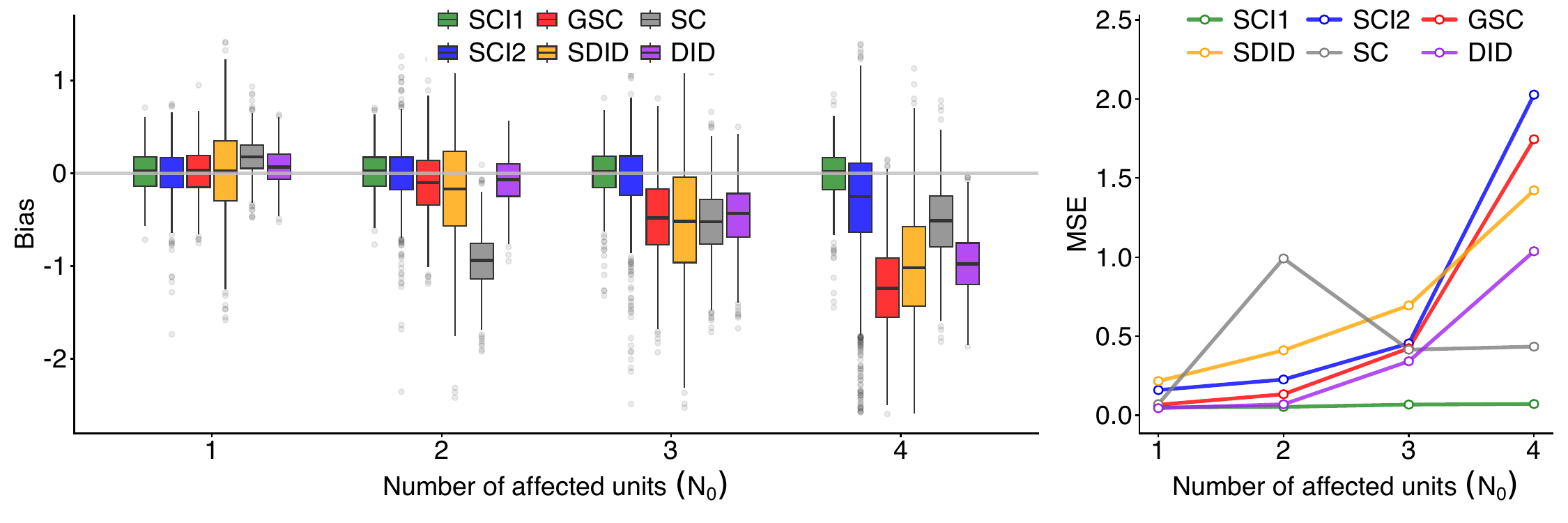} 
    	}\\
     \subfloat[$T_0=200, T=400$]{
  		 	\includegraphics[width=0.9\textwidth, height=0.3\textwidth]{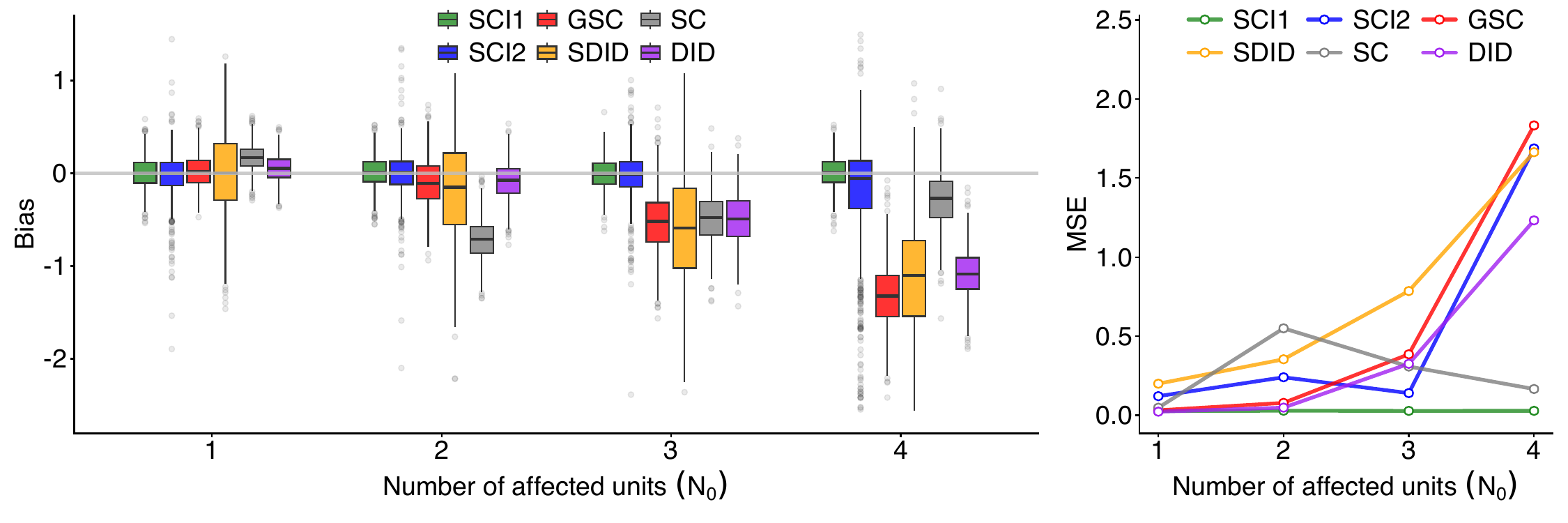}
    	}\\
         \captionsetup{skip=-5pt}
    \caption{Finite-sample performance of estimators for $\bar{\beta}_1$ under fixed-$N$ setting}
    \begin{minipage}{1\linewidth}
    \linespread{1.0}\selectfont 
    \footnotesize
    \textit{Note:} 
    Left panels show bias boxplots and right panels report MSE as varying the number of affected units $N_0=1,2,3,4$. 
    Panel (a) sets $T_0=100, T=200$ and (b) sets $T_0=200, T=400$. The proposed estimators (SCI1, SCI2) are compared against competing methods of GSC, SDID, SC, and DID.
\end{minipage}
 \label{fig:interfere1}
\end{figure}

Table \ref{tab:cvp} reports Monte Carlo coverage rates of 95\% 
confidence intervals for SCI$_1$ based on 500 circular block bootstrap 
replicates with block lengths $T^{1/4}, T^{1/3},$ and $T^{1/2}$. 
Empirical coverage is close to the nominal 95\% level for moderate $T$ 
across all block lengths.
\begin{table}[h]
\renewcommand{\arraystretch}{0.5} 
\caption{Monte Carlo coverage probabilities (\%) of the 95\% confidence intervals using the circular block bootstrap with different block lengths for $\bar{\beta}_{1}$ under fixed-$N$ setting}
\centering
\begin{tabular}[t]{ccccccccc}
\toprule
\multirow{2}{*}{Block length}& \multicolumn{4}{c}{$T_0 = T/2=100$} & \multicolumn{4}{c}{$T_0 =T/2=200$} \\
\cmidrule(lr){2-5} \cmidrule(lr){6-9}
 & $N_0=1$ & 2 & 3 & 4 & 1 & 2 & 3 & 4 \\
\midrule
$T^{1/4}$ & 93.8 & 94.8 & 96.3 & 97.3 & 93.2 & 94.3 & 95.2 & 95.1 \\
$T^{1/3}$ & 94.1 & 95.2 & 96.7 & 97.6 & 93.5 & 94.5 & 95.4 & 94.9 \\
$T^{1/2}$ & 92.6 & 94.2 & 95.7 & 96.7 & 92.4 & 93.8 & 94.5 & 94.1 \\
\bottomrule
\end{tabular}
\label{tab:cvp}
\end{table}

\subsection{Simulation for the large-$N$ setting}
For the large-$N$ setting, we similarly evaluate the finite-sample performance of the proposed method.
Outcomes are generated from the same model as in Section~\ref{subsec:simufix}, 
except that $\bm{\Lambda}$ is the orthonormal QR factor of an $N\times r$ matrix with i.i.d. standard Gaussian entries, scaled by $\sqrt{N}/4$ so that $\bm{\Lambda}^\top\bm{\Lambda}/N=\bm{I}_r/16$ for all $N$.
The effects are set as periodic: 
$\beta_{1t} = \{3 + \sin(\pi t/12)\} \mathbbm{1}(t>T_0),$ $\beta_{it} = 0.75\beta_{1t}\mathbbm{1}(t>T_0) \text{, for } 1< i \leq N_0 \text{ and }  \beta_{it}=10/(N-N_0) \text{ for } i>N_0$, where $N_0=20$ units exhibit large interference effects and all remaining units have weak interference effects, aligned with Assumption \ref{asmp: largeNsparse}. 
We vary $T_0 = T/2 \in \{100, 200\}$ and $N \in \{100, 200, 400, 600\}$. 
Each setting was replicated 1000 times.

\begin{figure}[h]

    \captionsetup[subfloat]{captionskip=-1pt}
    \subfloat[$T_0=100, T=200$]{
  		 	\includegraphics[width=0.9\textwidth, height=0.3\textwidth]{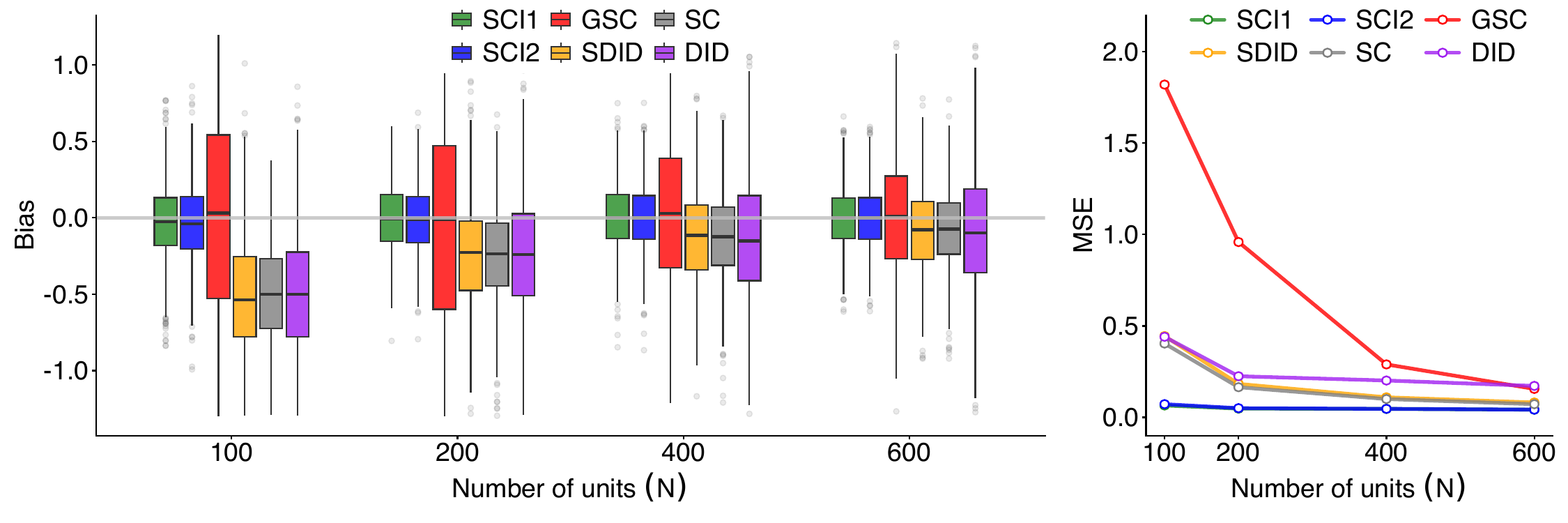}} \\
     \subfloat[$T_0=200, T=400$]{
  		 	\includegraphics[width=0.9\textwidth, height=0.3\textwidth]{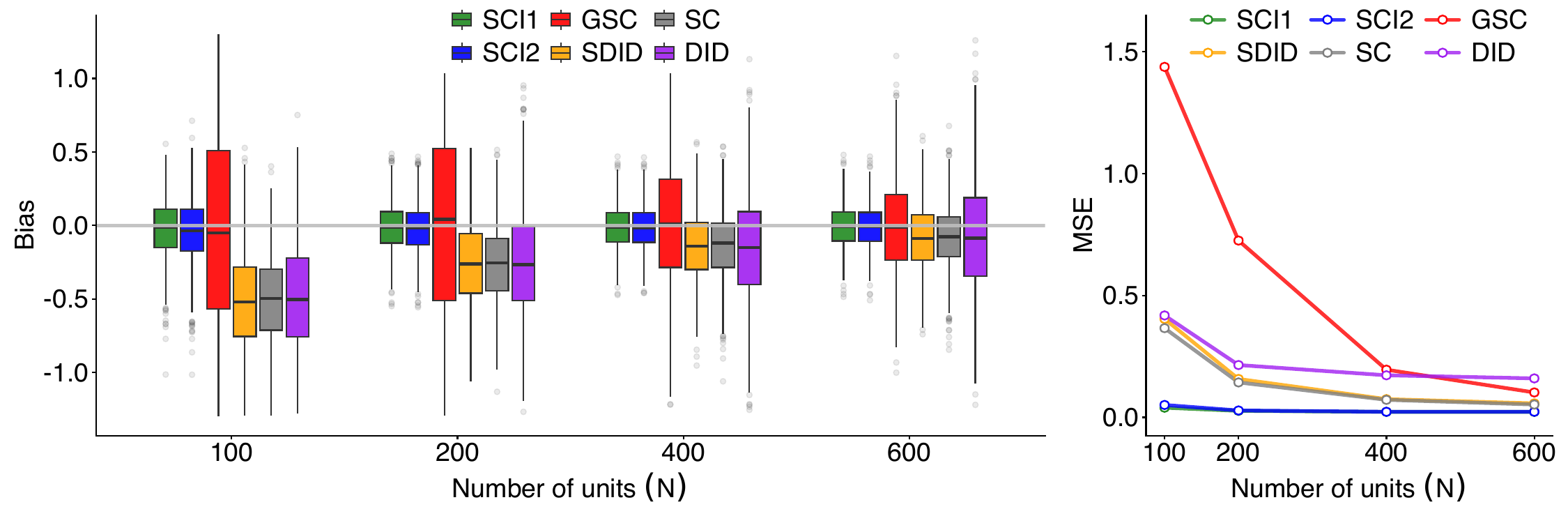}
    	}
    \caption{Finite-sample performance of estimators for $\bar{\beta}_1$ under large-$N$ setting}
    \begin{minipage}{1\linewidth}
    \linespread{1.0}\selectfont 
    \footnotesize
    \textit{Note:} 
    Left panels show bias boxplots and right panels report MSE as varying the number of units $N=100,200,400,600$. 
    Panel (a) sets $T_0=100, T=200$ and (b) sets $T_0=200, T=400$. The proposed estimators (SCI1, SCI2) are compared against competing methods of GSC, SDID, SC, and DID.
\end{minipage}
 \label{fig:interfere2}
\end{figure}

Here, SCI$_1$ and SCI$_2$ denote the proposed large-$N$ method in Section \ref{subs: largeNest} with a correct and misspecified number of factors, respectively.
Huber-type and log-cosh losses yield nearly identical results, so we report Huber results only.
Figure~\ref{fig:interfere2} compares the bias and MSE of different methods for estimating the average direct effect
$\bar{\beta}_{1}$ under different panel sizes.
SCI$_1$ achieves the smallest bias and MSE across all scenarios, while SCI$_2$ has slightly larger bias and MSE but still outperforms other methods. In contrast, the biases of SC, DID, and SDID remain non-negligible across all settings. 
As $N$ increases, interference contamination is diluted, reducing their bias and MSE, yet they still perform substantially worse than the proposed method. 
GSC exhibits significantly larger MSE in all settings; values exceeding the y-axis range are not displayed in the right panel of Figure~\ref{fig:interfere2}.

Table~\ref{tab:cvp2} reports Monte Carlo coverage rates of nominal 95\% circular block bootstrap confidence intervals for the average direct effect, using the proposed method SCI$_1$ and 500 bootstrap replications.
Coverage rates are close to the nominal level once the panel is sufficiently large.
When $T_0 =T/2 = 100$, all three block-length choices perform well, with mild over-coverage as $N$ grows.
When $T_0 = T/2 = 200$, shorter block lengths ($T^{1/3}$ and $T^{1/4}$) maintain satisfactory coverage, while $T^{1/2}$ exhibits some undercoverage, particularly for smaller $N$.
In both cases, coverage improves as $N$ increases, reflecting the importance of large $N$ and consistent with theoretical intuition.
Similar patterns hold for average interference effects $\bar{\beta}_i$ with $i>1$, see the Supplementary Material.
We also include simulations under alternative data-generating processes in the Supplementary Material.

\begin{table}[H]
\renewcommand{\arraystretch}{0.5} 
\caption{Monte Carlo coverage probabilities (\%) of the 95\% confidence intervals using the circular block bootstrap with different block lengths for $\bar{\beta}_{1}$ under large-$N$ setting}
\centering
\begin{tabular}[t]{ccccccccc}
\toprule
\multirow{2}{*}{Block length}& \multicolumn{4}{c}{$T_0 = T/2=100$} & \multicolumn{4}{c}{$T_0 =T/2=200$} \\
\cmidrule(lr){2-5} \cmidrule(lr){6-9}
 & $N=100$ & 200 & 400 & 600 & 100 & 200 & 400 & 600 \\
\midrule
$T^{1/4}$ & 93.6 & 93.4 & 94.2 & 94.7 & 92.3 & 95.8 & 95.7 & 96.3 \\
$T^{1/3}$ & 94.2 & 94.8 & 96.0 & 96.0 & 93.4 & 96.0 & 96.9 & 97.1 \\
$T^{1/2}$ & 92.4 & 93.8 & 94.9 & 95.5 & 87.2 & 90.6 & 91.5 & 92.1 \\
\bottomrule
\end{tabular}
\label{tab:cvp2}
\end{table}

Beyond asymptotic inference, Table~\ref{tab:perm} illustrates the nominal 5\% pointwise conformal permutation test for the first post-intervention direct effect, $\beta_{1,T_0+1}$, under three null hypotheses:
the true effect $H_0: \beta_{1,T_0+1} = 3 + \sin\{\pi (T_0+1)/12\}$ (True); a misspecified effect $H_0: \beta_{1,T_0+1} = 2$ (Mis); and no effect $H_0: \beta_{1, T_0+1} = 0$ (Null).
Each cell reports the empirical rejection rate over 5,000 replications.
Under True, rejection rates slightly exceed 5\%, indicating mild finite-sample over-rejection in the presence of interference.
Under Mis, they are substantial, reflecting growing power to detect the misspecification.
Under Null, they range from 93.9\% to 96.9\%, demonstrating high power.

\begin{table}[h]
\renewcommand{\arraystretch}{0.5} 
\centering
\caption{Empirical rejection rates (\%) of the nominal 5\% conformal permutation test for $\beta_{1,T_0+1}$ under three null hypotheses in the large-$N$ setting}
\begin{tabular}{lcccccccc}
\toprule
\multirow{2}{*}{$H_0$} &  \multicolumn{4}{c}{$T_0 = T/2=100$} & \multicolumn{4}{c}{$T_0 =T/2=200$} \\
\cmidrule(lr){2-5} \cmidrule(lr){6-9}
 & $N=100$ & 200 & 400 & 600 & $N=100$ & 200 & 400 & 600\\
\midrule
True  &  5.60 & 5.30 & 5.54 & 6.66  &  5.64 & 5.48 & 5.66 & 5.56 \\
Mis  &  46.26 & 47.36 & 45.82 & 47.62  &  38.30 & 37.16 & 36.44 & 38.42 \\
Null  &  95.62 & 96.44 & 96.88 & 96.38  &  93.88 & 94.62 & 94.50 & 94.28 \\
\bottomrule
\end{tabular}
\label{tab:perm}
\end{table}

\section{Applications} \label{sec:application}
\subsection{The US embassy relocation policy}
We apply the proposed fixed-$N$ method to evaluate the average direct and interference effects of the US embassy relocation from Tel Aviv to Jerusalem on conflicts in the Middle East.
On December 6, 2017, the US government announced the move of the embassy in Israel, sparking considerable global debate about its impact on regional stability.
We use data from the Armed Conflict Location and Event Data Project \citep{raleigh2010introducing}, with weekly country-level conflict counts from December 28, 2015 to November 3, 2018 as the outcome.
Treating the announcement as the intervention yields 101 pre-intervention and 48 post-intervention weeks. 
We exclude weeks 29--31 due to a military coup that caused a notable spike in conflicts in Turkey \citep{muhlbach2021tree}, leaving 98 pre-intervention weeks. 
Nine units (country or region) are included in our analysis: Israel-Palestine (treated), Bahrain, Iran, Iraq, Jordan, Lebanon, Saudi Arabia, Turkey, and Yemen. Israel and Palestine are combined as one unit since some conflicts occurred on their border. No other covariates are available at a weekly frequency.

\begin{figure}[h]
\centering
  		 \includegraphics[width=0.65\textwidth, height=0.43\textwidth]{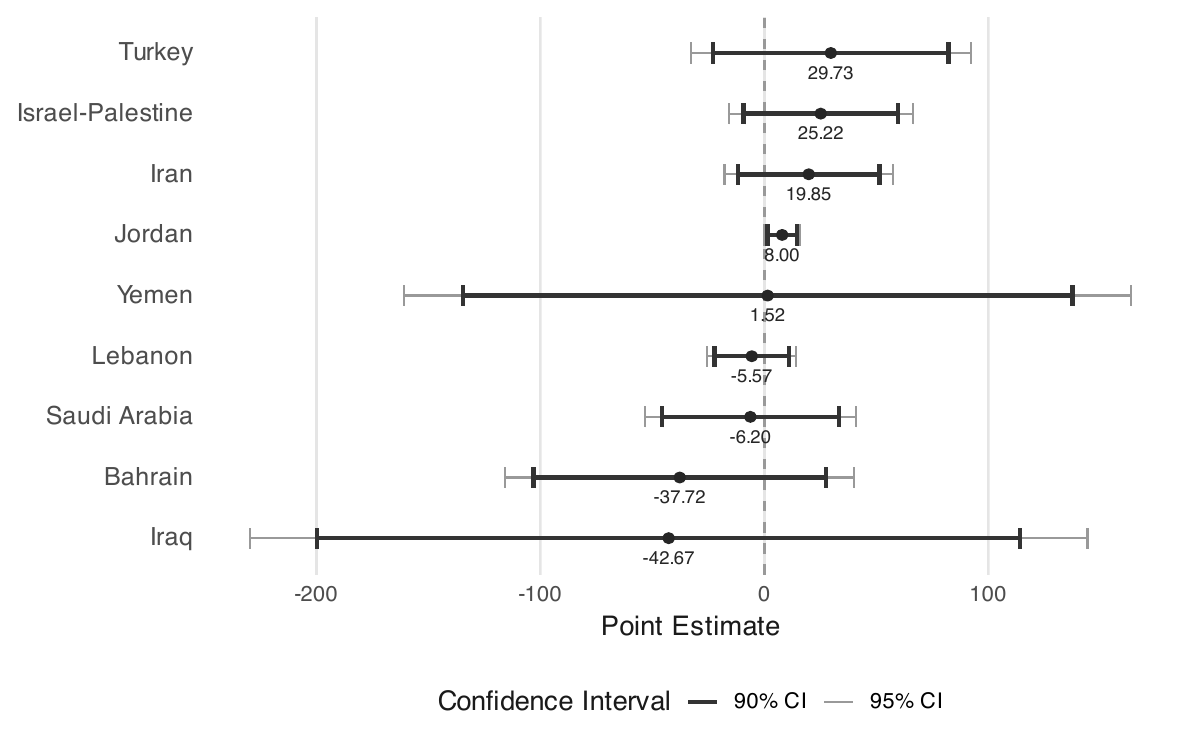} 
	\caption{Estimated average direct and interference effects on weekly conflict numbers.} 
    \begin{minipage}{1\linewidth}
    \linespread{1.0}\selectfont 
    \footnotesize
    \textit{Note:} 
    Dots indicate point estimates and bars indicate 90\% and 95\% confidence intervals. Israel–Palestine is the treated unit (average direct effect); remaining units correspond to average interference effects.
\end{minipage}
 \label{fig:appbase}
\end{figure}

Previously,  \cite{muhlbach2021tree} analyzed this dataset using a variant of the synthetic control method to estimate the effect on conflicts in Israel-Palestine, excluding Iran from the donor pool due to potential interference from US-Iran relations.
Their analysis did not consider interference effects on other Middle Eastern countries.
However, given the geographical, religious, and political complexities of the region, the embassy relocation may plausibly affect other Middle Eastern countries to produce interference effects.
We apply our method to estimate the direct effect on Israel-Palestine and potential interference effects on other countries, where we re-include Iran to examine possible interference. 
We fit the proposed method with two factors ($r=2$), assuming no more than three units affected by the intervention with Assumption \ref{asmp:sparsity}.
The block length for bootstrap inference is set to four weeks.
The results are summarized in Figure \ref{fig:appbase}.

Our analysis suggests that the US embassy relocation increased conflicts in Israel-Palestine by an average of 25.22 per week, closely aligned with the estimate of 26.10 reported by \cite{muhlbach2021tree}.
However, the confidence interval (90\% CI: $-$9.24, 59.68) does not indicate statistical significance, possibly because other long-standing factors driving conflicts in this region overwhelm the impact of the relocation policy. 
Among other countries, we detect a significant positive interference effect on Jordan, with an estimated average increase of 8.00 conflicts per week (95\% CI: 0.12, 15.88). 
Interference effects on other countries are not statistically significant at the 0.05 level.
In particular, the estimated average interference effect on Iran is 19.85 with a $p$-value of 0.301.
These findings suggest that Jordan should be excluded from the donor pool, highlighting the challenge of donor pool selection for researchers applying synthetic control methods.

\subsection{Beijing’s air pollution control policy}
To illustrate the proposed large-$N$ method, we study the effect of Beijing’s orange alert policy on air pollution control.
This example also demonstrates the applicability of our approach to settings with multiple treated units.
Beijing's color-coded pollution alerts, in use since 2013, trigger emergency measures intended to reduce emissions and improve air quality.  
We evaluate the effects of one orange alert, issued on November 17, 2016, on the reduction of PM$_{2.5}$ concentration.
The dataset, previously analyzed by \cite{zheng2024dynamic}, comprises hourly PM$_{2.5}$ measurements from $N=108$ air-quality stations in and around Beijing, with 48 pre-intervention and 24 post-intervention hours.
The alert targeted Beijing's core urban area, encompassing 20 stations as treated units. 
The remaining stations include 74 in neighboring provinces, where no alert was issued during the study period, and 14 in Beijing's suburban areas.
Interference effects on these stations are plausible due to pollutant transmission through the meteorological system.
In their analysis, \cite{zheng2024dynamic} excluded these 14 stations from the donor pool due to potential interference.

\begin{figure}[h]
\centering
  		 \includegraphics[width=0.9\textwidth, height=0.45\textwidth]{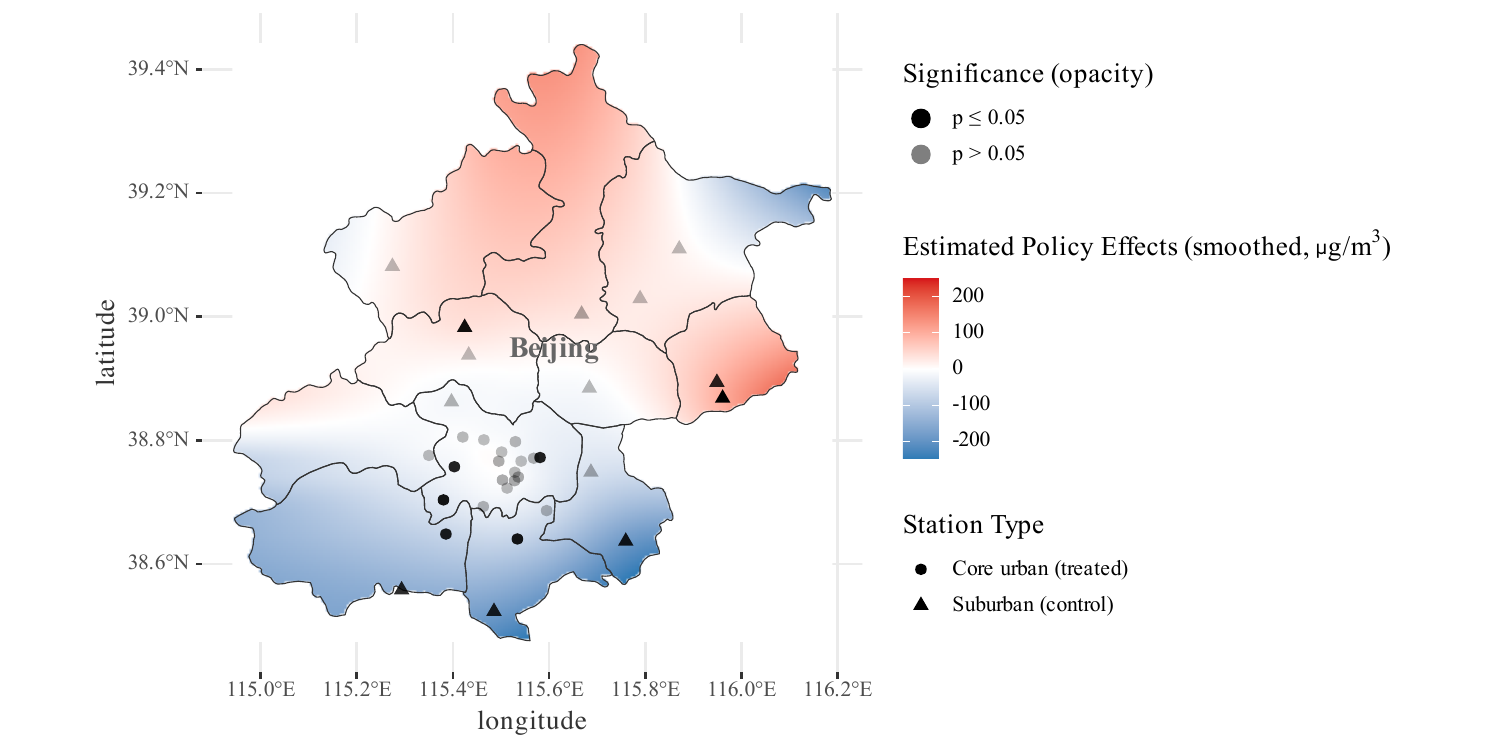} 
	\caption{Spatial pattern of estimated   effects on PM$_{2.5}$ reduction in Beijing.}
    \begin{minipage}{1\linewidth}
    \linespread{1.0}\selectfont 
    \footnotesize
    \textit{Note:} The map displays estimated average effects on PM$_{2.5}$ concentration ($\mu\mathrm{g}/\mathrm{m}^3$) for 20 core urban (treated) and 14 suburban (control) air-quality stations, where the station-level point estimates are smoothed over the Beijing area. 
    Circles (triangles) indicate the locations of core urban (suburban) stations.
    Solid points indicate statistical significance at the 5\% level.
\end{minipage}
 \label{fig:beijingmap}
\end{figure}
We determine the number of latent factors as $r=7$ using the information criteria of \cite{bai2002determining}, and estimate average direct and interference effects using the large-$N$ method.
Confidence intervals are constructed via the circular block bootstrap with block size $72^{1/2}\approx 9$.
Five of the 20 treated 
stations exhibit statistically significant PM$_{2.5}$ reductions attributed to the policy, with 
estimated average decreases of 73.06 (SE: 24.96), 88.43 (SE: 33.33), 
54.46 (SE: 21.51), 29.30 (SE: 9.63), and 53.42 (SE: 22.23) 
$\mu\mathrm{g}/\mathrm{m}^3$. Averaged across treated units, the 
estimated reduction in the core urban area is 15.80 
$\mu\mathrm{g}/\mathrm{m}^3$, close to \cite{zheng2024dynamic}. 
Three suburban Beijing stations also show statistically significant negative 
interference effects, with estimated average reductions of 177.25 (SE: 
80.66), 214.27 (SE: 82.04), and 147.38 (SE: 75.08) 
$\mu\mathrm{g}/\mathrm{m}^3$. 
Spatially, the treated stations with significant effects lie in the southern core, and the suburban stations with significant reductions are all located to the south; see Figure \ref{fig:beijingmap} for a map illustration.
This spatial pattern, combined with the prevailing northerly winds during the post-intervention period, provides a plausible mechanism for the interference effects.
A few suburban stations 
display significant increases, possibly reflecting atmospheric 
complexity and limited power to separate local effects from a modest 
overall PM$_{2.5}$ trend during the study period.

\section{Discussion}\label{sec:discussion}
We develop a robust regression perspective for synthetic control with interference. 
Under certain sparsity conditions on interference effects, we establish consistent and asymptotically normal estimation of the average direct and interference effects, complementing comparative case studies with potential interference features. 
The proposed approaches require only coarse information on the extent of interference, rather than a fully specified interference model or prior identification of valid control units, although the sparsity conditions are not directly testable from the observed data. 
Our framework covers two complementary regimes: a fixed-$N$ setting with long pre- and post-intervention periods, and a large-$N$ setting with a long pre-intervention period and an arbitrary post-intervention period.
The US embassy relocation and Beijing air pollution control applications illustrate the framework's wide applicability.
We focus on a practically relevant case of interference, comparing the 
realized and no-intervention regimes and accommodating general 
interventions.
Extensions to richer policy regimes (e.g., staggered adoption) are possible by indexing potential outcomes by regime and comparing each to the no-intervention regime, provided each regime offers sufficiently long post-intervention periods for estimation. 
Our framework relies crucially on the stability of the 
factor loadings (or their column space) throughout the study period.
The fixed-$N$ setting with fixed $T - T_0$ remains challenging: latent factors are generally not consistently estimable \citep{bai2003inferential}, complicating the construction of counterfactual proxies without additional structure. We leave these extensions to future work.

\section*{Supplementary material}
The Supplementary Material contains extensions on dynamic effects, regularity conditions and proofs for both the fixed-$N$ and large-$N$ results; details of the circular block bootstrap; additional simulations; and supporting information for the applications.

\section*{Acknowledgments}
We are grateful for the valuable comments from the editor, associate editor, and two anonymous reviewers.
We thank Dr. Xiangyu Zheng at Tsinghua University for providing the dataset on Beijing’s air pollution control policy.

\putbib
\end{bibunit}

%%%%%%%%%%%%%%%%%%%%%%%%%%%%%%%%%%%%%%%%%%%%%%%%%%%%%%%%%%%%%%%%%%%%%%%%%%%%%%
% Supplementary Material
%%%%%%%%%%%%%%%%%%%%%%%%%%%%%%%%%%%%%%%%%%%%%%%%%%%%%%%%%%%%%%%%%%%%%%%%%%%%%%
\newpage
\setcounter{equation}{0}
\setcounter{section}{0}
\setcounter{figure}{0}
\setcounter{table}{0}
\setcounter{example}{0}
\setcounter{proposition}{0}
\setcounter{corollary}{0}
\setcounter{theorem}{0}
\setcounter{condition}{0}
\setcounter{assumption}{0}
\setcounter{lemma}{0}
\setcounter{remark}{0}

\newtheorem{thm}{\sc Theorem S.\ignorespaces}
\newtheorem{lem}{\sc Lemma S.\ignorespaces}
\newtheorem{cond}{\sc Condition \thesection.\ignorespaces}
\theoremstyle{remark}
\newtheorem{asmp}{Assumption A.\ignorespaces}
\newtheorem{exam}{\sc Example S.\ignorespaces}
\newtheorem{coro}{\sc Corollary S.\ignorespaces}
\makeatletter
\renewcommand{\thesection}{\@Alph\c@section}
\renewcommand\thetable{\@Alph\c@section\@arabic\c@table}
\renewcommand\thefigure{\@Alph\c@section\@arabic\c@figure}
\renewcommand{\theequation}{S.\arabic{equation}}
% unique hyperref anchors for the supplement
\renewcommand{\theHsection}{S\thesection}
\renewcommand{\theHtable}{S\thetable}
\renewcommand{\theHfigure}{S\thefigure}
\renewcommand{\theHequation}{\theequation}
\makeatother
\counterwithin{condition}{section}
\counterwithin{assumption}{section}
\def\iii{\vert\!\vert\!\vert}
\def\pr{{P}}

\setcounter{page}{1}
\begin{center}
    {\LARGE Supplementary Material for \\[0.2em]
    \bfseries A robust regression approach to synthetic control with interference}
    \vspace{0.5cm}

    {\large Peiyu He$^{1,\star}$, Yilin Li$^{2,\star}$, Xu Shi$^3$, and Wang Miao$^2$}
    \vspace{0.3cm}

    {Center for Data Science, Peking University$^1$ \\
    Department of Probability and Statistics, Peking University$^2$\\
    Department of Biostatistics, University of Michigan$^3$}
\end{center}
\vspace{0.5cm}

The Supplementary Material contains extensions on dynamic effects, regularity conditions and proofs for both the fixed-$N$ and large-$N$ results; details of the circular block bootstrap; additional simulations; and supporting information for the applications.

\begin{bibunit}
\section{Estimating dynamic effects}\label{sec:dynamic}
\subsection{Fixed-$N$ setting}
In general, consistent estimation of the dynamic causal effects $\bm{\beta}_t$ is impossible without additional restrictions on temporal trends, because only a single post-intervention realization $\bm{Y}_t$ is available to estimate $\bm{\beta}_t$ at each time point $t$. 
To identify $\bm{\beta}_t$, one needs to impose certain smoothness conditions on the data-generating process that permit pooling information across time. 
Our approach treats $\{\bm{Y}_t\}_{t>T_0}$ as a time series with a deterministic trend $\bm{\mu}_t=\bm{\beta}_t+\bm{\Lambda}\bm{f}_t$ and weakly dependent error $\bm{\varepsilon}_t$. 
Under standard smoothness conditions, $\bm{\mu}_t$ can be consistently estimated using established time-series trend estimation methods. 
Conditioning on this trend estimate, recovering $\bm{\beta}_t$ reduces to an exact-fit robust regression problem analogous to the average effects estimation in Section 3. 
This idea of integrating both pre- and post-intervention data to estimate the time series trend of the outcomes under certain smoothness conditions has been used in synthetic control in parallel to that of \cite{shi2023theory}.
We next formalize the required conditions and, under these conditions, extend the method in Section 3 to obtain a consistent estimator of $\bm{\beta}_t$ for each $t>T_0$.

\begin{assumption}\label{asmp:stationarywdynamic}
(i) $T_0^{-1}\sum_{t=1}^{T_0}\bm{f}_t=\bm{\alpha}_0+o(1)$; (ii) $\operatorname{Var}(\bm{\varepsilon}_t) = \bm{\Sigma} = \operatorname{diag}(\sigma_1^2, \dots, \sigma_N^2)$ and $\bm{M}_f = \mathbf{I}_r$; (iii) $\bm{\Lambda}^{\t}\bm{\Sigma}^{-1}\bm{\Lambda}$ 
is a positive diagonal matrix with distinct diagonal entries arranged in decreasing order.
\end{assumption}

\begin{assumption}\label{asmp:dynamicnull}
    Let $\mathcal{C}_t = \{i : \beta_{it} = 0\}$. We assume $|\mathcal{C}_t|\geq \lfloor N/2\rfloor+r$ for all $t>T_0$. 
\end{assumption}

\begin{assumption}\label{asmp:trend}
There exists an estimator $\hat{\bm{\mu}}_t$ such that $\hat{\bm{\mu}}_t -  \bm{\mu}_t =O_p\{(T-T_0)^{-d}\}$, where $0<d\leq 1/2$ is some constant that depends on the modeling and estimating strategy.

\end{assumption}

Assumptions \ref{asmp:stationarywdynamic} and \ref{asmp:dynamicnull}  are parallel to Assumptions \ref{asmp:stationary} and \ref{asmp:sparsity}.
In contrast to  Assumption \ref{asmp:stationary},  the mean stationary condition for post-intervention factors is no longer required in Assumption \ref{asmp:stationarywdynamic}.
Assumption \ref{asmp:dynamicnull} states that, at each post-intervention time point, at least $\lfloor N/2 \rfloor + r$ units are valid controls, and we no longer consider the average effects as in Assumption \ref{asmp:sparsity}. 
Assumption \ref{asmp:trend} is a high-level condition stating that the deterministic trend $\bm{\mu}_t$ can be consistently estimated with certain convergence rates using post-intervention data.

As noted above, primitive smoothness conditions on the deterministic trend  $\bm \mu_t$ are indispensable for  Assumption \ref{asmp:trend} to hold; such conditions are readily available in the  time-series analysis literature, 
including the fact that  $(T-T_0)^{1/2}$-consistent estimation can be achieved  under  parametric modeling of $\bm {\mu}_t$.

\begin{exam}[Parametric model]\label{ex:p}
Suppose $\bm{\mu}_t = \bm{\mu}(t/T;\bm{\gamma})$ for some finite-dimensional parameter $\bm{\gamma}$; then  $\bm{\gamma}$ can be estimated with standard methods such as generalized estimating equations with time series dependence under mild conditions \citep{hamilton2020time}. 
The resulting estimator satisfies $\hat{\bm{\mu}}_t - \bm{\mu}_t = O_p\{(T-T_0)^{-1/2}\}$.
\end{exam}

In the nonparametric model for $\bm{\mu}_t$, the following result established by \cite{dong2018additive} describes how well it can be estimated.
\begin{exam}[Nonparametric model]\label{ex:np}
Consider the nonparametric trend model $\bm{\mu}_{t}=\bm{\mu}\{(t-T_0)/(T-T_0)\}$ for $t=T_0+1,\ldots,T$.
For each index $1\leq i\leq N$, the corresponding component function $\mu_{i}(\cdot)$ is assumed to belong to the Sobolev-type class on $[0,1]$: $\mu_{i}(\cdot)$ is continuously differentiable up to order $s$, and its $s$th derivative $\mu_{i}^{(s)}(\cdot)$ lies in the Hilbert space $L^2[0,1]=\{u:\int_0^1 u^2(r)\,dr<\infty\}$ equipped with inner product $\langle u_1,u_2\rangle=\int_0^1 u_1(r)u_2(r)\,dr$.
Applying the nonparametric sieve method to estimate $\bm{\mu}(\cdot)$,
with  the   number of sieve basis functions   $k$ satisfying $k(T-T_0)^{-1} = o(1)$ and $k^{-2s+1}(T-T_0) = o(1)$, the resultant estimator satisfies $\hat{\bm{\mu}}_{t}-\bm{\mu}_{t} = O_p\{k^{1/2}/(T-T_0)^{1/2}\} = O_p\{(T-T_0)^{-(s-1)/(2s-1)}\}$ for $t > T_0$. 
\end{exam}

More generally, the estimation of $\bm {\mu}_t$ can be formalized as a detrending problem,
which opens the way to leverage the rich literature on time-series modeling and estimation to facilitate the construction of synthetic controls.
In the following, we start with a trend estimator such that $\hat{\bm{\mu}}_t -  \bm{\mu}_t =O_p\{(T-T_0)^{-d}\}$ for each $t>T_0$,
and describe the estimation procedure for the dynamic effects  $\bm{\beta}_t$ based on the hard-thresholding that is parallel to Section 3. 
We denote $r_{it}({\bm{\alpha}}_t)= (\hat{\mu}_{it} - \hat{\bm{\lambda}}_i^{\t}\bm{\alpha}_t)^2$ and $\hat{\bm{V}} =  T_0^{-1}\sum_{t=1}^{T_0} (\bm{Y}_t- \bar{\bm{Y}})(\bm{Y}_t- \bar{\bm{Y}})^{\t}$ here with a slight abuse of notation.

\begin{itemize}
\item Step 1. Obtain an estimator of  time trend   such that $\hat{\bm{\mu}}_t -  \bm{\mu}_t =O_p\{(T-T_0)^{-d}\}$ for each $t>T_0$ with post-intervention data $\{\bm{Y}_t\}_{t>T_0}$.
\item Step 2. Obtain $\hat{\bm{\Lambda}}$ as the estimated factor loading by factor analysis with pre-intervention data $\{\bm{Y}_t\}_{t\leq T_0}$.
\item Step 3. For each $t>T_0$, given $\hat{\bm{\mu}}_t$ and $\hat{\bm{\Lambda}}$, solve the following robust linear regression problem to obtain $\hat{\bm{\alpha}}_t$,
\begin{equation}\label{eq:ltst}
    \hat{{\bm{\alpha}}}_t = \mathop{\arg\min_{\bm{\alpha}}} \sum_{i=1}^{\lfloor N/2\rfloor+1} r_{(i),t}({\bm{\alpha}}),
\end{equation}
where $r_{(i),t}({\bm{\alpha}})$ is the $i$th smallest value among $\{r_{it}({\bm{\alpha}}) = (\hat{\mu}_{it} - \bm{\hat\lambda}_i^{\t}{\bm{\alpha}})^2; 1\leq i\leq N\}$.
\item Step 4. For each $t> T_0$,  obtain $\hat{\bm{\beta}}_t = \hat{\bm{\mu}}_{t} - \hat{\bm{\Lambda}}\hat{\bm{\alpha}}_t$ and $\hat{\mathcal{C}}_t = \{i:|\hat{\mu}_{it} - \hat{\bm{\lambda}}_i^{\t}\hat{\bm{\alpha}}_t|\leq T_*^{-d}\{2\log(NT_*)\}^{1/2}\hat\phi\}$ as the estimated set of units with no interference, where $\hat{\phi}^2 = N^{-1}\operatorname{tr}(\hat{\bm{V}})$, and $\hat{\bm{V}} = T_*T_0^{-1}(T-T_0)^{-1}\{T/T_0\sum_{t=1}^{T_0} (\bm{Y}_t- \bar{\bm{Y}}_{\rm pre})(\bm{Y}_t- \bar{\bm{Y}}_{\rm pre})^{\t}\}.$
\end{itemize}

\begin{thm}\label{thm:fixedNdynamic}
Under Assumptions \ref{asmp:consistency}, \ref{asmp:model1},  \ref{asmp:stationarywdynamic}--\ref{asmp:trend} and regularity conditions \ref{asmp:techfactor}--\ref{asmp: fixNdep}, as $T_*\to\infty$, we have $\hat{\bm{\beta}}_t - \bm{\beta}_t = O_p(T_*^{-d})$ and $\pr(\hat{\mathcal{C}_t} = \mathcal{C}_t)\to1$ for each $t>T_0$.
\end{thm} 

The convergence rate of $\hat{\bm{\beta}}_t$ is related to  the convergence rates of the trend estimator $\hat{\bm{\mu}}_t$,
for example, $d = 1/2$ in Example S.\ref{ex:p} and $d = (s-1)/(2s-1)$ in Example S.\ref{ex:np}. In addition, selection consistency is achieved by hard-thresholding as in Section 3 of the main text.

As a concluding remark, estimating dynamic effects is inherently more challenging than estimating average post-intervention effects, because each $t>T_0$ provides only a single realization $\bm{Y}_t$ for learning $\bm{\beta}_t$ without further assumptions.
Accordingly, our framework imposes restrictions on the post-intervention trend to enable pooling across time, either through parametric specifications or through nonparametric modeling under certain smoothness conditions. 
This entails a transparent trade-off: parametric trends may be misspecified, whereas nonparametric approaches typically converge more slowly and therefore require a longer post-intervention span $T-T_0$ for reliable estimation. 
This limitation reflects the intrinsic difficulty of the problem and should be assessed carefully in applications. 
In practice, we recommend examining the fit of the estimated post-intervention trend as a diagnostic before further analysis.

\subsection{Large-$N$ setting}
As noted in Section~4 of the main text, robust regression can remain consistent even when the post-intervention period is short. 
A direct implication is that we can estimate the dynamic direct and interference effects by replacing the post-intervention period with a single post-intervention time point $t>T_0$.
Specifically, analogous to \eqref{eq: rridenlargeN}, the decomposition
\[\bm{Y}_t = \bm{\beta}_t + \bm{\Lambda}\bm{f}_t + \bm{\varepsilon}_t\]
can be viewed as a robust linear regression in which $\bm{f}_t$ plays the role of the latent regression coefficient and $\bm{\beta}_t$ captures sparse interference effects at time $t$. We therefore impose a time-specific sparsity condition on $\bm{\beta}_t$ parallel to  Assumption \ref{asmp: largeNsparse}.
\begin{assumption}\label{asmp:largeNdynamicSparse}
   For each $t>T_0$, there exists some set $\mathcal{S}\subset\{1,\ldots,N\}$ such that $|\mathcal{S}| = o(N)$ and $\|\bm{\beta}_{t,\mathcal{S}^c}\|_{1} = o(N)$ as $N\to\infty$, where $\mathcal{S}^c = \{1,\ldots, N\}\backslash \mathcal{S}$ and $\bm{\beta}_{t,\mathcal{S}^c}$ is the subvector of $\bm{\beta}_t$ with index $\mathcal{S}^c$.
\end{assumption}
In parallel with Section~4, for each $t>T_0$ we first obtain $\hat{\bm{\Lambda}}$ and $\hat{\bm{\Sigma}}$ from pre-intervention factor analysis, and then estimate the post-intervention factors via robust regression:
\begin{equation}\label{eq: rrobjlargeN}
    \hat{\bm{f}}_t = \arg\min_{\bm{f}} \sum_{i=1}^N\rho\left(\frac{Y_{it} - \hat{\bm{\lambda}}_i^{\t}\bm{f}}{\hat{\sigma}_i}\right).
\end{equation}
We then estimate the dynamic effects by $\hat{\bm{\beta}}_t=\bm{Y}_t-\hat{\bm{\Lambda}}\hat{\bm{f}}_t$ for each $t>T_0$. 
The corresponding asymptotic properties follow directly from Theorem~\ref{thm: largeNasymptotics}.
\begin{coro}\label{coro: dynamicEastimationLargeN}
 Under Assumptions \ref{asmp:consistency}, \ref{asmp:model1}, \ref{asmp: AFM}, \ref{asmp:largeNdynamicSparse}, \ref{asmp: largeNrrloss}, and regularity conditions \ref{asmp: AFMdep}--\ref{asmp: lossmatrix}, as $N,T_0\to\infty$ with $T_0^{1/2}N^{-1} = o(1)$ and $(\log N)^2/T_0 = o(1)$, we have $\hat{\bm{f}}_t - \bm{f}_{t} = o_p(1)$, and $\hat{\bm{\beta}}_t - \bm{\beta}_t = \bm{\varepsilon}_t + o_p(1)$ for each $t>T_0$.
\end{coro}

\section{Discussion of the factor loading stability condition}
\label{sec:loading-space}

Assumption~\ref{asmp:model1} stipulates time-invariant factor loadings
throughout the study period. In practice, the factor structure may evolve
over time, particularly when $T$ is large. 
This section discusses two complementary relaxations of loading stability that preserve identification: systematic shifts that remain within the column space of pre-intervention loadings, and mean-zero perturbations of post-intervention loadings.

\paragraph{Weak column-space stability.}
Because factor models are identified only up to a common rotation, our methods do not require the stability of any particular parameterization of the
loadings; the same observation extends to accommodate systematic shifts in the loadings after the intervention, provided the column space remains stable so that no new directions are introduced. Specifically, consider the post-intervention outcome model $Y_{it}(0)=\bm{\lambda}_{\mathrm{post},i}^{\t}\bm{f}_t+\varepsilon_{it}$ or, in matrix form, $\bm{Y}_t = \bm{\Lambda}_{\rm post}\bm{f}_t+\bm{\varepsilon}_t$ for $t>T_0$.
Suppose the post-intervention loading matrix $\bm{\Lambda}_{\rm post} = (\bm{\lambda}_{\mathrm{post},1},\ldots, \bm{\lambda}_{\mathrm{post},N})^{\t}$
admits the representation
\[
\bm{\Lambda}_{\mathrm{post}} = \bm{\Lambda}\bm{H}
\]
for some fixed matrix $\bm{H} \in \mathbb{R}^{r\times r}$,
where $\bm{\Lambda}$ is the pre-intervention loading matrix. In the
fixed-$N$ setting, the temporal mean difference becomes
\[
\bm{D}
= \bar{\bm{\beta}}
  + \bm{\Lambda}_{\mathrm{post}}\bm{\alpha}_1 - \bm{\Lambda}\bm{\alpha}_0
  + \tilde{\bm{\varepsilon}}
= \bar{\bm{\beta}} + \bm{\Lambda}\bm{\theta} + \tilde{\bm{\varepsilon}},
\]
where the nuisance coefficient $\bm{\alpha}$ in the original derivation is
replaced by $\bm{\theta} = \bm{H}\bm{\alpha}_1 - \bm{\alpha}_0$. The robust
regression argument therefore goes through unchanged with $\bm{\theta}$ in
place of $\bm{\alpha}$, and $\bar{\bm{\beta}}$ can be identified as
$\bm{D} - \bm{\Lambda}\bm{\theta}$. The same conclusion holds in the
large-$N$ setting: the post-intervention mean admits the decomposition
\[
\bar{\bm{Y}}_{\mathrm{post}}
   = \bar{\bm{\beta}}
     + \bm{\Lambda}_{\rm post}\bar{\bm{f}}_{\mathrm{post}}
     + \bar{\bm{\varepsilon}}_{\mathrm{post}} = \bar{\bm{\beta}}
     + \bm{\Lambda}\left(\bm{H}\bar{\bm{f}}_{\mathrm{post}}\right)
     + \bar{\bm{\varepsilon}}_{\mathrm{post}},
\]
so robust M-estimation recovers the rotated nuisance coefficient
$\bm{H}\bar{\bm{f}}_{\mathrm{post}}$ and $\bar{\bm{\beta}}$ can be identified
by the same residual argument. More generally, matrix $\bm{H}\in\mathbbm{R}^{r\times r}$ can be arbitrary and identification requires
only that the column space of $\bm{\Lambda}_{\mathrm{post}}$ be contained
in that of $\bm{\Lambda}$, a weak column space stability condition, accommodating both rotation-type
reparameterizations of the factor structure and systematic shifts that
remain within the column space of $\bm{\Lambda}$.

By contrast, unrestricted systematic shifts in loadings fall outside the
scope of our framework. If $\bm{\Lambda}_{\mathrm{post}}$ contains
directions outside the column space of $\bm{\Lambda}$, the fixed-$N$
temporal difference becomes
\[
\bm{D} = \bar{\bm{\beta}} + \bm{\Lambda}\bm{\theta} + \bm{b}
       + \tilde{\bm{\varepsilon}},
\]
where $\bm{b}$ captures the orthogonal component of the loading drift and
cannot be separated from $\bar{\bm{\beta}}$ without further restrictions; an analogous decomposition holds in the large-$N$ setting. We therefore
view the weak loading column-space stability as an explicit assumption and
a potential limitation of our approach.

\paragraph{Post-intervention mean-zero perturbations.}
A second, complementary relaxation allows the loadings to fluctuate within
the post-intervention period around a stable mean. Consider the model
\[
Y_{it}(0) = (\bm{\lambda}_i + \bm{\xi}_{it})^{\t}\bm{f}_t + \varepsilon_{it}
\quad \text{for } t > T_0,
\]
where the shocks $\bm{\xi}_{it}$ are mean zero across the post-intervention
period and uncorrelated with the latent factors,
\[
(T-T_0)^{-1}\sum_{t=T_0+1}^{T} \bm{\xi}_{it} = o_p(1),
\qquad
(T-T_0)^{-1}\sum_{t=T_0+1}^{T} \bm{\xi}_{it}^{\t}\bm{f}_t = o_p(1)
\]
for all $i$ as $T-T_0 \to \infty$. 
In both the fixed-$N$ and large-$N$ settings with a long post-intervention period, both terms vanish
asymptotically, so the temporal difference $\bm{D}$ and the post-intervention mean $\bar{\bm{Y}}_{\rm post}$ identify the same
probability limit as in the unperturbed model, and the robust regression estimators
consistently recover the regression coefficients. 
In the large-$N$ setting with fixed $T-T_0$, the perturbation term
$(T-T_0)^{-1}\sum_{t=T_0+1}^{T} \bm{\xi}_{it}^{\t}\bm{f}_t$ does not
vanish but can be absorbed into the error term of the robust regression;
under mild regularity conditions on $\bm{\xi}_{it}$ (e.g., symmetric
distribution, moment and dependence restrictions), robust M-estimation
still identifies $\bar{\bm{f}}_{\mathrm{post}}$. 
In all asymptotic regimes, the
average direct and interference effects remain consistently estimable. A
simulation assessing the impact of such perturbations on estimation and
inference is provided in Section~\ref{sec:suppsimupert}.

The two departures from stable factor loadings can be accommodated simultaneously by jointly reparametrizing the nuisance regression coefficients and absorbing the residual perturbation into the regression error.
Together, for $t>T_0$, we allow post-intervention outcome model $Y_{it}(0) = \bm{\lambda}_{\mathrm{post},it}^{\t}\bm{f}_t+\varepsilon_{it}$ with 
$\bm{\lambda}_{\mathrm{post},it}
=\bm{\lambda}_{\mathrm{post},i}+\bm{\xi}_{it}$, where
$\bm{\Lambda}_{\mathrm{post}}
=(\bm{\lambda}_{\mathrm{post},1},\ldots,\bm{\lambda}_{\mathrm{post},N})^{\t}
=\bm{\Lambda}\bm H$ for some fixed matrix
$\bm H\in\mathbb R^{r\times r}$, and
$\{\bm{\xi}_{it}\}_{t>T_0}$ is mean-zero and uncorrelated with
$\{\bm f_t\}_{t>T_0}$.

\section{Fixed-$N$ setting: Regularity conditions and proof of theoretical results}\label{sec:suppA}
\subsection{A useful lemma for fixed-$N$ robust regression and its proof}
The following lemma provides the general asymptotic results for the least trimmed squares (LTS) considered in the paper, as well as the least median of squares (LMS) estimators.

\begin{lem}\label{lem:ltslms}
For $1\leq i\leq p$, suppose $Y_i = \bm{X}_{i}^{\t}\bm{c} + d_i$ for some fixed values $Y_i\in\mathbb{R}, d_i\in \mathbb{R}, \bm{X}_{i}\in\mathbb{R}^k$, and $\bm{c}\in\mathbb{R}^k$. 
Let $\hat{Y}_i$ and   $\hat{\bm{X}}_i$ be   some estimators   such that $\hat{Y}_i - Y_i = O_p(n^{-1/2}), \hat{\bm{X}}_i - \bm{X}_i = O_p(n^{-1/2})$ as $n\to\infty$ for $1\leq i\leq p$. Let $r_{(i)}({\bm{c}})$ be the $i$th smallest among $\{(\hat{Y}_{i} - \hat{\bm{X}}_{i}^{\t}\bm{c})^2: i=1,\dots, p\}$. For the least trimmed squares
    \[
    \hat{\bm{c}}_{\rm LTS} \in \arg\min_{\bm{c}\in \mathbb{R}^{k}} \sum_{i=1}^{\lfloor p/2 \rfloor +1} r_{(i)}({\bm{c}});
    \]
and for the least median of squares
    \[
    \hat{\bm{c}}_{\rm LMS} \in \arg\min_{\bm{c}\in \mathbb{R}^{k}} {\rm median}_{1\leq i\leq p} \left(\hat{Y}_{i} - \hat{\bm{X}}_{i}^{\t}\bm{c}\right)^2.
    \]
Let $\bm{X} = (\bm{X}_1^{\t}, \ldots, \bm{X}_p^{\t})^{\t}$ and $\bm{X}_{\mathcal{S}}$ be the submatrix with rows indexed by $\mathcal{S}$. If $|\{d_i=0: i=1\dots,p\}| \geq \lfloor p/2\rfloor+k$, and $\bm{X}_{\mathcal{S}}$ is full-rank for every $\mathcal{S}\subset \{i:d_i=0\}$ with $|\mathcal{S}| = k$, then $$\hat{\bm{c}}_{\rm LTS} - \bm{c} =O_p(n^{-1/2}) \quad\text{and}\quad  \hat{\bm{c}}_{\rm LMS} - \bm{c} =O_p(n^{-1/2}).$$
\end{lem}

\begin{proof}[Proof of Lemma S.\ref{lem:ltslms}]
Let $\bm{Y} = (Y_1,\ldots,Y_p)^{\t}, \bm{X} = (\bm{X}_1^{\t}, \ldots, \bm{X}_p^{\t})^{\t},$ $\bm{d} = (d_1,\ldots,d_p)^{\t}$ and the estimators are correspondingly collected as $\hat{\bm{Y}}, \hat{\bm{X}}$.
Let $l_{(i)}$ stand for the $i$th smallest entry of vector $\bm{l}$ with non-negative entries.
By the definition of the least trimmed squares, we have
\begin{eqnarray*}
    \sum_{i=1}^{\lfloor p/2\rfloor+1} r_{(i)}(\hat{\bm{c}}_{\rm LTS})  
 &\leq& \sum_{i=1}^{\lfloor p/2\rfloor+1} r_{(i)}({\bm{c}}) \\
 &=& \sum_{i=1}^{\lfloor p/2\rfloor+1} \left\{\left(\hat{\bm{Y}} - \hat{\bm{X}} {\bm{c}}\right)_{(i)}\right\}^2\\
    &= & \sum_{i=1}^{\lfloor p/2\rfloor+1} \left[\left\{\hat{\bm{Y}} - \bm{Y} + \bm{Y} -  (\hat{\bm{X}} - \bm{X} + \bm{X}){\bm{c}}\right\}_{(i)}\right]^2\\
    &= & \sum_{i=1}^{\lfloor p/2\rfloor+1}\left[\left\{\hat{\bm{Y}} - \bm{Y} + \bm{d} -  (\hat{\bm{X}} - \bm{X}){\bm{c}}\right\}_{(i)}\right]^2.
\end{eqnarray*}

For any $i\in  \{i: d_i=0\}$ and $j \in \{j: d_j\neq 0\}$, asymptotically we have 
\begin{eqnarray*}
\big\{\hat{Y}_i- {Y}_i  - (\hat{\bm{X}}_i - \bm{X}_i)^{\t}{\bm{c}} \big\}^2 &=& O_p(n^{-1}), \\
\big\{\hat{Y}_j- {Y}_j  - (\hat{\bm{X}}_j - \bm{X}_j)^{\t}{\bm{c}}\big\}^2 &=& d_j^2 + O_p(n^{-1}).
\end{eqnarray*}

Since $|\{i: d_i = 0\}| \geq \lfloor p/2 \rfloor+k \geq \lfloor p/2\rfloor+1$, the $\lfloor p/2\rfloor+1$ smallest units among $\{r_i({\bm{c}}): i=1,\dots, p \}$ are attained among units in $\{i:d_i=0\}$ asymptotically, that is,
\[
\begin{aligned}
    \sum_{i=1}^{\lfloor p/2\rfloor+1}\left[\left\{\hat{\bm{Y}} - \bm{Y} + \bm{d} -  (\hat{\bm{X}} - \bm{X}){\bm{c}}\right\}_{(i)}\right]^2
    \leq \sum_{i=1}^{p} \big\{\hat{Y}_i- Y_i  - (\hat{\bm{X}}_i - \bm{X}_i)^{\t}{\bm{c}}\big\}^2.
\end{aligned}
\]
Thus for sufficiently large $n$, we have
\begin{equation}\label{eq:ltsleft}
	\sum_{i=1}^{\lfloor p/2\rfloor+1} r_{(i)}(\hat{\bm{c}}_{\rm LTS}) \leq \sum_{i=1}^{p} \big\{\hat{Y}_i- {Y}_i  - (\hat{\bm{X}}_i - \bm{X}_i)^{\t}{\bm{c}}\big\}^2.
\end{equation}
Next, we give a lower bound for the least trimmed squares objective function. 
Use $|\{i:d_i=0\}|\geq  \lfloor p/2\rfloor+k$ again, the inequality that 
\begin{equation}\label{eq:ltsright}
	\sum_{i=1}^{\lfloor p/2\rfloor+1} r_{(i)}(\hat{\bm{c}}_{\rm LTS}) \geq r_j(\hat{{\bm{c}}}_{\rm LTS})
\end{equation}
holds for at least $(\lfloor p/2\rfloor+1) + (\lfloor p/2\rfloor+k)-p \geq k$ units with $d_j=0$. 
Letting $\Delta = \hat{\bm{c}}_{\rm LTS} - {\bm{c}}$, for $i\in\{1,2,\dots, p\}$, the right term has the following expression:
\[
	r_i(\hat{{\bm{c}}}_{\rm LTS}) = \left(\hat{Y}_i - \hat{\bm{X}}_i^{\t}\hat{\bm{c}}_{\rm LTS}\right)^2 =\left\{\hat{Y}_i- {Y}_i + d_{i} -(\hat{\bm{X}}_i - \bm{X}_i)^{\t}({\bm{c}}+\Delta)- \bm{X}_i^{\t}\Delta\right\}^2.
\]
For each unit that satisfies \eqref{eq:ltsright} and in $\{i:d_i=0\}$, we get the following inequality by combining \eqref{eq:ltsleft} and \eqref{eq:ltsright}:
\begin{equation}\label{eq: lem1key}
    \left\{\hat{Y}_i- {Y}_i -(\hat{\bm{X}}_i - \bm{X}_i)^{\t}({\bm{c}}+\Delta)- \bm{X}_i^{\t}\Delta\right\}^2\leq \sum_{i=1}^{p}  \big\{\hat{Y}_i- {Y}_i  - (\hat{\bm{X}}_i - \bm{X}_i)^{\t}{\bm{c}}\big\}^2,
\end{equation}
where at least $k$ units satisfy the inequality as we asserted.
As the right hand is $O_p(n^{-1})$, we have $\bm{X}_i^{\t}\Delta = O_p(n^{-1/2})$ for at least $k$ units, i.e., there exists $\mathcal{S}$ such that $|\mathcal{S}|\geq k$ and $\bm{X}_{\mathcal{S}}\Delta = O_p(n^{-1/2})$. Since $\bm{X}$ is of full rank, the submatrix $\bm{X}_{\mathcal{S}}$ with $|\mathcal{S}|\geq k$ has full rank, then $\Delta =  \hat{\bm{c}}_{\rm LTS} - {\bm{c}}=O_p(n^{-1/2})$.

For the least median of squares, we have
\begin{eqnarray*}
    {\rm median}_{1\leq i\leq p} \left(\hat{Y}_{i} - \hat{\bm{X}}_{i}^{\t}\bm{c}_{\rm LMS}\right)^2 
 &\leq& {\rm median}_{1\leq i\leq p} \left(\hat{Y}_{i} - \hat{\bm{X}}_{i}^{\t}\bm{c}\right)^2 \\
 &\leq& \sum_{i=1}^{\lfloor p/2\rfloor+1} \left(\hat{\bm{Y}} - \hat{\bm{X}} {\bm{c}}\right)^2_{(i)}\\
    &= & \sum_{i=1}^{\lfloor p/2\rfloor+1}\left\{\hat{\bm{Y}} - \bm{Y} + \bm{d} -  (\hat{\bm{X}} - \bm{X}){\bm{c}}\right\}^2_{(i)}.
\end{eqnarray*}

Replicating the previous arguments, for sufficiently large $n$,
\begin{equation}\label{eq:lmsleft}
	 {\rm median}_{1\leq i\leq p} \left(\hat{Y}_{i} - \hat{\bm{X}}_{i}^{\t}\bm{c}_{\rm LMS}\right)^2 \leq \sum_{i=1}^{p} \big\{\hat{Y}_i- {Y}_i  - (\hat{\bm{X}}_i - \bm{X}_i)^{\t}{\bm{c}}\big\}^2.
\end{equation}
Next, we show that the objective function of the least median of squares is also bounded by some quantities below. 
Use $|\{i:d_i =0\}|\geq  \lfloor p/2\rfloor+k$, the following inequality holds 
\begin{equation}\label{eq:lmsright}
	{\rm median}_{1\leq i\leq p} \left(\hat{Y}_{i} - \hat{\bm{X}}_{i}^{\t}\bm{c}_{\rm LMS}\right)^2 \geq \left(\hat{Y}_j - \hat{\bm{X}}_j^{\t}\hat{\bm{c}}_{\rm LMS}\right)^2,
\end{equation}
for at least $(\lfloor (p+1)/2\rfloor) + (\lfloor p/2\rfloor+k) -p\geq k$ units with $d_j=0$. 
Letting $\Delta = \hat{\bm{c}}_{\rm LMS} - {\bm{c}}$, note that for $i\in\{1,2,\dots, p\}$,
\[
\left(\hat{Y}_i - \hat{\bm{X}}_i^{\t}\hat{\bm{c}}_{\rm LMS}\right)^2 =\left\{\hat{Y}_i- {Y}_i + d_{i} -(\hat{\bm{X}}_i - \bm{X}_i)^{\t}({\bm{c}}+\Delta)- \bm{X}_i^{\t}\Delta\right\}^2.
\]
Then by equations \eqref{eq:lmsleft} and \eqref{eq:lmsright}, there are at least $k$ units in $\{i:d_i=0\}$ such that
\[
\left\{\hat{Y}_i- {Y}_i -(\hat{\bm{X}}_i - \bm{X}_i)^{\t}({\bm{c}}+\Delta)- \bm{X}_i^{\t}\Delta\right\}^2\leq \sum_{i=1}^{p}  \big\{\hat{Y}_i- {Y}_i  - (\hat{\bm{X}}_i - \bm{X}_i)^{\t}{\bm{c}}\big\}^2,
\]
where the right hand is $O_p(n^{-1})$ as assumed.
Hence, $\bm{X}_i^{\t}\Delta = O_p(n^{-1/2})$ for at least $k$ units in $\{i:d_i=0\}$, i.e., there exists $\mathcal{S}$ such that $|\mathcal{S}|= k, \mathcal{S}\subset \{i:d_i=0\}$ such that $\bm{X}_{\mathcal{S}}\Delta = O_p(n^{-1/2})$. 
By condition, the submatrix $\bm{X}_{\mathcal{S}}$ has full rank, then $\Delta =  \hat{\bm{c}}_{\rm LMS} - {\bm{c}}=O_p(n^{-1/2})$.

\end{proof}
Let $\hat{\bm{c}}$ be either $\hat{\bm{c}}_{\rm LTS}$ or $\hat{\bm{c}}_{\rm LMS}$. 
As a remark, the proof implies a more general result from \eqref{eq: lem1key} that
\[\hat{\bm{c}} - \bm{c} = O_p\left(\max_{1\leq i\leq p}\left|\hat{Y}_i-Y_i\right|\right) +O_p\left(\max_{1\leq i\leq p}\left\|\hat{\bm{X}}_i-\bm{X}_i\right\|_2\right).\]
As an implication, for any $d<0$, if $\hat{Y}_i - Y_i = O_p(n^{d}), \hat{\bm{X}}_i - \bm{X}_i = O_p(n^{d})$ as $n\to\infty$ for all $i$, then $\hat{\bm{c}}-\bm{c} = O_p(n^{d})$; if $\hat{Y}_i - Y_i = o_p(1), \hat{\bm{X}}_i - \bm{X}_i = o_p(1)$ as $n\to\infty$ for all $i$, then $\hat{\bm{c}}-\bm{c} = o_p(1)$.
\subsection{Details for fixed-$N$ factor analysis}\label{suppsubsec:factanal}

In this subsection, we justify the use of classical factor analysis \citep{anderson1956factor} with strongly mixing errors. 
Define the strong mixing coefficients of $\{\bm{\varepsilon}_t\}_{t\geq 1}$ by $\alpha(0) = 1/2$, and for $k\geq 1$,
    \[ \alpha(k) = \sup_{n\in\mathbbm{Z}}\sup\left\{\left|\pr\left(A\cap B\right) - \pr\left(A\right)\pr\left(B\right)\right|:A\in\sigma(\bm{\varepsilon}_t:t\leq n), B\in\sigma(\bm{\varepsilon}_t:t\geq n+k)\right\}.\]

The conditions are listed as follows.
\begin{condition}\label{asmp:techfactor}
    (i) $\Psi_{ij}\neq 0 $ for all $1\leq i\leq N, 1\leq j\leq N$, where $\bm{\Psi} = \bm{\Sigma} - \bm{\Lambda}(\bm{\Lambda}^{\t}\bm{\Sigma}^{-1}\bm{\Lambda})^{-1}\bm{\Lambda}^{\t}$; 
    (ii) The submatrix $\bm{\Lambda}_{\mathcal{S}}$ has full rank for every $\mathcal{S}\subset \{1,\ldots,N\}$ with $|\mathcal{S}| = r$.
\end{condition}
\begin{condition}\label{asmp: partialsumfactors}
    As $T_* = \min\{T_0,T-T_0\}\to\infty$, the sequence of factors satisfies that $\max_{1\leq k\leq T_0}T_0^{-1/2}\sum_{t=1}^k\|\bm{f}_t-\bm{\alpha}_0\|_2 =O(1)$ and $(T-T_0)^{-1/2}\sum_{t=T_0+1}^{T}(\bm{f}_t-\bm{\alpha}_1) = O(1)$.
\end{condition}

\begin{condition}\label{asmp: fixNdep}
The error process $\{\bm{\varepsilon}_t\}_{t\geq 1}$ is a strictly stationary and strong mixing process with $E(\|\bm{\varepsilon}_t\|_2^{4+\delta}) <\infty$ and mixing coefficients $\{\alpha(k)\}_{k\geq 0}$ satisfying $\sum_{k= 0}^{\infty}\alpha(k)^{1-2/(4+\delta)}< \infty$ for some $\delta>0$.
\end{condition}

Condition \ref{asmp:techfactor}(i) is a standard regularity condition for factor analysis listed in \cite{anderson1956factor}.
Condition \ref{asmp:techfactor}(ii) requires nondegenerate factor loadings, serving as a factor model regularity condition and also implying the robust regression design matrix condition in Lemma S.\ref{lem:ltslms}.
Condition \ref{asmp: partialsumfactors} rules out dominant factors and controls for dependence. 
Pre-intervention factor residuals satisfy a standard bounded partial sums condition, ensuring that no single time segment dominates the asymptotics.
The requirement on post-intervention factors is weaker.
This condition can be met if the factors are the realization of a stationary and strong mixing process.
Condition \ref{asmp: fixNdep} requires the error process $\{\bm{\varepsilon}_t\}_{1\leq t\leq T_0}$ to be stationary and strongly mixing, which is a standard condition to ensure the central limit theorem (CLT) \citep{rio2017asymptotic}.
Independent observations and a broad class of weakly dependent processes, including ARMA \citep{tuan1985some} and GARCH models \citep{carrasco2002mixing} under standard regularity conditions, satisfy these requirements; see \citet{doukhan2012mixing, hamilton2020time} for further examples. 
Similar stationarity and mixing assumptions are standard in the synthetic control literature for developing theoretical results for estimation and inference \citep{carvalho2018arco, li2020istatistical, chernozhukov2021exact,cattaneo2021prediction, ferman2021imperfect, viviano2023synthetic}. 

\begin{lem}\label{lemma: factor loading}
    Under Assumptions \ref{asmp:consistency}--\ref{asmp:stationary}, and regularity conditions \ref{asmp:techfactor}--\ref{asmp: fixNdep}, for the estimator of factor loading $\hat{\bm{\Lambda}}$ in \cite{anderson1956factor}, we have $T_0^{1/2}(\hat{\bm{\Lambda}} - \bm{\Lambda}\bm{R})$ is asymptotically normally distributed, where $\bm{R}$ is some $r\times r$ orthogonal matrix.
\end{lem}

\begin{proof}[Proof of Lemma S.\ref{lemma: factor loading}]
Recall we use $\bm{M}_Y$ and $\bm{M}_f$ to denote the pre-intervention outcome and factor empirical covariance, respectively. 
Recall we use $\bar{\bm{Y}}_{\rm pre}$, $\bar{\bm{f}}_{\rm pre}$, and  $\bar{\bm{\varepsilon}}_{\rm pre}$ to denote the pre-intervention outcome, factor, and error means, respectively. 
Under the assumed factor structure, we have
\begin{equation}
    \begin{aligned}
    \bm{M}_Y &= T_0^{-1}\sum_{t=1}^{T_0}\left(\bm{Y}_t - \bar{\bm{Y}}_{\rm pre}\right)\left(\bm{Y}_t - \bar{\bm{Y}}_{\rm pre}\right)^{\t}\\ 
    &= T_0^{-1}\sum_{t=1}^{T_0}\left\{\bm{\Lambda}\left(\bm{f}_t - \bar{\bm{f}}_{\rm pre}\right) +\bm{\varepsilon}_{t} - \bar{\bm{\varepsilon}}_{\rm pre}\right\}\left\{\bm{\Lambda}\left(\bm{f}_t - \bar{\bm{f}}_{\rm pre}\right) +\bm{\varepsilon}_{t} - \bar{\bm{\varepsilon}}_{\rm pre}\right\}^{\t}\\
    &= T_0^{-1}\sum_{t=1}^{T_0}\left\{\bm{\Lambda}\left(\bm{f}_t - \bar{\bm{f}}_{\rm pre}\right)\left(\bm{\varepsilon}_t - \bar{\bm{\varepsilon}}_{\rm pre}\right)^{\t}+\left(\bm{\varepsilon}_t - \bar{\bm{\varepsilon}}_{\rm pre}\right)\left(\bm{f}_t - \bar{\bm{f}}_{\rm pre}\right)^{\t}\bm{\Lambda}^{\t}\right\}\\
    &\quad+  \bm{\Lambda}\bm{M}_f\bm{\Lambda}^{\t}+ T_0^{-1}\sum_{t=1}^{T_0}\left(\bm{\varepsilon}_t - \bar{\bm{\varepsilon}}\right)\left(\bm{\varepsilon}_t - \bar{\bm{\varepsilon}}\right)^{\t}\\
    & = T_0^{-1}\sum_{t=1}^{T_0}\left\{\bm{\Lambda}\left(\bm{f}_t -\bm{\alpha}_0\right)\bm{\varepsilon}_t^{\t}+\bm{\varepsilon}_t\left(\bm{f}_t - \bm{\alpha}_0\right)^{\t}\bm{\Lambda}^{\t} +  \bm{\Lambda}\bm{\Lambda}^{\t}+ T_0^{-1}\sum_{t=1}^{T_0}\bm{\varepsilon}_t \bm{\varepsilon}_t^{\t}\right\} + o_p(T_0^{-1/2}),
\end{aligned}
\end{equation}
where we use $\bar{\bm{f}}_{\rm pre} = \bm{\alpha}_0 + O(T_0^{-1/2})$, $\bm{M}_f = \bm{I}_r$ by Assumption \ref{asmp:stationary}, $\bar{\bm{\varepsilon}}_{\rm pre} = O_p(T_0^{-1/2})$ by Condition \ref{asmp: fixNdep} and mixing CLT \citep{rio2017asymptotic}.
Clearly $E(\bm{M}_Y) = \bm{\Lambda}\bm{\Lambda}^{\t} + \bm{\Sigma}$, next we prove $T_0^{1/2}(\bm{M}_Y - \bm{\Lambda}\bm{\Lambda}^{\t} - \bm{\Sigma})$ is asymptotically normal. 
Fix arbitrary deterministic vectors $u$, $v$. 
Consider the scalar projection $T_0^{1/2}\bm{u}^{\t}(\bm{M}_Y - \bm{\Lambda}\bm{\Lambda}^{\t} - \bm{\Sigma})\bm{v}$, which can be written as $
T_0^{-1/2}\sum_{t=1}^{T_0} H_{t}$,
where $H_{t}$ is a scalar measurable function of $\bm{\varepsilon}_t$ with deterministic coefficients involving $\bm{u}^{\t}\bm{\Lambda}(\bm{f}_t - \bm{\alpha}_0)$ and $\bm{v}^{\t}\bm{\Lambda}(\bm{f}_t - \bm{\alpha}_0)$, plus the centered quadratic term $(\bm{u}^{\t}\bm{\varepsilon}_t)(\bm{v}^{\t}\bm{\varepsilon}_t) - E\{(\bm{u}^{\t}\bm{\varepsilon}_t)(\bm{v}^{\t}\bm{\varepsilon}_t)\}$. Since $H_t$ is a measurable transform of $\bm{\varepsilon}_t$, the array $\{H_{t}\}$ is strongly mixing with the same mixing coefficients as $\{\bm{\varepsilon}_t\}$. 
The $(4 + \delta)$ moment implies $H_{t}$ has a finite $(2 + \delta/2)$ moment, which is sufficient for mixing CLT.
Condition \ref{asmp: partialsumfactors} ensures the deterministic coefficients have stabilized second moments and bounded partial sums, so no short time segment dominates the normalization.
Therefore, a standard strong-mixing CLT for triangular arrays with deterministic coefficients applies to $T_0^{-1/2}\sum_{t\le T_0} H_t$, yielding convergence to a centered normal distribution \citep{davidson1994stochastic}.
Since this holds for every fixed $\bm{u}$, $\bm{v}$, the Cramér--Wold device implies the desired asymptotic normality.

Combining Condition \ref{asmp:techfactor} with Assumptions \ref{asmp:consistency}--\ref{asmp:stationary},
note Assumption \ref{asmp:stationary}(iii) implies an identification condition that the two submatrices after deleting a row of the loading matrix are still of full rank, by theorem 12.1 of \cite{anderson1956factor}, we have the estimated factor loading $\hat{\bm{\Lambda}}$ satisfies that $T_0^{1/2}(\hat{\bm{\Lambda}}-\bm{\Lambda}\bm{R})$ is asymptotically normal for some orthogonal rotation matrix $\bm{R}$.
\end{proof}

As a concluding remark, under mild time-series conditions, the empirical covariance matrix remains asymptotically normal, ensuring the same large-sample properties as the classical factor-loading estimator \citep{anderson1956factor} and justifying its implementation via the standard R package \texttt{factanal}. 
Without loss of generality, the orthogonal rotation matrix $\bm{R}$ can be set to the identity $\bm{I}_r$, since all results are invariant by replacing $\bm{\Lambda}$ and $\bm{f}_t$ as $\bm{\Lambda}\bm{R}$ and $\bm{R}^{\t}\bm{f}_t$.

\subsection{The choice of hard-threshold}
Recall that the estimator $D_{i} -\hat{\bm{\lambda}}^{\t}_i\tilde{{\bm{\alpha}}}$ is $T_*^{1/2}$-consistent for $\bar{\beta}_i$.  
Consequently, any threshold $\gamma$ satisfying $ 
\gamma T_*^{1/2}\to\infty$ and $\gamma\to 0$ as $T_*\to\infty$ is theoretically sufficient for consistent recovery of the zero coordinates of $\bar{\bm{\beta}}$ as $T_*\to\infty$.
Nevertheless, an adaptive threshold can substantially improve finite-sample performance.
Rather than fully tuning $\gamma$, we adopt a threshold with explicit form inspired by \cite{donoho1994ideal}, and we found it performs well in both simulations and empirical applications.
\cite{donoho1994ideal} proposed a hard threshold in the form of $\{2\log(n)\}^{1/2}\phi$ to separate deterministic signals from noises, where $n$ is the number of observations and $\phi^2$ is the variance of error. 
In our context, signal recovery is based on the model
\[\bm{D}_i - \bm{\lambda}_i^{\t}\bm{\alpha} = \bar{\bm{\beta}}_i + \tilde{e}_i, \quad 1 \leq i \leq N,\]
where $\tilde{{\bm{e}}} = (T-T_0)^{-1}\sum_{t=T_0+1}^T \bm{e}_t -T_0^{-1}\sum_{t=1}^{T_0}\bm{e}_t$, $\bm{e}_t = \bm{\Lambda}\bm{w}_t + \bm{\varepsilon}_t$, and $\bm{w}_t = \bm{f}_t - \bm{\alpha}_0$ for $t\leq T_0$ and $\bm{w}_t = \bm{f}_t - \bm{\alpha}_1$ for $t> T_0$. 
Motivated by \cite{donoho1994ideal}, we define the threshold that accounts for data variability: 
\[\gamma = \{2T_*^{-1} \log \left(NT_*\right)\}^{1/2}\phi,\quad \phi^2 = N^{-1} \operatorname{tr}\{\operatorname{Var}(T_*^{1/2}\tilde{\bm{e}})\}.\]
Here we rescale the error term according to its rate so that $\hat{\phi}^2 = O_p(1)$ and make the rate of threshold explicit in the constant component.
In addition, we include $T_*$ into the $\log$ term to fulfill the general rate requirement of $\gamma\to0$.
To obtain a feasible estimate of $\phi^2$, write $\tilde{\bm{w}} = \bar{\bm{w}}_{\rm post} - \bar{\bm{w}}_{\rm pre} = (T-T_0)^{-1}\sum_{t=T_0+1}^{T}\bm{w}_t - T_0^{-1}\sum_{t=1}^{T_0}\bm{w}_t$, $\tilde{\bm{\varepsilon}} = \bar{\bm{\varepsilon}}_{\rm post} - \bar{\bm{\varepsilon}}_{\rm pre} = (T-T_0)^{-1}\sum_{t=T_0+1}^{T}\bm{\varepsilon}_t - T_0^{-1}\sum_{t=1}^{T_0}\bm{\varepsilon}_t$.
Then
\begin{eqnarray*}
    \operatorname{Var}\left(\tilde{\bm{e}}\right) &=& E\left\{\left(\bm{\Lambda}\tilde{\bm{w}}+\tilde{\bm{\varepsilon}}\right)\left(\bm{\Lambda}\tilde{\bm{w}}+\tilde{\bm{\varepsilon}}\right)^{\t}\right\}\\
    &=& \bm{\Lambda}\tilde{\bm{w}}\tilde{\bm{w}}^{\t}\bm{\Lambda}^{\t}+ E(\tilde{\bm{\varepsilon}}\tilde{\bm{\varepsilon}}^{\t})\\
    &=& \bm{\Lambda}\bar{\bm{w}}_{\rm post}\bar{\bm{w}}_{\rm post}^{\t}\bm{\Lambda}^{\t} + \bm{\Lambda}\bar{\bm{w}}_{\rm pre}\bar{\bm{w}}_{\rm pre}^{\t}\bm{\Lambda}^{\t}+ E\left(\bar{\bm{\varepsilon}}_{\rm post}\bar{\bm{\varepsilon}}_{\rm post}^{\t}\right) + E\left(\bar{\bm{\varepsilon}}_{\rm pre}\bar{\bm{\varepsilon}}_{\rm pre}^{\t}\right) + o_p(1)\\
    &\approx& \left(T-T_0\right)^{-1}\bm{\Lambda}\bm{\Lambda}^{\t}+ T_0^{-1}\bm{\Lambda}\bm{\Lambda}^{\t}+ \left(T-T_0\right)^{-1}\bm{\Sigma} + T_0^{-1}\bm{\Sigma},
\end{eqnarray*}
where we ignore the small term and serial dependencies to approximate $\bar{\bm{w}}_{\rm pre}\bar{\bm{w}}_{\rm pre}^{\t}$ as $T_0^{-2}\sum_{t=1}^{T_0}\bm{w}_t\bm{w}_t^{\t} = T_0^{-1}\bm{I}_{r}$ (by Assumption \ref{asmp:stationary}) and similarly approximate other terms.
Therefore
\[\operatorname{Var}(T_*^{1/2}\tilde{\bm{e}}) \approx \frac{T_{*} T}{T_0\left(T-T_0\right)}\left(\bm{\Lambda}\bm{\Lambda}^{\t} + \bm{\Sigma}\right).
\]
We approximate $\bm{\Lambda}\bm{\Lambda}^{\t} + \bm{\Sigma}$ using the empirical covariance of either pre- or post-intervention outcomes, ignoring potential time variation in treatment effects.
Accordingly, the estimation formula of $\hat{\phi}^2$ is summarized as follows:
\[
\begin{aligned}
\hat{\phi}^2 &= N^{-1}\operatorname{tr}(\hat{\bm{V}}),\\ \hat{\bm{V}} &= \frac{T_*}{T_0\left(T-T_0\right)}\left\{\sum_{t=1}^{T_0} \left(\bm{Y}_t- \bar{\bm{Y}}_{\rm pre}\right)\left(\bm{Y}_t- \bar{\bm{Y}}_{\rm pre}\right)^{\t} + \sum_{t=T_0+1}^{T} \left(\bm{Y}_t- \bar{\bm{Y}}_{\rm post}\right)\left(\bm{Y}_t- \bar{\bm{Y}}_{\rm post}\right)^{\t}\right\}.
\end{aligned}
\]

As a concluding remark, multiplying the threshold by an additional constant remains theoretically valid, and such tuning may improve finite-sample performance. 
The proposed threshold is adaptive to unit-level heterogeneity, so a single universal tuning parameter would be sufficient. 
Moreover, when the pre- and post-intervention periods are of comparable order, that is, $T_0/T \to \kappa \in (0,1)$, all rate statements involving $T_*$ can be equivalently expressed in terms of $T$, and the threshold can be modified accordingly by replacing $T_*$ with $T$.
When the post-intervention variability is too large, we can only use the pre-intervention period to estimate the covariance matrix. See a discussion after Condition \ref{asmp: stableeffects}.
\subsection{Proof of Theorem \ref{thm:1}}\label{proof:thm1}

In this subsection, we prove the selection consistency of $\hat{\mathcal{C}}$ and the asymptotic normality of the estimator $\hat{\bm{\beta}}$. 
Recall that $T_0/T \to \kappa\in[0,1]$ as $T_* = \min\{T_0, T-T_0\}\to\infty$.
Before proving the main theorem, we provide a lemma for the convergence rate of the initial robust regression estimator.

\begin{lem}\label{lem:alpha}
Under Assumptions \ref{asmp:consistency}--\ref{asmp:stationary}, and regularity conditions \ref{asmp:techfactor}--\ref{asmp: fixNdep}, we have $\tilde{{\bm{\alpha}}} - {\bm{\alpha}} =O_p\{T_0^{-1/2}+(T-T_0)^{-1/2}\}$.
\end{lem}

\begin{proof}[Proof of Lemma S.\ref{lem:alpha}]

We start with $\bm{D} = \bar{\bm{\beta}} + \bm{\Lambda}\tilde{\bm{f}} + \tilde{\bm{\varepsilon}}$ 
where $\tilde{\bm{f}} = (T-T_0)^{-1}\sum_{t=T_0+1}^T\bm{f}_{t} -  T_0^{-1}\sum_{t=1}^{T_0}\bm{f}_{t}$ and $\tilde{\bm{\varepsilon}} = (T-T_0)^{-1}\sum_{t=T_0+1}^T\bm{\varepsilon}_{t} - T_0^{-1}\sum_{t=1}^{T_0}\bm{\varepsilon}_{t}$. 
By Condition \ref{asmp: partialsumfactors},
\[\tilde{\bm{f}}-{\bm{\alpha}} = (T-T_0)^{-1}\sum_{t=T_0+1}^T\left(\bm{f}_{t}-\bm{\alpha}_1\right) -  T_0^{-1}\sum_{t=1}^{T_0}\left(\bm{f}_{t}-\bm{\alpha}_0\right) = O\left\{\left(T-T_0\right)^{-1/2}+T_0^{-1/2}\right\}.\]
By Condition \ref{asmp: fixNdep} and mixing CLT \citep{rio2017asymptotic}, $\tilde{\bm{\varepsilon}} = O_p\{(T-T_0)^{-1/2}+T_0^{-1/2}\} = O_p(T_*^{-1/2})$.
As a result, $\bm{D} - \bar{\bm{\beta}}-\bm{\Lambda}\bm{\alpha} = O_p\{(T-T_0)^{-1/2}+T_0^{-1/2}\}$.
For the $\hat{\bm{\Lambda}}$ obtained from the pre-intervention periods, we have $\hat{\bm{\Lambda}} -{\bm{\Lambda}} = O_p(T_0^{-1/2})$ by Lemma S.\ref{lemma: factor loading}, where we set $\bm{R}$ as the identity matrix for notational simplicity and without loss of generality, see the remark in Section \ref{suppsubsec:factanal}.

Recall the $\tilde{\bm{\alpha}}$ solves the optimization problem for the least trimmed squares under our setting
\begin{equation*}
    \tilde{\bm{\alpha}} \in \arg\min_{{\bm{\alpha}}} \sum_{i=1}^{\lfloor N/2\rfloor+1} r_{(i)}({\bm{\alpha}}),
\end{equation*}
where $r_{(i)}({\bm{\alpha}})$ is the $i$th smallest among $\{(D_{i} - \hat{\bm{\lambda}}_i^{\t}{\bm{\alpha}})^2: i=1,\dots, N\}$. 

Applying Lemma S.\ref{lem:ltslms} by replacing $\hat{\bm{Y}},\hat{\bm{X}}$ with $\bm{D},\hat{\bm{\Lambda}}$ respectively, we have  
 $ \tilde{\bm{\alpha}} - {\bm{\alpha}}=O_p\{(T-T_0)^{-1/2}+T_0^{-1/2}\} = O_p(T_*^{-1/2})$.
 % $\tilde{\bm{\alpha}}_{\rm LMS} - {\bm{\alpha}}=O_p(T^{-1/2})$.
\end{proof}

We give a regularity condition on the growth of post-intervention dynamic effects that is purely for the validity of the hard-threshold that adjusts for whole-period outcome variance.
\begin{condition}\label{asmp: stableeffects}
    $(T-T_0)^{-1}\sum_{t=T_0+1}^T\|\bm{\beta}_t - \bar{\bm{\beta}}\|_2^2 = O\{(T-T_0)^{1/2-\epsilon}\}$ for some $\epsilon>0$.
\end{condition}
This condition assumes no significant trend in the dynamic effects.
It is easy to be satisfied with bounded dynamic effects or dynamic effects with uniformly bounded second moments as trivial examples. 
If the variability of outcomes is too large in the post-intervention periods, we can use the threshold with $\hat{\phi}^2 = N^{-1}\operatorname{tr}(\hat{\bm{V}})$ and $\hat{\bm{V}} = T_*T_0^{-1}(T-T_0)^{-1}\{T/T_0\sum_{t=1}^{T_0} (\bm{Y}_t- \bar{\bm{Y}}_{\rm pre})(\bm{Y}_t- \bar{\bm{Y}}_{\rm pre})^{\t}\}$.
\begin{proof}[Proof of Theorem \ref{thm:1}]

We first prove the selection consistency of $\hat{\mathcal{C}}$. For unit $i$, 
\[
\begin{aligned}
    D_i - \hat{\bm{\lambda}}_i^{\t}\tilde{{\bm{\alpha}}} =&  \bar{\beta}_i + \bm{\lambda}_i\tilde{\bm{f}} + \tilde{\bm{\varepsilon}} - \hat{\bm{\lambda}}_i^{\t}\tilde{{\bm{\alpha}}}\\
    =&\bar{\beta}_i  + \tilde{\bm{\varepsilon}} + \bm{\lambda}_i^{\t}(\tilde{\bm{f}} - {\bm{\alpha}}) + (\bm{\lambda}_i - \hat{\bm{\lambda}}_i)^{\t}{\bm{\alpha}} + \hat{\bm{\lambda}}_i^{\t}(\bm{\alpha} - \tilde{{\bm{\alpha}}})\\
    =&\bar{\beta}_i + O_p(T_*^{-1/2}),
\end{aligned}
\]
where $\tilde{\bm{f}}-{\bm{\alpha}} = (T-T_0)^{-1}\sum_{t=T_0+1}^T(\bm{f}_{t}-\bm{\alpha}_1) -  T_0^{-1}\sum_{t=1}^{T_0}(\bm{f}_{t}-\bm{\alpha}_0) = O\{(T-T_0)^{-1/2}+T_0^{-1/2}\} = O(T_*^{-1/2})$ by Assumption \ref{asmp: fixNdep}, $\hat{\bm{\lambda}}_i - \bm{\lambda}_i = O_p(T_0^{-1/2}) = O_p(T_*^{-1/2})$ by factor analysis, $\tilde{\bm{\alpha}} - \bm{\alpha} = O_p(T_*^{-1/2})$ by Lemma S.\ref{lem:alpha}.
It is easy to verify that
$\hat{\bm{V}} = T_*T_0^{-1}(T-T_0)^{-1}\{\sum_{t=1}^{T_0} (\bm{Y}_t- \bar{\bm{Y}}_{\rm pre})(\bm{Y}_t- \bar{\bm{Y}}_{\rm pre})^{\t} + \sum_{t=T_0+1}^{T} (\bm{Y}_t- \bar{\bm{Y}}_{\rm post})(\bm{Y}_t- \bar{\bm{Y}}_{\rm post})^{\t}\} = \min\{1/\kappa, 1/(1-\kappa)\}(\bm{\Lambda}\bm{\Lambda}^{\t}+\bm{\Sigma}) + (T-T_0)^{-1}\sum_{t=T_0+1}^T(\bm{\beta}_t -\bar{\bm{\beta}})(\bm{\beta}_t - \bar{\bm{\beta}})^{\t} + o_p(1)$, where the first term is a positive definite and the second is the empirical covariance of dynamic effects, which is at least semi-definite. 
Therefore, $\operatorname{tr}(\hat{\bm{V}})$ is bounded away from zero and controlled with $O(T_*^{1/2-\epsilon})$ with Condition \ref{asmp: stableeffects}.
Therefore $\pr(\hat\phi^2>c)\to 1$ for some $c>0$ and $\hat\phi^2 = O_p(T_*^{1/2-\epsilon})$ for some $\epsilon>0$.
The selection consistency is guaranteed by
\begin{eqnarray*}
\pr\left(\hat{\mathcal{C}}\neq \mathcal{C}\right) &=& \pr\left(\bigcup_{i=1}^N\{i \in \mathcal{C}, i \not\in \hat{\mathcal{C}}\} \cup \{i \not\in \mathcal{C}, i \in \hat{\mathcal{C}}\}\right) \\
&\leq & \sum_{i=1}^N \pr\left(i \in \mathcal{C}, i \not\in \hat{\mathcal{C}}\right) + \sum_{i=1}^N \pr\left(i \not\in \mathcal{C}, i \in \hat{\mathcal{C}}\right)\\
&\leq& \sum_{i=1}^N \pr\left(i \in \hat{\mathcal{C}} \mid i \not\in \mathcal{C}\right) + \sum_{i=1}^N \pr\left(i \not\in \hat{\mathcal{C}}\mid i \in \mathcal{C}\right)\\
&\leq& \sum_{i=1}^N \pr\left(|D_{i} - \hat{\bm{\lambda}}_i^{\t}\tilde{{\bm{\alpha}}}| > \{2T_*^{-1}\log(NT_*)\}^{1/2}\hat\phi \mid \bar{\beta}_i = 0\right) \\ && + \sum_{i=1}^N \pr\left(|D_{i} - \hat{\bm{\lambda}}_i^{\t}\tilde{{\bm{\alpha}}}|\leq \{2T_*^{-1}\log(NT_*)\}^{1/2}\hat\phi\mid \bar{\beta}_i \neq 0 \right)\\
&\to& 0, \qquad \text{as } T_*\to \infty.
\end{eqnarray*}

Next, we prove the consistency and asymptotic normality of the updated estimator $\hat{\bm{\beta}}$. 
We have
\begin{eqnarray}\label{eq:eqsplit}
T_*^{1/2}\left(\hat{\bm{\beta}}-\bar{\bm{\beta}}\right) & = & T_*^{1/2}\left\{\bar{\bm{Y}}_{{\rm post}} -\hat{\bm{\Lambda}}\left(\hat{\bm{\Lambda}}_{\hat{\mathcal{C}}}^{\t} \hat{\bm{\Lambda}}_{\hat{\mathcal{C}}}\right)^{-1} \hat{\bm{\Lambda}}_{\hat{\mathcal{C}}}^{\t} \bar{\bm{Y}}_{{\rm post},\hat{\mathcal{C}}} - \bar{\bm{\beta}}\right\} \nonumber \\
&=&T_*^{1/2}\left\{ \bm{\Lambda} \bar{\bm{f}}_{\rm post} + \bar{\bm{\varepsilon}}_{\rm post} -\hat{\bm{\Lambda}}\left(\hat{\bm{\Lambda}}_{\hat{\mathcal{C}}}^{\t} \hat{\bm{\Lambda}}_{\hat{\mathcal{C}}}\right)^{-1} \hat{\bm{\Lambda}}_{\hat{\mathcal{C}}}^{\t} \left(\bar{\bm{\beta}}_{\hat{\mathcal{C}}} + \bm{\Lambda}_{\hat{\mathcal{C}}} \bar{\bm{f}}_{\rm post} + \bar{\bm{\varepsilon}}_{\rm post,\hat{\mathcal{C}}}\right) \right\}\nonumber \\
&=&-T_*^{1/2}\hat{\bm{\Lambda}}\left(\hat{\bm{\Lambda}}_{\hat{\mathcal{C}}}^{\t} \hat{\bm{\Lambda}}_{\hat{\mathcal{C}}}\right)^{-1} \hat{\bm{\Lambda}}_{\hat{\mathcal{C}}}^{\t} \bar{\bm{\beta}}_{\hat{\mathcal{C}}} 
 + T_*^{1/2} \left\{\bm{\Lambda} -\hat{\bm{\Lambda}}\left\{\hat{\bm{\Lambda}}^{\t}_{\hat{\mathcal{\mathcal{C}}}}\hat{\bm{\Lambda}}_{\hat{\mathcal{\mathcal{C}}}}\right)^{-1}\hat{\bm{\Lambda}}^{\t}_{\hat{\mathcal{\mathcal{C}}}}\bm{\Lambda}_{\hat{\mathcal{\mathcal{C}}}}\right\}\left(\bar{\bm{f}}_{\rm post}-\bm{\alpha}_1\right) \nonumber  \\
&& 
+ T_*^{1/2}\left\{\bm{\Lambda} -\hat{\bm{\Lambda}}\left(\hat{\bm{\Lambda}}^{\t}_{\hat{\mathcal{\mathcal{C}}}}\hat{\bm{\Lambda}}_{\hat{\mathcal{\mathcal{C}}}}\right)^{-1}\hat{\bm{\Lambda}}^{\t}_{\hat{\mathcal{\mathcal{C}}}}\bm{\Lambda}_{\hat{\mathcal{\mathcal{C}}}}\right\}\bm{\alpha}_1 \\
&& +  T_*^{1/2}\left\{\bar{\bm{\varepsilon}}_{\rm post} 
- \hat{\bm{\Lambda}}\left(\hat{\bm{\Lambda}}_{\hat{\mathcal{C}}}^{\t} \hat{\bm{\Lambda}}_{\hat{\mathcal{C}}}\right)^{-1} \hat{\bm{\Lambda}}_{\hat{\mathcal{C}}}^{\t} \bar{\bm{\varepsilon}}_{{\rm post},\hat{\mathcal{C}}}\right\},\nonumber
\end{eqnarray}
where $\bar{\bm{Y}}_{\rm post}, \bar{\bm{f}}_{\rm post}, \bar{\bm{\varepsilon}}_{\rm post}$ are the post-intervention mean of $\bm{Y}_t$, $\bm{f}_t$, $\bm{\varepsilon}_t$, respectively.

Since $\pr(\hat{\mathcal{C}} =  \mathcal{C})\to 1$ and $\bar{\bm{\beta}}_{\mathcal{C}} = \bm{0}$, the first term of equation \eqref{eq:eqsplit}
\[T_*^{1/2}\hat{\bm{\Lambda}}\left(\hat{\bm{\Lambda}}_{\hat{\mathcal{C}}}^{\t} \hat{\bm{\Lambda}}_{\hat{\mathcal{C}}}\right)^{-1} \hat{\bm{\Lambda}}_{\hat{\mathcal{C}}}^{\t} \bar{\bm{\beta}}_{\hat{\mathcal{C}}} =o_p(1).
\]

By Lemma S.\ref{lemma: factor loading}, $T_0^{1/2} (\hat{\bm{\Lambda}} - {\bm{\Lambda}}) \stackrel{d}{\to} \bm{U}$ and $T_0^{1/2}(\hat{\bm{\Lambda}} - {\bm{\Lambda}}) = O_p(1)$, where we set the rotation matrix $\bm{R} = \bm{I}_r$ without loss of generality and $\bm{U}$ is a mean zero normally distributed random vector after vectorization. 
By Condition \ref{asmp: fixNdep}, $\bar{\bm{f}}_{\rm post}-{\bm{\alpha}}_1 = O\{(T-T_0)^{-1/2}\} = O(T_*^{-1/2})$ and 
by continuous mapping theorem, $\bm{\Lambda} -\hat{\bm{\Lambda}}(\hat{\bm{\Lambda}}^{\t}_{\hat{\mathcal{C}}}\hat{\bm{\Lambda}}_{\hat{\mathcal{C}}})^{-1}\hat{\bm{\Lambda}}^{\t}_{\hat{\mathcal{C}}}\bm{\Lambda}_{\hat{\mathcal{C}}} = o_p(1)$,
together, the second term of equation \eqref{eq:eqsplit}
\[T_*^{1/2} \left\{\bm{\Lambda} -\hat{\bm{\Lambda}}\left(\hat{\bm{\Lambda}}^{\t}_{\hat{\mathcal{C}}}\hat{\bm{\Lambda}}_{\hat{\mathcal{C}}}\right)^{-1}\hat{\bm{\Lambda}}^{\t}_{\hat{\mathcal{C}}}\bm{\Lambda}_{\hat{\mathcal{C}}}\right\}\left(\bar{\bm{f}}_{\rm post}-\bm{\alpha}_1\right) = o_p(1).
\]
Consider the third term of equation \eqref{eq:eqsplit}, as $\pr(\hat{\mathcal{C}}=\mathcal{C})\to 1$, we have 
\[
T_*^{1/2}\left\{\bm{\Lambda}-\hat{\bm{\Lambda}}\left(\hat{\bm{\Lambda}}_{\hat{\mathcal{C}}}^{\t}\hat{\bm{\Lambda}}_{\hat{\mathcal{C}}}\right)^{-1}\hat{\bm{\Lambda}}_{\hat{\mathcal{C}}}^{\t}\bm{\Lambda}_{\hat{\mathcal{C}}}\right\}\bm{\alpha}_1 = T_*^{1/2}\left\{\bm{\Lambda}-\hat{\bm{\Lambda}}\left(\hat{\bm{\Lambda}}_{\mathcal{C}}^{\t}\hat{\bm{\Lambda}}_{\mathcal{C}}\right)^{-1}\hat{\bm{\Lambda}}_{\mathcal{C}}^{\t}\bm{\Lambda}_{\mathcal{C}}\right\}\bm{\alpha}_1
+o_p(1).
\]
Write $\hat{\bm{\Lambda}}=\bm{\Lambda}+\bm{\Delta}$ and $\hat{\bm{\Lambda}}_\mathcal{C}=\bm{\Lambda}_\mathcal{C}+\bm{\Delta}_\mathcal{C}$, where $\bm{\Delta}=O_p(T_0^{-1/2})$ and $\bm{\Delta}_\mathcal{C}=O_p(T_0^{-1/2})$. Define
\[
\bm{M}(\hat{\bm{\Lambda}}_\mathcal{C}):=\left(\hat{\bm{\Lambda}}_{\mathcal{C}}^{\t}\hat{\bm{\Lambda}}_{\mathcal{C}}\right)^{-1}\hat{\bm{\Lambda}}_{\mathcal{C}}^{\t}\bm{\Lambda}_{\mathcal{C}}.
\]
A first-order expansion around $\bm{\Lambda}_\mathcal{C}$ yields
\[
\bm{M}(\hat{\bm{\Lambda}}_\mathcal{C})
=\bm{I}_r-\bm{A}^{-1}\bm{\Lambda}_\mathcal{C}^{\t}\bm{\Delta}_\mathcal{C}+o_p\left(T_0^{-1/2}\right),
\]
where $\bm{A} = \bm{\Lambda}_\mathcal{C}^{\t}\bm{\Lambda}_\mathcal{C}$, $\hat{\bm{\Lambda}}_\mathcal{C}^{\t}\hat{\bm{\Lambda}}_\mathcal{C}=\bm{A}+O_p(T_0^{-1/2})$, and $\hat{\bm{\Lambda}}_\mathcal{C}^{\t}\bm{\Lambda}_\mathcal{C}=\bm{A}+O_p(T_0^{-1/2})$. Consequently,
\[
\hat{\bm{\Lambda}}\bm{M}\left(\hat{\bm{\Lambda}}_\mathcal{C}\right)
=\left(\bm{\Lambda}+\bm{\Delta}\right)\left(\bm{I}_r-\bm{A}^{-1}\bm{\Lambda}_\mathcal{C}^{\t}\bm{\Delta}_\mathcal{C}\right)+o_p(T_0^{-1/2})
=\bm{\Lambda}+\bm{\Delta}-\bm{\Lambda}\bm{A}^{-1}\bm{\Lambda}_\mathcal{C}^{\t}\bm{\Delta}_\mathcal{C}+o_p(T_0^{-1/2}),
\]
where the cross-term $\bm{\Delta}\bm{A}^{-1}\bm{\Lambda}_\mathcal{C}^{\t}\bm{\Delta}_\mathcal{C}=O_p(T_0^{-1})$ is negligible.
Therefore,
\[
\bm{\Lambda}-\hat{\bm{\Lambda}}\bm{M}(\hat{\bm{\Lambda}}_\mathcal{C})
=-\bm{\Delta}+\bm{\Lambda}\bm{A}^{-1}\bm{\Lambda}_\mathcal{C}^{\t}\bm{\Delta}_\mathcal{C}+o_p(T_0^{-1/2}),
\]
and multiplying by $T_*^{1/2}$ gives
\[
T_*^{1/2}\left\{\bm{\Lambda}-\hat{\bm{\Lambda}}\left(\hat{\bm{\Lambda}}_{\hat{\mathcal{C}}}^{\t}\hat{\bm{\Lambda}}_{\hat{\mathcal{C}}}\right)^{-1}\hat{\bm{\Lambda}}_{\hat{\mathcal{C}}}^{\t}\bm{\Lambda}_{\hat{\mathcal{C}}}\right\}\bm{\alpha}_1
\stackrel{d}{\to} \min\left(1,\frac{1-\kappa}{\kappa}\right)^{1/2}
\left\{-\bm{U}+\bm{\Lambda}\left(\bm{\Lambda}_\mathcal{C}^{\t}\bm{\Lambda}_\mathcal{C}\right)^{-1}\bm{\Lambda}_\mathcal{C}^{\t}\bm{U}_\mathcal{C}\right\}\bm{\alpha}_1,
\]
which is normally distributed.

For the last term of \eqref{eq:eqsplit}, note that $(T-T_0)^{1/2} \bar{\bm{\varepsilon}}_{\rm post} \stackrel{d}{\to} \bm{Z}_{1} $ for a normally distributed mean zero vector $\bm{Z}_1$ by mixing CLT. 
By Slutsky's theorem,
\[
 T_*^{1/2}\left\{\bar{\bm{\varepsilon}}_{\rm post}  
- \hat{\bm{\Lambda}}\left(\hat{\bm{\Lambda}}_{\hat{\mathcal{C}}}^{\t} \hat{\bm{\Lambda}}_{\hat{\mathcal{C}}}\right)^{-1} \hat{\bm{\Lambda}}_{\hat{\mathcal{C}}}^{\t} \bar{\bm{\varepsilon}}_{{\rm post},\hat{\mathcal{C}}}\right\} \stackrel{d}{\to}
\min\left(1,\frac{\kappa}{1-\kappa}\right)^{1/2}
\left\{\bm{Z}_1  
- \bm{\Lambda}\left(\bm{\Lambda}_{\mathcal{C}}^{\t} \bm{\Lambda}_{\mathcal{C}}\right)^{-1} \bm{\Lambda}_{\mathcal{C}}^{\t} \bm{Z}_{1,\mathcal{C}}\right\}
% \left\{\left(\bm{I}_{|\mathcal{C}|} - \bm{\Lambda}_{\mathcal{C}}\left(\bm{\Lambda}_{\mathcal{C}}^{\t} \bm{\Lambda}_{\mathcal{C}}\right)^{-1} \bm{\Lambda}_{\mathcal{C}}^{\t}\right)\bm{Z}_{1,\mathcal{C}}^{\t},  \bm{Z}_{1,\mathcal{C}^c}^{\t}\right\}^{\t}.
\]
A byproduct is that $T_*^{1/2}(\hat{\bm{\beta}} - \bar{\bm{\beta}}) = O_p(1)$.

For the asymptotic normality, recall that $\bm{Z}_{1}, \bm{U}$ are normally distributed with mean zero from $\hat{\bm{\Lambda}}$ and $\bar{\bm{\varepsilon}}_{\rm post}$.
Since $\hat{\bm{\Lambda}},\bar{\bm{\varepsilon}}_{\rm post}$ are derived from the pre-intervention and post-intervention periods, the correlation between $\bm{Z}_{1}, \bm{U}$ vanishes as $T\to\infty$ under Assumption \ref{asmp: fixNdep}. 
Combining all the above, we have
\[
\begin{aligned}
    T_*^{1/2}\left(\hat{\bm{\beta}}-\bar{\bm{\beta}}\right) &\stackrel{d}{\to} \min\left(1,\frac{1-\kappa}{\kappa}\right)^{1/2}
\left\{-\bm{U}+\bm{\Lambda}\left(\bm{\Lambda}_\mathcal{C}^{\t}\bm{\Lambda}_\mathcal{C}\right)^{-1}\bm{\Lambda}_\mathcal{C}^{\t}\bm{U}_\mathcal{C}\right\}\bm{\alpha}_1\\
&\quad+\min\left(1,\frac{\kappa}{1-\kappa}\right)^{1/2}
\left\{\bm{Z}_1  
- \bm{\Lambda}\left(\bm{\Lambda}_{\mathcal{C}}^{\t} \bm{\Lambda}_{\mathcal{C}}\right)^{-1} \bm{\Lambda}_{\mathcal{C}}^{\t} \bm{Z}_{1,\mathcal{C}}\right\},
\end{aligned}
\]
which is normally distributed.

The asymptotic normality holds for all $\kappa\in[0,1]$ and the asymptotic covariance does not degenerate even on the boundary cases $\kappa = 0,1$.
\end{proof}

\subsection{Proof of Proposition \ref{prop:weight}}\label{proof:prop1}
Recall our estimator $
\hat{\bm{\beta}} = \bar{\bm{Y}}_{\rm post} - \bm{\hat{\Lambda}}(\bm{\hat{\Lambda}}_{\hat{\mathcal{C}}}^{\t}\bm{\hat{\Lambda}}_{\hat{\mathcal{C}}})^{-1}\bm{\hat{\Lambda}}_{\hat{\mathcal{C}}}^{\t}\bar{\bm{Y}}_{{\rm post}, \hat{\mathcal{C}}},
$ the implied weight for estimating the average direct effect $\hat{\beta}_1$ that applied to unit $j\in\hat{\mathcal{C}}$ is $\hat{w}_j = \bm{\hat{\lambda}}_{1}^{\t}(\bm{\hat{\Lambda}}_{\hat{\mathcal{C}}}^{\t}\bm{\hat{\Lambda}}_{\hat{\mathcal{C}}})^{-1}\bm{\hat{\lambda}}_j$.

\begin{proof}[Proof of Proposition \ref{prop:weight}.]
Let $w_j= \bm{\lambda}_{1}^{\t}(\bm{\Lambda}_{\mathcal{C}}^{\t}\bm{\Lambda}_{\mathcal{C}})^{-1}\bm{\lambda}_j$ be the designed weights.
Although the factor loadings are identified only up to an orthogonal rotation, the weights are uniquely identified.
To see this, consider a rotation $\bm{\Lambda}\to\bm{\Lambda\bm{R}}$, where $\bm{R}$ is an orthogonal matrix. 
Under this transformation, the weights become
\[ \left(\bm{R}\bm{\lambda}_{1}\right)^{\t}(\bm{R}^{\t}\bm{\Lambda}_{\mathcal{C}}^{\t}\bm{\Lambda}_{\mathcal{C}}\bm{R})^{-1}\bm{R}\bm{\lambda}_j = \bm{\lambda}_{1}^{\t}\bm{R}^{\t}\bm{R}^{-1}(\bm{\Lambda}_{\mathcal{C}}^{\t}\bm{\Lambda}_{\mathcal{C}})^{-1}\left(\bm{R}^{\t}\right)^{-1}\bm{R}\bm{\lambda}_j = \bm{\lambda}_{1}^{\t}(\bm{\Lambda}_{\mathcal{C}}^{\t}\bm{\Lambda}_{\mathcal{C}})^{-1}\bm{\lambda}_j,\]
which is identical to $w_j$.

Moreover, these weights arise as the probability limits of the estimated weights, since Lemma S.\ref{lemma: factor loading} implies
$\hat{\bm{\Lambda}}-\bm{\Lambda}\bm{R} = o_p(1)$ for some orthogonal matrix $\bm{R}$, and the same calculation and continuous mapping theorem apply.
For the following, we take $\bm{R}$ as $\bm{I}_r$ to simplify the notation without loss of generality.

First, by definition, 
 \[ \sum_{j\in\mathcal{C}}w_j\bm{\lambda}_j =\left(\sum_{j\in\mathcal{C}}\bm{\lambda}_j\bm{\lambda}_{j}^{\t}\right) \left(\bm{\Lambda}_{\mathcal{C}}^{\t}\bm{\Lambda}_{\mathcal{C}}\right)^{-1}\bm{\lambda}_1 = \bm{\lambda}_1,\]
which indicates that the designed weights exactly match the factor loadings of the treated unit using non-interfered control units.

Note that $\bm{\hat{\Lambda}}_{\hat{\mathcal{C}}}^{\t}\bm{\hat{\Lambda}}_{\hat{\mathcal{C}}} = (\sum_{j\in\hat{\mathcal{C}}}\hat{\bm{\lambda}}_j\hat{\bm{\lambda}}_j^{\t} - \sum_{j\in\mathcal{C}}\hat{\bm{\lambda}}_j\hat{\bm{\lambda}}_j^{\t}) + \sum_{j\in\mathcal{C}}(\hat{\bm{\lambda}}_j\hat{\bm{\lambda}}_j^{\t} - \bm{\lambda}_j\bm{\lambda}_j^{\t}) + \sum_{j\in\mathcal{C}}\bm{\lambda}_j\bm{\lambda}_j^{\t},$ where the first term is zero in high probability since $\pr(\hat{\mathcal{C}}=\mathcal{C})\to 1$ by Theorem \ref{thm:1}. The second summation is $O_p(T_0^{-1/2})$ since $\hat{\bm{\lambda}}_j - \bm{\lambda}_j=O_p(T_0^{-1/2})$ for all $j$ by Lemma S.\ref{lemma: factor loading}.
Together we have $\bm{\hat{\Lambda}}_{\hat{\mathcal{C}}}^{\t}\bm{\hat{\Lambda}}_{\hat{\mathcal{C}}} = \bm{\Lambda}_{\mathcal{C}}^{\t}\bm{\Lambda}_{\mathcal{C}} + O_p(T_0^{-1/2})$. Using the fact that $(\bm{A}+\bm{B})^{-1} = \bm{A}^{-1}-\bm{A}^{-1}\bm{B}(\bm{A}+\bm{B})^{-1}$ for nonsingular matrices $\bm{A}$ and $\bm{A}+\bm{B}$, let $\bm{A} = \bm{\Lambda}_{\mathcal{C}}^{\t}\bm{\Lambda}_{\mathcal{C}}, \bm{B} = \bm{\hat{\Lambda}}_{\hat{\mathcal{C}}}^{\t}\bm{\hat{\Lambda}}_{\hat{\mathcal{C}}}-\bm{\Lambda}_{\mathcal{C}}^{\t}\bm{\Lambda}_{\mathcal{C}} = O_p(T_0^{-1/2})$, we have $(\bm{\hat{\Lambda}}_{\hat{\mathcal{C}}}^{\t}\bm{\hat{\Lambda}}_{\hat{\mathcal{C}}})^{-1} = (\bm{\Lambda}_{\mathcal{C}}^{\t}\bm{\Lambda}_{\mathcal{C}})^{-1} + O_p(T_0^{-1/2})$. 
Therefore, 
\[\hat{w}_j  = \left\{\bm{\lambda}_1 + O_p(T_0^{-1/2})\right\}^{\t}\left\{(\bm{\Lambda}_{\mathcal{C}}^{\t}\bm{\Lambda}_{\mathcal{C}})^{-1} + O_p(T_0^{-1/2})\right\}\left\{\bm{\lambda}_j +O_p(T_0^{-1/2})\right\} = w_j +O_p(T_0^{-1/2}).\]
Combining it with the fact that the difference between two summations in $\hat{\mathcal{C}}$ and $\mathcal{C}$ is $o_p(1)$ since $\pr(\hat{\mathcal{C}}=\mathcal{C})\to 1$, we have
\[\bm{\lambda}_1 - \sum_{j\in\hat{\mathcal{C}}}\hat{w}_j\bm{\lambda}_j = \sum_{j\in\hat{\mathcal{C}}}(w_j - \hat{w}_j)\bm{\lambda}_j + \left(\sum_{j\in\mathcal{C}}w_j\bm{\lambda}_j - \sum_{j\in\hat{\mathcal{C}}}w_j\bm{\lambda}_j\right) = O_p(T_0^{-1/2}).\] 

Next, consider
\[\begin{aligned}
     \left|E\left\{\bar{Y}_{\mathrm{post},1}(0) - \sum_{j\in\hat{\mathcal{C}}}\hat{w}_j\bar{Y}_{\mathrm{post},j}\right\}\right| 
    &= \left|E\left(\bm{\lambda}_1 - \sum_{j\in\hat{\mathcal{C}}}\hat{w}_j\bm{\lambda}_{j}\right)^{\t}\bar{\bm{f}}_{\rm post} 
    -E\left(  \sum_{j\in\hat{\mathcal{C}}}\hat{w}_j\bar{\varepsilon}_{\mathrm{post},j}\right)-E\left(\sum_{j\in \hat{\mathcal{C}}}\hat{w}_j\bar{\beta}_{j}\right)\right| \\
    &\leq \left\|E\left(\bm{\lambda}_1 - \sum_{j\in\hat{\mathcal{C}}}\hat{w}_j\bm{\lambda}_{j}\right) \right\|_2\left\|\bar{\bm{f}}_{\rm post}\right\|_2+\left|E\left(  \sum_{j\in\hat{\mathcal{C}}}\hat{w}_j\bar{\varepsilon}_{\mathrm{post},j}\right)\right| \\
    &\quad + \left|E\left(\sum_{j\in \hat{\mathcal{C}}}\hat{w}_j\bar{\beta}_{j}\right)\right|
\end{aligned}\]
We bound the three terms separately. 
For the first term, since $\hat{w}_j - w_j = O_p(T_0^{-1/2})$ for all $j$, in particular $\hat{w}_j = w_j +o_p(1)$. 
Moreover, by assumption, $|\hat{w}_j - w_j| \leq |\hat{w}_j| + |w_j| \leq C + |w_j|$ a.s., by dominate convergence theorem, we have 
$E|\hat{w}_j - w_j| =o(1)$ for all $j$.
Using the identity $\bm{\lambda}_1 = \sum_{j\in\mathcal{C}}w_j\bm{\lambda}_j$, it follows that
\[\begin{aligned}
    \left\|E\left(\bm{\lambda}_1 - \sum_{j\in\hat{\mathcal{C}}}\hat{w}_j\bm{\lambda}_{j}\right) \right\|_2 
    &= \left\|E\left(\sum_{j\in\mathcal{C}}\left(w_j-\hat{w}_j\right)\bm{\lambda}_{j}- \sum_{j\in\hat{\mathcal{C}}\backslash\mathcal{C}}\hat{w}_j\bm{\lambda}_{j}\right) \right\|_2\\
    &\leq \sum_{j\in\mathcal{C}}E\left|w_j-\hat{w}_j\right|\left\|\bm{\lambda}_{j}\right\|_2
+E\left|\sum_{j\in\hat{\mathcal{C}}\backslash\mathcal{C}}\hat{w}_j\bm{\lambda}_{j}\right|\\
&= O\left\{ \sum_{j\in\mathcal{C}}E\left|w_j-\hat{w}_j\right|+ \pr \left(\hat{\mathcal{C}}\neq\mathcal{C}\right)\right\}\\
&= o(1),
\end{aligned}
\]
where we use the boundedness of $w_j, \hat{w}_j$, and $\bm \lambda_j$ either induced by fixed $N$ or by assumed condition, and the consistency of $\hat{\mathcal{C}}$.
The norm of factors $\|\bar{\bm{f}}_t\|_2\leq (T-T_0)^{-1}\sum_{t=T-T_0+1}^T\|\bm f_t\|_2\leq C$.

For the second term, since $E|\bar{\varepsilon}_{\mathrm{post},j}|\leq \{E(\bar{\varepsilon}_{\mathrm{post},j}^2)\}^{1/2}<\infty$, the product $(\hat{w}_j - w_j)\bar{\varepsilon}_{\mathrm{post},j}$ is $o_p(1)$ and is dominated by the integrable random variable $ (C + |w_j|)|\bar{\varepsilon}_{\mathrm{post},j}|$.
By the dominated convergence theorem,
$E\{(\hat{w}_j - w_j)\bar{\varepsilon}_{\mathrm{post},j}\} = o(1)$  and hence $E(\hat{w}_j\bar{\varepsilon}_{\mathrm{post},j})=E(w_j\bar{\varepsilon}_{\mathrm{post},j})+o(1) = o(1)$.
Therefore,
\[\left|E\left(  \sum_{j\in\hat{\mathcal{C}}}\hat{w}_j\varepsilon_{jt}\right)\right|\leq\sum_{j=1}^NE\left|\hat{w}_j\varepsilon_{jt}\right| = o(1).\]

For the third term, note that $\bar{\beta}_{\mathrm{post},j} = 0$ for $j\in \mathcal{C}$, then
\[
\begin{aligned}
    \left|E\left(\sum_{j\in \hat{\mathcal{C}}}\hat{w}_j\bar{\beta}_{j}\right)\right| &= \left|E\left(\sum_{j\in \hat{\mathcal{C}}\backslash\mathcal{C}}\hat{w}_j\bar{\beta}_{j}\right)\right|\\
    &=O\left\{P\left(\hat{\mathcal{C}}\neq \mathcal{C}\right)\right\}\\
    & = o_p(1).
\end{aligned} \]
Combining the three bounds completes the proof.

\end{proof}
As a remark on the proof, we also have $E\{Y_{1t}(0) - \sum_{j\in\hat{\mathcal{C}}}\hat{w}_jY_{jt}\} = o(1)$ for each $1\leq t \leq T_0$ analogously using the first two bounds; that is, the estimated weights also yield asymptotically unbiased counterfactual predictions for each pre-intervention time point.
Asymptotic unbiasedness for each post-intervention time point, $E\{Y_{1t}(0) - \sum_{j\in\hat{\mathcal{C}}}\hat{w}_jY_{jt}\} = o(1)$ for $t>T_0$, can be obtained analogously if we slightly strengthen Assumption~\ref{asmp:sparsity} to $|\{i: \beta_{it} = 0 \text{ for all }t>T_0 \}|\geq \lfloor N/2\rfloor+r$.
\subsection{Proof of Theorem S.\ref{thm:fixedNdynamic}}
Suppose that for each $t>T_0$, the estimator $\hat{\bm{\mu}}_t$ satisfies
$\hat{\bm{\mu}}_t-\bm{\mu}_t=O_p\{(T-T_0)^{-d}\}$ for some $0<d\le 1/2$.
For each such $t$, the least trimmed squares estimator is defined as
\[
\hat{\bm{\alpha}}_t
=\arg\min_{\bm{\alpha}}
\sum_{i=1}^{\lfloor N/2\rfloor+1}
r_{(i),t}(\bm{\alpha}),
\]
where $r_{(i),t}(\bm{\alpha})$ denotes the $i$th smallest among $\{(\hat\mu_{it}-\hat{\bm{\lambda}}_i^{\t}\bm{\alpha})^2:i=1,\ldots,N\}$.

Let $\hat{\bm{\Lambda}}$ be the factor loading estimator obtained from the
pre-intervention periods. 
By Lemma S.\ref{lemma: factor loading},
$\hat{\bm{\Lambda}}-\bm{\Lambda}=O_p(T_0^{-1/2})$, where, without loss of generality, the orthogonal rotation matrix is taken to be the identity matrix.
Applying Lemma S.\ref{lem:ltslms} with $(\hat{\bm Y},\hat{\bm X})$ replaced by
$(\hat{\bm{\mu}}_t,\hat{\bm{\Lambda}})$ and replacing the rate
$n^{-1/2}$ with $T_*^{-d}$ everywhere (see the remark after the proof of Lemma S.\ref{lem:ltslms}), we have
\[
\hat{\bm{\alpha}}_t-\bm{\alpha}_t=O_p(T_*^{-d}).
\]
Then,
\[
\hat{\bm{\beta}}_t-\bm{\beta}_t
=\hat{\bm{\mu}}_t-\bm{\beta}_t-\hat{\bm{\Lambda}}\hat{\bm{\alpha}}_t
=O_p(T_*^{-d}),
\]
which establishes $T_*^{d}$-consistency of $\hat{\bm{\beta}}_t$.
Finally, recall that the selected unit set is defined as
\[
\hat{\mathcal C}_t
=\left\{i:\,
\left|\hat\mu_{it}-\hat{\bm{\lambda}}_i^{\t}\hat{\bm{\alpha}}_t\right|
\le T_*^{-d}\{2\log(NT_*)\}^{1/2}\hat\phi
\right\}.
\]

As $T_*\to\infty$, we have $\hat{V} = \min\{1/\kappa, 1/(1-\kappa)\}(\bm{\Lambda}\bm{\Lambda}^{\t}+\bm{\Sigma}) + o_p(1)$ and $\hat{\phi} = \phi_0+o_p(1)$ for some $\phi_0>0$, then
\begin{eqnarray*}
\pr\left(\hat{\mathcal{C}}_t \neq \mathcal{C}_t\right) &\leq & \sum_{i=1}^N \pr\left(i \in \mathcal{C}_t, i \not\in \hat{\mathcal{C}}_t\right) + \sum_{i=1}^N \pr\left(i \not\in \mathcal{C}_t, i \in \hat{\mathcal{C}}_t\right)\\
&\leq& \sum_{i=1}^N \pr\left(i \in\hat{\mathcal{C}}_t\mid i \not\in \mathcal{C}_t\right) + \sum_{i=1}^N \pr\left(i \not\in \hat{\mathcal{C}}_t\mid i \in \mathcal{C}_t\right)\\
& = &  \sum_{i=1}^N \pr\left(\left|\hat\mu_{it}-\hat{\bm{\lambda}}_i^{\t}\hat{\bm{\alpha}}_t\right|\leq T_*^{-d}\{2\log(NT_*)\}^{1/2}\hat\phi \mid \beta_{it}\neq 0 \right) \\ &\quad& + \sum_{i=1}^N \pr\left(\left|\hat\mu_{it}-\hat{\bm{\lambda}}_i^{\t}\hat{\bm{\alpha}}_t\right|\geq T_*^{-d}\{2\log(NT_*)\}^{1/2}\hat\phi\mid \beta_{it} =  0\right)\\
&=& o(1).
\end{eqnarray*}
\qed

\section{Large-$N$ setting: Regularity conditions and proof of theoretical results}
\subsection{Some technical conditions related to factor analysis}
We adopt a high-dimensional approximate factor model under the quasi–maximum likelihood framework of \cite{bai2016maximum}.
Their model allows the idiosyncratic covariance matrix $E(\bm{\varepsilon}_t\bm{\varepsilon}_t^{\t})$ to be non-diagonal and time-varying, and focuses on estimating the average diagonal variances $\operatorname{diag}\{T_0^{-1}\sum_{t=1}^{T_0}E(\bm{\varepsilon}_t\bm{\varepsilon}_t^{\t})\}$.
We simplify this specification by imposing time-invariant variances and targeting $\bm{\Sigma} = \operatorname{diag}\{E(\bm{\varepsilon}_t\bm{\varepsilon}_t^{\t})\} = \operatorname{diag}(\sigma^2_1,\ldots, \sigma^2_N)$.

The asymptotic theory for factor model estimation under this framework is well established in \cite{bai2016maximum}. 
For completeness, we summarize below the technical conditions required for our theoretical analysis, including weak dependence across time and units, moment restrictions, and central limit theorems.
Most of these conditions are identical to those required in \cite{bai2016maximum}; we only add two analogous conditions: an additional moment control in Condition \ref{asmp: AFMdep}(iv) and additional dependence restrictions on post-intervention periods and between pre- and post-intervention periods in Condition \ref{asmp: AFMmoments}(i).

\begin{condition}[Dependence and heteroskedasticity]\label{asmp: AFMdep}
    For a constant $C$ large enough, 
(i) $E(\varepsilon_{it}) = 0,\; E(\varepsilon_{it}^{8}) \leq C$;
(ii) $E(\varepsilon_{it} \varepsilon_{is}) = \varpi_{i,ts}$ with $|\varpi_{i,ts}| \leq \varpi_{ts}$ for some $\varpi_{ts} > 0$ and for all $i$.  
In addition, for $1\leq s \leq T_0$, $\sum_{t=1}^{T} \varpi_{ts} \leq C$;
(iii) For all $1\leq i,j \leq N$,  
$
E | T_0^{-1/2} \sum_{t=1}^{T_0} \{ \varepsilon_{it} \varepsilon_{jt} - E(\varepsilon_{it} \varepsilon_{jt}) \} |^4 \leq C
$;
(iv) For all $1\leq j \leq N$ and coordinate $1\leq m\leq r$,
$E|N^{-1/2}\sum_{i=1}^{N}\lambda_{im}(\varepsilon_{it}\varepsilon_{jt} - E(\varepsilon_{it}\varepsilon_{jt}))|^4 \leq C.$
\end{condition}

\begin{condition}\label{asmp: AFMsupport}
    The diagonal elements of $\bm{\Sigma}$ are estimated in the compact set 
$[C^{-2}, C^{2}]$ for some large enough constant $C$.
\end{condition}

\begin{condition}[Moment conditions]\label{asmp: AFMmoments}
    For some constant $C$ large enough, we have (i) $E(\varepsilon_{it} \varepsilon_{js}) = \gamma_{ij,ts}$ with constraints $N^{-1}T_0^{-1} \sum_{i=1}^N \sum_{j=1}^N \sum_{t=1}^{T_0} \sum_{s=1}^{T_0} |\gamma_{ij,ts}| \leq C$ in pre-intervention period, $N^{-1} (T-T_0)^{-1} \sum_{i=1}^N \sum_{j=1}^N \sum_{t=T_0+1}^{T} \sum_{s=T_0+1}^{T} 
|\gamma_{ij,ts}| \leq C$ in post-intervention period, and $N^{-1}(T-T_0)^{-1}\sum_{i=1}^N\sum_{j=1}^N\sum_{t=T_0+1}^{T}\sum_{s=1}^{T_0}|\gamma_{ij,ts}|\leq C$, $N^{-1}T_0^{-1}\sum_{i=1}^N\sum_{j=1}^N\sum_{t=1}^{T_0}\sum_{s=T_0+1}^{T}|\gamma_{ij,ts}|\leq C$ between pre-intervention and post-intervention periods;
(ii) 
$
E (\| (NT_0)^{-1/2} \sum_{i=1}^N \sum_{t=1}^{T_0} 
\sigma_i^{-2} \bm{\lambda}_i
\{ \varepsilon_{it} \varepsilon_{jt} - E(\varepsilon_{it} \varepsilon_{jt}) \} \|_2^2) \leq C
$ for all $j$;
(iii)
$
E \| (NT_0)^{-1/2} \sum_{i=1}^N \sum_{t=1}^{T_0} 
\sigma_i^{-4} \bm{\lambda}_i \bm{\lambda}_i^{\t} 
( \varepsilon_{it}^2 - \sigma_i^2 ) \|_2^2 \leq C
$;
(iv) For $1\leq t \leq T_0$, $E(\|(NT_0)^{-1/2} \sum_{i=1}^N \sum_{s=1}^{T_0} \sigma_i^{-2} \bm{f}_s[\varepsilon_{it}\varepsilon_{is} - E(\varepsilon_{it}\varepsilon_{is})]\|^2) \leq C$;
(v) For $1\leq t\leq T_0$, $E(N^{-1} \sum_{i=1}^N \|T_0^{-1/2} \sum_{s=1}^T \bm{f}_s\{\varepsilon_{it}\varepsilon_{is} - E(\varepsilon_{it}\varepsilon_{is})\}\|^2) \leq C$.
\end{condition}

\begin{condition}[Central limit theorems]\label{asmp: AFMCLT}
    (i) For each $i$, as $T_0 \to \infty$,
$
T_0^{-1/2} \sum_{t=1}^{T_0} \bm{f}_t \varepsilon_{it} 
\;\; \overset{d}{\to} \;\; 
N\!(0, \; \lim_{T_0 \to \infty} T_0^{-1} 
\sum_{t=1}^{T_0} \sum_{s=1}^{T_0} \bm{f}_t \bm{f}_s^{\t} \varpi_{i,ts} )
$;
(ii) For each $i$, as $T_0 \to \infty$,
$
T_0^{-1/2} \sum_{t=1}^{T_0} (\varepsilon_{it}^2 - \sigma_i^2) 
\;\; \overset{d}{\to} \;\; 
N(0, \lim_{T \to \infty} T_0^{-1} 
\sum_{t=1}^{T_0} \sum_{s=1}^{T_0} E\{(\varepsilon_{it}^2 - \sigma_i^2)(\varepsilon_{is}^2 - \sigma_i^2)\})
$;
(iii) For each $t$, as $N \to \infty$, $N^{-1/2} \sum_{i=1}^N \sigma_i^{-2} \bm{\lambda}_i \varepsilon_{it} \xrightarrow{d} N(0, \lim_{N \to \infty} N^{-1} \sum_{i=1}^N \sum_{j=1}^N \sigma_i^{-2} \sigma_j^{-2} \bm{\lambda}_i \bm{\lambda}_j^{\prime} \tau_{ij,t})$.
\end{condition}

Under Assumptions \ref{asmp:consistency}, \ref{asmp:model1}, \ref{asmp: AFM}, and regularity conditions \ref{asmp: AFMdep}--\ref{asmp: AFMCLT}, as $N, T_0\to\infty$ and $T_0^{1/2}N^{-1}= o(1)$, the estimated factor loading and individual covariances are $T_0^{1/2}$-consistent and asymptotically normally distributed.
See \cite{bai2016maximum} for details.

\subsection{High-dimensional time series conditions and lemmas}
We assume the random errors $\{\bm{\varepsilon}_t\}_{t\geq 1}$ admit an exponential tail and a strong mixing coefficient with exponential decay for high-dimensional analysis.

We begin by defining the relevant Orlicz norms. Let $\psi: [0, \infty) \to [0, \infty)$ be a convex, non-decreasing function satisfying $\psi(0) = 0$ and $\lim_{x \to \infty} \psi(x) = \infty$. 
The $\psi$-Orlicz norm of a real-valued random variable $X$ is defined as
$$\iii X\iii_{\psi} := \inf\left\{ C > 0 : E\left( \psi\left\{ |X|/C \right\} \right) \leq 1 \right\}.$$
In the following, we use the exponential Orlicz norm, corresponding to the function $\psi_\theta$ with parameter $\theta > 0$. 
For $\theta \geq 1$, we define $\psi_\theta(x) = \exp(x^\theta) - 1$. 
For $\theta \in (0, 1)$, $\psi_\theta$ is defined as the convex hull of the function $x \mapsto \exp(x^\theta) - 1$.
And we denote the exponential Orlicz norm of a random variable $X$ as $\iii X\iii_{\exp({\theta})}$.
For random vector $\bm{X} \in \mathbb{R}^p$, $\iii\bm{X}\iii_{\exp({\theta})} := \sup_{\|\bm{u}\|_2 \leq 1} \iii\bm{u}^{\t}\bm{X}\iii_{\exp({\theta})}$.
Recall that $\{\alpha(m)\}_{m\geq 0}$ denotes the strong mixing coefficients of the process $\{\bm{\varepsilon}_t\}_{t\geq 1}$, which are defined in Section \ref{suppsubsec:factanal} of this Supplementary Material.

\begin{condition}[Mixing time series with exponential tails]\label{asmp: expmixing}
    There are universal constants $C, K_1, \theta_1, \theta_2 > 0$ such that, for all $1\leq t, s \leq T$ and $1\leq i\leq N$, (i) $\iii\bm{\varepsilon}_t\iii_{\exp(\theta_2)} \leq C$; (ii) $\alpha (m) \leq \exp(-K_1 m^{\theta_1})$ for $1 \leq m \leq T$; (iii) $\theta < 1$ where $\theta$ is defined by $1/\theta = 1/\theta_1 + 2/\theta_2$;
    (iv) $\iii N^{-1/2} \sum_{i=1}^{N} \lambda_{ji} \varepsilon_{it}\iii_{\exp(\theta_2)} \leq C$;
    (v) $(\log N)^{(2/\theta)-1}T_0^{-1} = o_p(1)$ as $N$, $T_0\to\infty$.
\end{condition}
Similar mixing and tail conditions are made in \cite{fan2023bridging}.

We need several lemmas from \cite{fan2023supplement} to characterize the concentration behavior under such mixing and exponential tails; we list them as follows for completeness.
\begin{lem}\label{lem: mixingconcentration}
    Let $S_{\tilde{T}} = \sum_{t=1}^{\tilde{T}} X_t$ where $\{X_t : 1\leq t\leq \tilde{T}\}$ is a sequence of zero-mean real-valued random variables such that
\begin{itemize}
    \item[(i)] There exist two positive constants $\theta_1$ and $K_1$ such that the strong mixing coefficients of the sequence satisfy $\alpha(m) \leq \exp(-K_1 m^{\theta_1})$ for any $m$,
    \item[(ii)] There exist two positive constants $\theta_2$ and $K_2$ such that $\max_{1 \leq t \leq \tilde{T}} \iii X_t\iii_{\exp(\theta_2)} \leq K_2$,
    \item[(iii)] $\theta < 1$ where $\theta$ is defined by $1/\theta = 1/\theta_1 + 1/\theta_2$.
\end{itemize}

Then, there exist positive constants $C_1$, $C_2$, $C_3$ depending only on $K_2, K_1, \theta$ and $\theta_1$ such that, for $x > 0$ and $\tilde{T} \geq 4$,
\[
\pr(|S_{\tilde{T}}| \geqslant x) \leqslant \tilde{T} \exp\left( -\frac{x^\theta}{C_1} \right) + \exp\left( -\frac{x^2}{C_2(1+\tilde{T}V)} \right) + \exp\left( -\frac{x^2}{C_3 \tilde{T}} \right),
\]
where $V$ is a finite constant.
\end{lem}

\begin{lem}\label{lem: mixingsumrate}
    Let $\bm{S}_{\tilde{T}} = \sum_{t=1}^{\tilde{T}} \bm{X}_t$ where $\{\bm{X}_t := (X_{1t}, \ldots, X_{nt})^{\t} : 1 \leq t \leq \tilde{T}\}$ be a sequence of zero mean $n$-dimensional random variables with strong mixing coefficients $\{\alpha(m)\}_{m\geq 0}$ and write $\bm{S}_{\tilde{T}} = (S_{\tilde{T},1},\ldots, S_{\tilde{T},n})^{\t}$ for its sum. Assume:
\begin{itemize}
    \item[(i)] There exist positive constants $\theta_1$ and $K_1$ such that the strong mixing coefficients $\alpha(m) \leq \exp(-K_1 m^{\theta_1})$ for any $m$;
    \item[(ii)] There exist two positive constants $\theta_2$ and $K_2$ such that $\max_{1 \leq t \leq \tilde{T}} \iii{X_{it}}\iii_{\exp(\theta_2)} \leq K_2$,
    \item[(iii)] $\theta < 1$ where $\theta$ is defined by $1/\theta = 1/\theta_1 + 1/\theta_2$.
\end{itemize}
Then $\max_{1\leq i\leq n}|S_{\tilde{T},i}| = O_p\{ \tilde{T}^{1/2} (\log n)^{1/2}\}$ whenever $(\log n)^{(2/\theta)-1}\tilde{T}^{-1} = o(1)$.
\end{lem}

\begin{lem}\label{lem: orlicacontrol}
    There exists some constant $C_4$ such that $\iii{XY}\iii_{\exp(\theta/2)} \leq C_4 \max\{ \iii{X}\iii_{\exp(\theta)}^2, \iii{Y}\iii_{\exp(\theta)}^2\}$.

\end{lem}
We refer to \cite{fan2023supplement} for the proof of the above lemmas.

\subsection{A lemma for high-dimensional rate results for factor analysis}
We first prove a lemma on maximum rate control of factor analysis results parallel to Lemma A.1 of \cite{wang2017confounder}.
Let $\hat{\bm{\Lambda}},\hat{\bm{\Sigma}}$ be the quasi-maximum likelihood estimator in \cite{bai2016maximum}.
\begin{lem}\label{lemma: maxrate}
Under Assumptions \ref{asmp:consistency}, \ref{asmp:model1}, \ref{asmp: AFM} and regularity conditions \ref{asmp: AFMdep}--\ref{asmp: expmixing}, as $N,T_0\to\infty$ and $(\log N)^2T_0^{-1}=o(1)$ and $T_0^{1/2}N^{-1} = o(1)$, we have
\begin{equation}\label{eq: maxratefactormodel}
    \max_{1\leq i\leq N} \left|\hat{\sigma}_i^2-\sigma_i^2\right| = O_p\left\{\left(\log N\right)^{1/2}T_0^{-1/2}\right\}, \max_{1\leq i\leq N} \left|\hat{\bm{\lambda}}_i-\bm{\lambda}_i\right| = O_p\left\{\left(\log N\right)^{1/2}T_0^{-1/2}\right\},
\end{equation}
and 
\begin{equation}\label{eq: maxrateloading}
    \max_{1\leq i\leq N} \left|\hat{\bm{\lambda}}_i-\bm{\lambda}_i - T_0^{-1}\sum_{t=1}^{T_0}\bm{f}_t\varepsilon_{it}\right| = O_p\left(T_0^{-1/2}\right).
\end{equation}
\end{lem}
\begin{proof}[Proof of Lemma S.\ref{lemma: maxrate}]
Decompose $\hat{\bm{\lambda}}_i-\bm{\lambda}_i = \sum_{k=1}^{13}\bm{b}_{ki}$, where $\bm{b}_{ki}$ is the $k$th term in the right side of equation (S.14) in \cite{bai2016supplement}.
Similarly, let $\hat{\sigma}_i^2-\sigma_i^2 = \sum_{k=1}^{13}a_{ki}$ and $a_{ki}$ is the $k$th term on the right-hand of equation (S.25) of \cite{bai2016supplement}.
Using a similar argument in \cite{wang2017confounder}, it is easy to obtain that $\max_{1\leq i\leq N}|\bm{b}_{ki}| = o_p(T_0^{-1/2})$ for $k\neq 6,8,10,11,12,13$ and $\max_{1\leq i\leq N}|a_{ki}| = o_p(T_0^{-1/2})$ for $k\neq 1,2,6,9,10,11,12,13$ by Lemma S.11 of \cite{bai2016supplement}. 
Let $\bm{H} = (\bm{\Lambda}^{\t}\bm{\Sigma}^{-1}\bm{\Lambda})^{-1}$, $\hat{\bm{H}} = (\hat{\bm{\Lambda}}^{\t}\hat{\bm{\Sigma}}^{-1}\hat{\bm{\Lambda}})^{-1}$, and write $\bar{\bm{\varepsilon}}_{\rm pre} = T_0^{-1}\sum_{t=1}^{T_0}\bm{\varepsilon}_t$ as $\bar{\bm{\varepsilon}}$ in this proof for simplicity.

First, consider the summation $
\bm{A}_i := T_0^{-1}\sum_{t=1}^{T_0} \bm{f}_t \varepsilon_{it}, i=1,\dots,N$.
As $
\iii\bm{f}_t \varepsilon_{it}\iii_{\exp(\theta_2)}
\le
\|\bm{f}_t\|_2\,\iii\varepsilon_{it}\iii_{\exp(\theta_2)}
\le
C^2
$
and the sequence $\{\bm{f}_t \varepsilon_{it}\}_{t\geq 1}$ shares strong mixing property as $\{\varepsilon_{it}\}_{t\geq 1}$ with the same mixing rate.
By Lemma S.\ref{lem: mixingsumrate}, $\max_{1\leq i\leq N}|\bm{A}_i|=O_p\{(\log N)^{1/2}T_0^{-1/2}\}$ given $(\log N)^{(2/\theta)-1}T_0^{-1} = o_p(1)$.
Combined with Lemma S.11 of \cite{bai2016supplement}, we have $\max_{1\leq i\leq N}|\bm{b}_{8i}| = O_p\{(\log N)^{1/2}T_0^{-1/2}\}$ and $\max_{1\leq i\leq N}|a_{12,i}| = O_p\{(\log N)^{1/2}T_0^{-1/2}\}$.
Similarly, by Lemma S.\ref{lem: orlicacontrol}, $\iii\varepsilon_{it}^2\iii_{\exp(\theta_2/2)}\leq C_4\iii\varepsilon_{it}\iii_{\exp(\theta_2)}^4\leq C_4C^4$, combined with $\{\varepsilon_{it}^2-\sigma^2_i\}_{t\geq 1}$ is strongly mixing with the same mixing coefficients as $\{\bm{\varepsilon}_{t}\}_{t\geq 1}$, we have $\max_{1\leq i\leq N}|a_{1i}| = O_p\{(\log N)^{1/2}T_0^{-1/2}\}$ by Lemma S.\ref{lem: mixingsumrate}.

We next prove for each $1\leq s\leq r$,
\begin{equation}\label{eq: A7}
    \max_{1\leq i\leq N} N^{-1}T_0^{-1}\left|\sum_{j=1}^{N}\lambda_{js}\sum_{t=1}^{T_0}\left\{\varepsilon_{it}\varepsilon_{jt} - E\left(\varepsilon_{it}\varepsilon_{jt}\right)\right\}\right| = o_p(T_0^{-1/2}),
\end{equation}
and 
\begin{equation}\label{eq: A8}
    \max_{1\leq i\leq N} N^{-1}T_0^{-2}\sum_{j=1}^{N}\left(\sum_{t=1}^{T_0}\left\{\varepsilon_{it}\varepsilon_{jt} - E(\varepsilon_{it}\varepsilon_{jt})\right\}\right)^2 = o_p(T_0^{-1/2}).
\end{equation}

Fix $ s$, for each $1\leq i\leq N$, define
\[
\tilde{M}_{it} =\sum_{j=1}^N \lambda_{js}\left\{\varepsilon_{it}\varepsilon_{jt}-E(\varepsilon_{it}\varepsilon_{jt})\right\},
\qquad
M_i=\sum_{t=1}^{T_0} \tilde{M}_{it}.
\]
By Condition \ref{asmp: expmixing}(iv), $\iii \tilde{M}_{it}\iii_{\exp(\theta)}\leq CN^{-1/2}$, then by Lemma S.\ref{lem: mixingsumrate} we have $\max_{1\leq i\leq N}|M_i| = O_p\{N^{1/2}(\log N) T_0^{1/2}\} = o_p(NT_0^{1/2})$, which proves \eqref{eq: A7}.

For \eqref{eq: A8}, define
\[
\tilde{U}_{ij}
=
\sum_{t=1}^{T_0}
\left\{
\varepsilon_{it}\varepsilon_{jt}
-
E(\varepsilon_{it}\varepsilon_{jt})
\right\}, \quad U_i
=
% N^{-1} T_0^{-2}
\sum_{j=1}^N
\tilde{U}_{ij}^2,
\]
The goal is to show $
\max_{1\le i\le N} U_i
=
o_p(NT_0^{3/2})$.

Fix $1\leq i,j\leq N$, the sequence
$\{\varepsilon_{it}\varepsilon_{jt}
-
E(\varepsilon_{it}\varepsilon_{jt})\}_{t\geq 1}$ is mean zero and strongly mixing with the same exponential rate as $\{\bm{\varepsilon}_t\}_{t\geq 1}$, and $\iii\varepsilon_{it}\varepsilon_{jt}\iii_{\exp(\theta_2/2)}\leq C_4\iii\varepsilon_{it}\iii_{\exp(\theta_2)}^2 \iii\varepsilon_{jt}\iii_{\exp(\theta_2)}^2 \leq C_4C^4$ by Lemma S.\ref{lem: orlicacontrol}.
Therefore, by Lemma S.\ref{lem: mixingconcentration}, for some universal constants $C_1,C_2,C_3$,
\[\begin{aligned}
    \pr\left(\max_{1\leq i\leq j}\left|\tilde{U}_{ij}\right|\geq x\right)
    &\leq N^2\pr\left(\left|\tilde{U}_{ij}\right|\geq x\right)\\
& \leq N^2\left\{T_0 \exp\left( -\frac{x^\theta}{C_1} \right) + \exp\left( -\frac{x^2}{C_2(1+T_0V)} \right) + \exp\left( -\frac{x^2}{C_3 T_0} \right)\right\}.
\end{aligned}\]
Setting $x = K\max\{(C_1\log\{N^2T_0\})^{1/\theta}, (C_5T_0\log N^2)^{1/2}\}$ for some $K>0$ and $C_5 = \max\{C_2+V, C_3\}$, for large enough $T_0$, we have $\pr\left(\max_{1\leq i\leq j}\left|\tilde{U}_{ij}\right|\geq x\right)\leq (N^2T_0)^{1-K^\theta}+2N^{2-2K^2}$.
Note that $(C_1\log\{N^2T_0\})^{1/\theta}\leq (C_5T_0\log N^2)^{1/2}$ for large $N,T_0$ provided $(\log N)^{(2/\theta)-1}T_0^{-1} = o_p(1)$, we have $\max_{1\leq i,j\leq N}\tilde{U}_{ij} = O_p\{(\log N)^{1/2}T_0^{1/2}\}$.
Therefore,
\[\max_{1\leq i\leq N}U_i\leq N\max_{1\leq i,j\leq N}\tilde{U}^2_{ij} = O_p\{N(\log N) T_0\} = o_p(NT_0^{3/2}),\]
which proves \eqref{eq: A8}. Here we use $(\log N)^{2}T_0^{-1} = o_p(1)$.

Follow a similar Cauchy--Schwarz inequality argument in \cite{wang2017confounder}, equations \eqref{eq: A7} and \eqref{eq: A8} imply 
\[\max_{1\leq i\leq N}\left|\hat{\bm{H}}\left(\sum_{j=1}^{N}\frac{1}{\hat{\sigma}_j^2}\hat{\bm{\lambda}}_jT_0^{-1}\sum_{t=1}^{T_0}\left\{\varepsilon_{it}\varepsilon_{jt} - E\left(\varepsilon_{it}\varepsilon_{jt}\right)\right\}\right)\right| = o_p(T_0^{-1/2}).\]
Along with the uniform boundedness of the factor loadings and noting that $\hat{\bm{M}}_f$ and $\bm{M}_f$ are exactly $\bm{I}_r$ in \cite{bai2016supplement}'s notation, we have $\max_{1\leq i\leq N}|a_{11,i}| = o_p(T_0^{-1/2})$ and $\max_{1\leq i\leq N}|\bm{b}_{10,i}| = o_p(T_0^{-1/2})$.

Using Lemma S.5 of \cite{bai2016supplement}, combining the fact that $\hat{\bm{H}} = O_p(N^{-1})$ and the uniform boundedness of $\bm{\lambda}_j$, we have $\max_{1\leq i\leq N}|a_{6i}| = o_p(T_0^{-1/2})$ and $\max_{1\leq i\leq N}|\bm{b}_{6i}| = o_p(T_0^{-1/2})$.

Next using Lemma S.10 of \cite{bai2016supplement} and Proposition 1 of \cite{bai2016maximum}, we have 
\[\begin{aligned}
    \left\|\bm{\lambda}_i^{\t}\hat{\bm{H}}T_0^{-1}\sum_{j=1}^{N}\sum_{t=1}^{T}\frac{1}{\hat{\sigma}_j^2}\hat{\bm{\lambda}}_jE(\varepsilon_{it}\varepsilon_{jt})\right\|_2 &\leq \sum_{i=1}^N\left\|\bm{\lambda}_i^{\t}\hat{\bm{H}}T_0^{-1}\sum_{j=1}^{N}\sum_{t=1}^{T_0}\frac{1}{\hat{\sigma}_j^2}\hat{\bm{\lambda}}_jE(\varepsilon_{it}\varepsilon_{jt})\right\|_2 \\
    &= O_p(N^{-1}) + O_p\left(N^{-1}\sum_{i=1}^N\frac{1}{\hat{\sigma}^2_i}\|\hat{\bm{\lambda}}_i-\bm{\lambda}_i\|_2^2\right)\\
    &= O_p(N^{-1}) + O_p(T_0^{-1})+ O_p(N^{-2}),
\end{aligned}\]
which is $o_p(T_0^{-1/2})$ under $T_0^{1/2}N^{-1}=o(1)$.
As a result, $\max_{1\leq i\leq N}|a_{10,i}| = o_p(T_0^{-1/2})$ and $\max_{1\leq i\leq N}|\bm{b}_{12,i}| = o_p(T_0^{-1/2})$.

Using $\hat{\bm{H}}\hat{\bm{\Lambda}}^{\t}\hat{\bm{\Sigma}}^{-1}\bar{\bm{\varepsilon}} = O_p(T_0^{-1/2})$ by Lemma S.5 (a) of \cite{bai2016supplement} and $\max_{1\leq i\leq N}|\bar{\varepsilon}_i| = O_p\{(\log N)^{1/2}T_0^{-1/2}\}$ by Lemma S.\ref{lem: mixingsumrate}, we have $\max_{1\leq i\leq N}|a_{13,i}| = o_p(T_0^{-1/2})$ and $\max_{1\leq i\leq N}|\bm{b}_{13,i}| = o_p(T_0^{-1/2})$.

Finally, $\max_{1\leq i\leq N}|\bm{b}_{11,i}| = O_p(N^{-1}\max_{1\leq i\leq N}|\hat{\bm{\lambda}}_i|)$, 
combined with previous results, we have $\max_{1\leq i\leq N}|\hat{\bm{\lambda}}_i - \bm{\lambda}_i| = O_p\{(\log N)^{1/2}T_0^{-1/2}\}+o_p(\max_{1\leq i\leq N}|\hat{\bm{\lambda}}_i|)$, which implies $\max_{1\leq i\leq N}|\hat{\bm{\lambda}}_i| = O_p(1)$.
Therefore $\max_{1\leq i\leq N}|\bm{b}_{11,i}| = O_p(N^{-1}\max_{1\leq i\leq N}|\hat{\bm{\lambda}}_i|) = O_p(N^{-1})$, and $\max_{1\leq i\leq N}|a_{9,i}| = O_p(N^{-1})\cdot O_p(1) = O_p(N^{-1})$, and $O_p(N^{-1}) = o_p(T_0^{-1/2})$ under $T_0^{1/2}N^{-1} = o(1)$.
So far we have finished the proof of $\max_{1\leq i\leq N} |\hat{\bm{\lambda}}_i-\bm{\lambda}_i| = O_p\{(\log N)^{1/2}T_0^{-1/2}\}$, in turn, we have $\max_{1\leq i\leq N}|a_{2,i}| = O_p\{\max_{1\leq i\leq N}\|\hat{\bm{\lambda}}_i - \bm{\lambda}\|_2^2\} = O_p\{(\log N)T_0^{-1}\} = O_p\{(\log N)^{1/2}T_0^{-1/2}\}$ given $(\log N)^{2}T_0^{-1} = o(1)$, which completes the proof of \eqref{eq: maxratefactormodel}.

For \eqref{eq: maxrateloading}, note we have proved that $\max_{1\leq i\leq N}|\hat{\bm{\lambda}}_i - \bm{\lambda}_i - \bm{b}_{8,i}| = o_p(T_0^{-1/2})$, then
\[\begin{aligned}
    &\quad \max_{1\leq i\leq N} \left|\hat{\bm{\lambda}}_i-\bm{\lambda}_i - T_0^{-1}\sum_{t=1}^{T_0}\bm{f}_t\varepsilon_{it}\right| \\
    &\leq \max_{1\leq i\leq N} \left|\hat{\bm{\lambda}}_i-\bm{\lambda}_i - \bm{b}_{8,i}\right|+\max_{1\leq i\leq N}\left|\bm{b}_{8i} -T_0^{-1}\sum_{t=1}^{T_0}\bm{f}_t\varepsilon_{it}\right|\\
    &= o_p\left(T_0^{-1/2}\right) + \left\|\hat{\bm{H}}\hat{\bm{\Lambda}}^{\t}\hat{\bm{\Sigma}}^{-1}\left(\hat{\bm{\Lambda}}-\bm{\Lambda}\right)\right\|_2\max_{1\leq i\leq N}\left|T_0^{-1}\sum_{t=1}^{T_0}\bm{f}_t\varepsilon_{it}\right|\\
    &=o_p(T_0^{-1/2}).
\end{aligned}\]
\end{proof}

\subsection{Proof of Theorem 2}
Before the proof, we need an additional regularity condition as follows.
\begin{condition}\label{asmp: lossmatrix}
    For each $i$, $E\{\varphi(\bar{\varepsilon}_{\mathrm{post},i}/\sigma_i)\} = 0$, and $N^{-1}\sum_{i=1}^N \sigma_i^{-2} \bm{\lambda}_i \bm{\lambda}_i^\top E\{\varphi'(\bar{\varepsilon}_{\mathrm{post},i}/\sigma_i)\} \to \tilde{\bm{Q}}$ as $N \to \infty$ for some positive definite matrix $\tilde{\bm{Q}}$, where $\varphi = \rho'$ and $\varphi' = \rho''$ denote the first and second derivatives of the robust loss function $\rho(\cdot)$.
\end{condition}
The mean-zero condition is standard for regression analysis. 
It can hold when the distribution of $\bar{\varepsilon}_{i}$ is symmetric around zero and the loss function is even that $\rho(x) = \rho(-x)$, which is commonly satisfied in standard error distributions and loss functions.
The matrix condition is imposed on factor loading and covariance matrices of approximate factor models. 
For many commonly used robust losses—including the Huber loss and smooth variants such as pseudo-Huber and the log-cosh loss—the second derivative of the loss function $\varphi^{\prime}$ is constant in a neighborhood of zero.
In the special case $\varphi^{\prime}\equiv 1$, the condition reduces to the standard factor model identification assumption \citep{bai2012Statistical,bai2016maximum}, which we have assumed in Assumption \ref{asmp: AFM}(iii).
A sufficient condition for Condition \ref{asmp: lossmatrix} is that the normalized errors $\bar{\varepsilon}_{\mathrm{post},i}/\sigma_i$ are identically distributed across all $i$, as such $E\{\varphi^{\prime}\left(\bar{\varepsilon}_{\mathrm{post},i}/\sigma_i\right)\}$ is a positive constant and this condition reduces to Assumption \ref{asmp: AFM}(iii).
Thus, this condition can be viewed as a natural extension of Assumption \ref{asmp: AFM}(iii) to accommodate robust regression estimation.
\begin{proof}[Proof of Theorem 2(i)]
For an arbitrary post-intervention length $T-T_0$, denote the objective function for some $r$-dimensional vector $\bm{f}$ as
\[
        L(\bm{f}) = \sum_{i=1}^N\rho\left(\frac{\bar{Y}_{\mathrm{post},i} - \hat{\bm{\lambda}}_i^{\t}\bm{f}}{\hat{\sigma}_i}\right).\]
Then $\hat{\bm{f}} = \arg\min_{\bm{f}}L(\bm{f})$.
Let $\varphi = \rho^{\prime}$, the derivative is
    \begin{equation}\label{eq: rrestiequation}
        \bm{\Phi}\left(\bm{f}\right) = N^{-1}\sum_{i=1}^N\frac{\hat{\bm{\lambda}}_i}{\hat{\sigma}_i}\varphi\left(\frac{\bar{Y}_{\mathrm{post},i} -\hat{\bm{\lambda}}_i^{\t}\bm{f}}{\hat{\sigma}_i}\right).
    \end{equation}
    By first-order condition, the estimator $\hat{\bm{f}}$ satisfies $\bm{\Phi}(\hat{\bm{f}}) = 0$.
    % \textcolor{blue}{Check this: As the $\rho(\cdot)$ is strictly convex, the solution of $\bm{\Phi}(\bm{f}) = 0$ is unique.}

    Let $\bm{\Delta} = \bm{f} - \bar{\bm{f}}_{\rm post}$ and reparametrize the function $\bm{\Phi}(\bm{f})$ as a function of $\bm{\Delta}$, we define
\[
        \bm{\Psi}\left(\bm{\Delta}\right) = N^{-1}\sum_{i=1}^N\frac{\hat{\bm{\lambda}}_i}{\hat{\sigma}_i}\varphi\left(\frac{\bar{\varepsilon}_{\mathrm{post},i} +\bar{\beta}_{i}-\hat{\bm{\lambda}}^{\t}_i\bm{\Delta} +\left( \bm{\lambda}_i-\hat{\bm{\lambda}}_i\right)^{\t}\bar{\bm{f}}_{\rm post}}{\hat{\sigma}_i}\right).
\]
    Then $\bm{\Psi}(\hat{\bm{\Delta}})=0$, where $\hat{\bm{\Delta}} = \hat{\bm{f}}- \bar{\bm{f}}_{\rm post}$.
    We consider the following decomposition:
\[
\bm{\Psi}\left(\bm{\Delta}\right)=\bm{\Psi}_0\left(\bm{\Delta}\right) + \bm{\Psi}_{\rm nuis,1}\left(\bm{\Delta}\right) + \bm{\Psi}_{\rm nuis,2}\left(\bm{\Delta}\right),
\]
    where the first component is the clean part
\[
        \bm{\Psi}_0\left(\bm{\Delta}\right) = N^{-1}\sum_{i=1}^N\frac{\bm{\lambda}_i}{\sigma_i}\varphi\left(\frac{\bar{\varepsilon}_{\mathrm{post},i}-\bm{\lambda}^{\t}_i\bm{\Delta}}{\sigma_i}\right),
\]
    and the remaining two terms are nuisance components
\[
        \bm{\Psi}_{\rm nuis,1}\left(\bm{\Delta}\right) = N^{-1}\sum_{i=1}^N\frac{\bm{\lambda}_i}{\sigma_i}\left\{\varphi\left(\frac{\bar{\varepsilon}_{\mathrm{post},i} +\bar{\beta}_{i}-\hat{\bm{\lambda}}^{\t}_i\bm{\Delta} +\left( \bm{\lambda}_i-\hat{\bm{\lambda}}_i\right)^{\t}\bar{\bm{f}}_{\rm post}}{\hat{\sigma}_i}\right)-\varphi\left(\frac{\bar{\varepsilon}_{\mathrm{post},i}-\bm{\lambda}^{\t}_i\bm{\Delta}}{\sigma_i}\right)\right\},
\]
\[
        \bm{\Psi}_{\rm nuis,2}\left(\bm{\Delta}\right) = N^{-1}\sum_{i=1}^N\left(\frac{\hat{\bm{\lambda}}_i}{\hat{\sigma}_i}-\frac{\bm{\lambda}_i}{\sigma_i}\right)\varphi\left(\frac{\bar{\varepsilon}_{\mathrm{post},i} +\bar{\beta}_{i}-\hat{\bm{\lambda}}^{\t}_i\bm{\Delta} +\left( \bm{\lambda}_i-\hat{\bm{\lambda}}_i\right)^{\t}\bar{\bm{f}}_{\rm post}}{\hat{\sigma}_i}\right).
\]
    In the following, we show that the two nuisance components $\bm{\Psi}_{\rm nuis,1}(\bm{\Delta})$ and $\bm{\Psi}_{\rm nuis,2}(\bm{\Delta})$ are uniformly controlled as small terms for any set ${\|\bm{\Delta}}\|_2\leq \delta$.
  Finally, we show that the clean part can consistently recover the true $\bar{\bm{f}}_{\rm post}$.

    According to Assumption \ref{asmp: largeNsparse} on sparse interference effects, we divide the sums in $\bm{\Psi}_{\rm nuis,1}$ as two parts, either in the set of $S$ or in the set of $S^c$, and bound the L$_2$ norm separately:
    \[\begin{aligned}
    	\left\|\bm{\Psi}_{\rm nuis,1}\left(\bm{\Delta}\right)\right\|_2 &= \left\|N^{-1}\sum_{i\in S}\left(\cdot\right) +N^{-1}\sum_{i\in S^c}\left(\cdot\right) \right\|_2\\
    	&\leq \left\|N^{-1}\sum_{i\in S}\left(\cdot\right)\right\|_2 +\left\| N^{-1}\sum_{i\in S^c} \left(\cdot\right)\right\|_2.
    \end{aligned}\]
    For the sum under the index set $S$, we have as $N, T_0\to \infty$
    \[\begin{aligned} 
    &\left\|N^{-1}\sum_{i\in S}\frac{\bm{\lambda}_i}{\sigma_i}\left\{\varphi\left(\frac{\bar{\varepsilon}_{\mathrm{post},i} +\bar{\beta}_{i}-\hat{\bm{\lambda}}^{\t}_i\bm{\Delta} +\left( \bm{\lambda}_i-\hat{\bm{\lambda}}_i\right)^{\t}\bar{\bm{f}}_{\rm post}}{\hat{\sigma}_i}\right)-\varphi\left(\frac{\bar{\varepsilon}_{\mathrm{post},i}-\bm{\lambda}^{\t}_i\bm{\Delta}}{\sigma_i}\right)\right\}\right\|_2\\ 
    &\leq N^{-1}\sum_{i\in S}\frac{\|\bm{\lambda}_i\|_2}{\sigma_i}\left|\varphi\left(\frac{\bar{\varepsilon}_{\mathrm{post},i} +\bar{\beta}_{i}-\hat{\bm{\lambda}}^{\t}_i\bm{\Delta} +\left( \bm{\lambda}_i-\hat{\bm{\lambda}}_i\right)^{\t}\bar{\bm{f}}_{\rm post}}{\hat{\sigma}_i}\right)-\varphi\left(\frac{\bar{\varepsilon}_{\mathrm{post},i}-\bm{\lambda}^{\t}_i\bm{\Delta}}{\sigma_i}\right)\right|\\
    &\leq N^{-1}2C^3|S|\\
    & =o_p(1),
    \end{aligned}\]
    where the second inequality uses $\|\bm{\lambda}_i\|_2\leq C, 1/\sigma_i\leq C$ and $|\varphi|\leq C$, and the last inequality holds by the construction of set $S$.
    
    For the sum under the index set $S^{c}$ and within $\|\bm{\Delta}\|_2\leq \delta$, we have as $N, T_0\to \infty$
    \[\begin{aligned} 
    &\left\|N^{-1}\sum_{i\in S^c}\frac{\bm{\lambda}_i}{\sigma_i}\left\{\varphi\left(\frac{\bar{\varepsilon}_{\mathrm{post},i} +\bar{\beta}_{i}-\hat{\bm{\lambda}}^{\t}_i\bm{\Delta} +\left( \bm{\lambda}_i-\hat{\bm{\lambda}}_i\right)^{\t}\bar{\bm{f}}_{\rm post}}{\hat{\sigma}_i}\right)-\varphi\left(\frac{\bar{\varepsilon}_{\mathrm{post},i}-\bm{\lambda}^{\t}_i\bm{\Delta}}{\sigma_i}\right)\right\}\right\|_2\\ 
    &\leq N^{-1}\sum_{i\in S^c}\frac{\|\bm{\lambda}_i\|_2}{\sigma_i}\left|\varphi\left(\frac{\bar{\varepsilon}_{\mathrm{post},i} +\bar{\beta}_{i}-\hat{\bm{\lambda}}^{\t}_i\bm{\Delta} +\left( \bm{\lambda}_i-\hat{\bm{\lambda}}_i\right)^{\t}\bar{\bm{f}}_{\rm post}}{\hat{\sigma}_i}\right)-\varphi\left(\frac{\bar{\varepsilon}_{\mathrm{post},i}-\bm{\lambda}^{\t}_i\bm{\Delta}}{\sigma_i}\right)\right|\\
    &\leq N^{-1}\sum_{i\in S^c} C^3 \left|\frac{\bar{\varepsilon}_{\mathrm{post},i} +\bar{\beta}_{i}-\hat{\bm{\lambda}}^{\t}_i\bm{\Delta} +\left( \bm{\lambda}_i-\hat{\bm{\lambda}}_i\right)^{\t}\bar{\bm{f}}_{\rm post}}{\hat{\sigma}_i}-\frac{\bar{\varepsilon}_{\mathrm{post},i}-\bm{\lambda}^{\t}_i\bm{\Delta}}{\sigma_i}\right|\\
    &\leq C^3\left\{N^{-1}\left\|\bar{\bm{\beta}}_{S^c}\right\|_1\max_{1\leq i\leq N}\left|\frac{1}{\hat{\sigma}_i}\right| + N^{-1}\sum_{i\in S^c}\left(\left|\bar{\varepsilon}_{\mathrm{post},i}\right|+\left\|\bm{\lambda}_i\right\|_2\left\|\bm{\Delta}\right\|_2\right)\max_{1\leq i\leq N}\left|\frac{1}{\hat{\sigma}_i} - \frac{1}{\sigma_i}\right|\right.\\
    &\qquad\left.+ N^{-1}\left(\sum_{i=1}^N \frac{1}{\hat{\sigma}_i^2}\left\|\bm{\lambda}_i-\hat{\bm{\lambda}}_i\right\|_2^2\right)^{1/2}\left(\left\|\bar{\bm{f}}_{\rm post}\right\|_2+\left\|\bm{\Delta}\right\|_2\right)\right\}\\
    & =o_p(1),
    \end{aligned}\]
    where the second inequality uses boundedness of $\bm{\Lambda}$ and $\varphi^{\prime}$, the third inequality applies the triangle inequality and the Cauchy--Schwarz inequality, and the last inequality is from Assumption \ref{asmp: largeNsparse} that $N^{-1}\|\bar{\bm{\beta}}_{S^c}\|_1=o(1)$, Condition \ref{asmp: AFMsupport} that $\hat{\sigma}_i\in[C^{-1},C]$, $N^{-1}\sum_{i=1}^N|\bar{\varepsilon}_{\mathrm{post},i}| = O_p(1)$ by Markov inequality, $\|\bar{\bm{f}}_{\rm post}\|_2\leq (T-T_0)^{-1}\sum_{t=T_0+1}^T\|\bm{f}_t\|_2\leq C$ by Assumption \ref{asmp: AFM}, $N^{-1}\sum_{i=1}^N \hat{\sigma}_i^{-2}\|\bm{\lambda}_i-\hat{\bm{\lambda}}_i\|_2^2 = O_p(T_0^{-1}) + O_p(N^{-2})= o_p(1)$ by Proposition 1 of \cite{bai2016maximum}, and Lemma S.\ref{lemma: maxrate}.
    Thus, we have
    \[\sup_{\|\bm{\Delta}\|\leq \delta}\|\bm{\Psi}_{\rm nuis,1}\|_2 = o_p(1). \]
    
    For $\bm{\Psi}_{\rm nuis,2}$, it is controlled by
    \[\begin{aligned}
        \sup_{\|\bm{\Delta}\| \leq \delta}\|\bm{\Psi}_{\rm nuis,2}\|_2
        &\leq N^{-1}\sum_{i=1}^N C\left\|\frac{\hat{\bm{\lambda}}_i}{\hat{\sigma}_i}-\frac{\bm{\lambda}_i}{\sigma_i}\right\|_2 = o_p(1),\\
        &\leq C \left(N^{-1}\sum_{i=1}^N\frac{1}{\hat{\sigma}_i^2}\left\|\hat{\bm{\lambda}}_i-\bm{\lambda}_i\right\|_2^2\right)^{1/2} + CN^{-1}\sum_{i=1}^N\left\|\bm{\lambda}_i\right\|_2\max_{1\leq i\leq N}\left|\frac{1}{\hat{\sigma}_i}-\frac{1}{\sigma_i}\right|\\
        &=o_p(1),
    \end{aligned}\]
    where we use the boundedness of $\varphi(\cdot)$ and the same arguments used in the last inequality of controlling $\|\bm{\Psi}_{\mathrm{nuis},1}\|_2$.
    Together, we have as $N, T_0\to\infty$,
    \begin{equation}\label{eq: rrcon1}
\sup_{\|\bm{\Delta}\|_2\leq\delta}\left\|\bm{\Psi}\left(\bm{\Delta}\right) -\bm{\Psi}_0\left(\bm{\Delta}\right)\right\|_2 = \sup_{\|\bm{\Delta}\|_2\leq\delta}\left\|\bm{\Psi}_{\rm nuis,1}\left(\bm{\Delta}\right) + \bm{\Psi}_{\rm nuis,2}\left(\bm{\Delta}\right)\right\|_2 = o_p(1).
    \end{equation}

    Let 
    \[\bm{g}_i(\bm{\Delta}) = \frac{1}{\sigma_i}\bm{\lambda}_i\varphi\left(\frac{\bar{\varepsilon}_{\mathrm{post},i} -\bm{\lambda}_i^{\t}\bm{\Delta}}{\sigma_i}\right).\] 
    Fix $\delta$, we next show that as $N\to\infty$
    \begin{equation}\label{eq: rrlln}
        \sup_{\|\bm{\Delta}\|_2\leq \delta}\left\|\bm{\Psi}_0\left(\bm{\Delta}\right) - N^{-1}\sum_{i=1}^NE\left\{\bm{g}_i\left(\bm{\Delta}\right)\right\}\right\|_2 = \sup_{\|\bm{\Delta}\|\leq \delta} \left\| N^{-1} \sum_{i=1}^N \left\{\bm{g}_i\left(\bm{\Delta}\right)-E\left\{\bm{g}_i\left(\bm{\Delta}\right)\right\}\right\} \right\|_2 = o_p(1).
    \end{equation}
    By Assumption \ref{asmp: AFM}, for two $\bm{\Delta}, \bm{\Delta}'$,
\[\begin{aligned}
\|\bm{g}_i(\bm{\Delta}) - \bm{g}_i(\bm{\Delta}')\|_2 &\leq \frac{\left\|\bm{\lambda}_i\right\|_2}{\sigma_i}\left\|\varphi\left(\frac{\varepsilon_{it} -\bm{\lambda}_i^{\t}\bm{\Delta}}{\sigma_i}\right) -\varphi\left(\frac{\varepsilon_{it} -\bm{\lambda}_i^{\t}\bm{\Delta}^{\prime}}{\sigma_i}\right)\right\|_2 \\
&\leq C^3 \left|\frac{\bm{\lambda}_i^{\t} (\bm{\Delta} - \bm{\Delta}')}{\sigma_i}\right| \\
&\leq C^5 \left\|\bm{\Delta} - \bm{\Delta}'\right\|_2.
\end{aligned}\]
As a result, $\bm{g}_i(\bm{\Delta})$ is Lipschitz-continuous in $\bm{\Delta}$ with Lipschitz-constant $C^5$.

The set $\|\bm{\Delta}\|\leq \delta$ is compact with respect to $\bm{\Delta}$, so there exists a finite $\eta$-net with centers of $\{\bm{\Delta}^{(1)}, \ldots, \bm{\Delta}^{(m)}\}$ covering the set. 
For each net point $1\leq j\leq m$, the standard Chebyshev inequality yields
\[N^{-1} \sum_{i=1}^N \left\{\bm{g}_i(\bm{\Delta}^{(j)}) - E\left\{\bm{g}_i(\bm{\Delta}^{(j)})\right\}\right\} = o_p(1),\]
as $N \to \infty$.
Below, we provide more details.
Note that for each coordinate $1\leq k\leq r$,
\[\begin{aligned}
    |\operatorname{Cov}\left\{g_{ik}\left(\bm{\Delta}\right),g_{jk}\left(\bm{\Delta}\right)\right\}| &= |\lambda_{ik}\lambda_{jk}|\operatorname{Cov}\left(\varphi\left\{\left(\bar{\varepsilon}_{\mathrm{post},i} -\bm{\lambda}_i^{\t}\bm{\Delta}\right)/\sigma_i\right\},\varphi\left\{\left(\bar{\varepsilon}_{\mathrm{post},j} -\bm{\lambda}_j^{\t}\bm{\Delta}\right)/\sigma_j\right\}\right)\\
    &\leq C^2 (C^5)^2|\operatorname{Cov}\left(\bar{\varepsilon}_{\mathrm{post},i},\bar{\varepsilon}_{\mathrm{post},j}\right)|\\
    &\leq C^{12} (T-T_0)^{-2}\sum_{t=T_0+1}^{T} \sum_{s=T_0+1}^{T} |\gamma_{ij,ts}|,
\end{aligned}\]
where the last inequality is implied using the Hoeffding's representation that $\operatorname{Cov}\{f(X), h(Y)\} = \int f^{\prime}(x)h^{\prime}(y)\{F_{X,Y}(x,y) - F_{X}(x)F_{Y}(y)\}\mathrm{d}x
\mathrm{d}y$ and $\varphi\{\left(\bar{\varepsilon}_{\mathrm{post},i} -\bm{\lambda}_i^{\t}\bm{\Delta}\right)/\sigma_i\}$ is Lipschitz-continuous with constant $C^5$.
Then, by Condition \ref{asmp: AFMmoments}, we have 
\[\operatorname{Var}\left(N^{-1}\sum_{i=1}^Ng_{ik}\left(\bm{\Delta}\right)\right) \leq C^{12}N^{-2}(T-T_0)^{-2}\sum_{i=1}^N\sum_{j=1}^N\sum_{t=T_0+1}^{T}\sum_{s=T_0+1}^{T} 
|\gamma_{ij,ts}|\leq C^{13}N^{-1}(T-T_0)^{-1},\]
which is $o(1)$ as $N\to\infty$ and justifies the Chebyshev claim.

Next, a finite union bound gives the same result simultaneously for all $j$. 
For arbitrary $\bm{\Delta}$ that $\|\bm{\Delta}\|_2\leq \delta$, choose $\bm{\Delta}^{(j)}$ with $\|\bm{\Delta} - \bm{\Delta}^{(j)}\|_2 \leq \eta$; then 
\[\left\|N^{-1} \sum_{i=1}^N \left\{\bm{g}_i(\bm{\Delta}) - E\left\{\bm{g}_i\left(\bm{\Delta}\right)\right\}\right\}\right\| \leq \left\|N^{-1} \sum_{i=1}^N \left\{\bm{g}_i(\bm{\Delta}^{(j)}) - E\left\{\bm{g}_i(\bm{\Delta}^{(j)})\right\}\right\}\right\| + 2C^5 \eta.\]
Thus, letting $N \to \infty$ then $\eta \downarrow 0$ gives the desired result of \eqref{eq: rrlln}.

Denote ${\bm{\Psi}}^*(\bm{\Delta}) = N^{-1}\sum_{i=1}^NE\{\bm{g}_i\left(\bm{\Delta}\right)\}$. 
Then $\bm{\Psi}^*(0) = 0$ by Condition \ref{asmp: lossmatrix}, and we have proved that $\sup_{\|\Delta\|_2\le \delta}\|\bm{\Psi}(\bm{\Delta})-\bm{\Psi}^*(\bm{\Delta})\|_2
= o_p(1)$.

Recall $\hat{\bm{\Delta}}=\hat{\bm f}-\bar{\bm f}_{\rm post}$ satisfies $\bm{\Psi}(\hat{\bm{\Delta}})=0$. 
Fix any $\varepsilon>0$. 
We show that for some (arbitrarily small) constant $\delta>0$ and  $N$ large enough,
$\pr(\|\hat{\bm{\Delta}}\|_2\ge \delta)\le \varepsilon$, which implies $\hat{\bm{\Delta}}=o_p(1)$.

For any constant $\eta>0$ and fixed $\delta>0$, define the event
\[
\mathcal E_N(\eta,\delta):=\left\{\sup_{\|\Delta\|_2\le \delta}\|\Psi(\Delta)-\Psi^*(\Delta)\|_2\le \eta\right\},
\]
then $P\{\mathcal E_N(\eta,\delta)\}\to 1$ as $N$, $T_0\to\infty$.
For $\bm{\Delta}\in\mathbb R^r$, define the Jacobian matrix as
\[
\bm{J}_N(\bm{\Delta}) =
N^{-1}\sum_{i=1}^N \sigma_i^{-2}\bm\lambda_i\bm\lambda_i^{\t}\,
E\!\left\{\varphi'\!\left(\frac{\bar\varepsilon_{{\rm post},i}-\bm{\lambda}_i^{\t}\bm{\Delta}}{\sigma_i}\right)\right\}.
\]
Since $\varphi$ has bounded derivatives, $\varphi'$ is Lipschitz; together with bounded $\|\bm\lambda_i\|_2$ and
$C^{-1}\le \sigma_i\le C$, there exists a constant $L<\infty$ such that for all sufficiently small $\delta>0$,
\[
\sup_{\|\bm{\Delta}\|_2\le \delta}\|\bm{J}_N(\bm{\Delta})-\bm{J}_N(\bm{0})\|_2\le L\delta .
\]
Condition \ref{asmp: lossmatrix} implies $\bm{J}_N(0)\to \tilde{\bm{Q}}$ for some positive definite $\tilde{\bm{Q}}$ as $N\to\infty$, hence there exist $N_0$ and a constant $c_0>0$ such that
$\lambda_{\min}\{\bm{J}_N(0)\}\ge 2c_0$ for all $N\ge N_0$. Choose and fix $\delta>0$ so that $L\delta\le c_0$.
Then for all $N\ge N_0$ and all $\|\bm{\Delta}\|_2\le \delta$,
\[
\lambda_{\min}\{\bm{J}_N(\bm{\Delta})\}\ge \lambda_{\min}\{\bm{J}_N(0)\}-\|\bm{J}_N(\Delta)-\bm{J}_N(0)\|_2\ge 2c_0-c_0=c_0.
\]
Differentiating under the expectation gives $\nabla \bm{\Psi}^*(\bm{\Delta})=-\bm{J}_N(\bm{\Delta})$, and thus
\[
\bm{\Psi}^*(\bm{\Delta})-\bm{\Psi}^*(0)=\int_0^1 \nabla\bm{\Psi}^*(t\bm{\Delta})\,\bm{\Delta}\,dt
=-\int_0^1 \bm{J}_N(t\bm{\Delta})\bm{\Delta}\,dt.
\]
Let $\|\bm{u}\|_2=1$. Setting $\bm{\Delta}=\delta \bm{u}$ and using $\bm{\Psi}^*(0)=0$ yields
\[
\bm{u}^{\t}\bm{\Psi}^*(\delta \bm{u})
=-\delta \int_0^1 \bm{u}^{\t} \bm{J}_N(\bm{\Delta})\bm{u}\,dt
\le -\delta\int_0^1 c_0\,dt
= -c_0\delta .
\]

On $\mathcal E_N(c_0\delta/2,\delta)$, for every $\|\bm{u}\|_2=1$,
\[
\bm{u}^{\t}\bm{\Psi}(\delta \bm{u})
\le \bm{u}^{\t}\bm{\Psi}^*(\delta \bm{u})+\|\bm{\Psi}(\delta \bm{u})-\bm{\Psi}^*(\delta \bm{u})\|_2
\le -c_0\delta + c_0\delta/2
= -c_0\delta/2<0.
\]
Because $\rho$ is convex, $L(\bar{\bm f}_{\rm post}+t\bm{u})$ is convex in $t$, and hence for each $\bm{u}$ the map
$t\mapsto \bm{u}^{\t}\bm{\Psi}(t\bm{u})$ is non-increasing on $t\ge 0$. Therefore, still on $\mathcal E_N(c_0\delta/2,\delta)$,
for all $t\ge \delta$,
\[
\bm{u}^{\t}\bm{\Psi}(t\bm{u})\le \bm{u}^{\t}\bm{\Psi}(\delta \bm{u})<0,
\]
which implies $\bm{\Psi}(t\bm{u})\neq 0$ for all $t\ge\delta$, i.e., $\bm{\Psi}(\bm{\Delta})\neq 0$ whenever $\|\bm{\Delta}\|_2\ge\delta$.
Since $\bm{\Psi}(\hat{\bm{\Delta}})=0$, we conclude that on $\mathcal E_N(c_0\delta/2,\delta)$ necessarily $\|\hat{\bm{\Delta}}\|_2<\delta$.
Letting $N$, $T_0\to\infty$, $\pr\{\mathcal E_N(c_0\delta/2,\delta)\}\to 1$ and $
\pr(\|\hat{\bm{\Delta}}\|_2\ge \delta)\to 0$.
As $\delta$ can be arbitrarily small, we have $\hat{\bm{f}} - \bar{\bm{f}}_{\rm post} = o_p(1)$.

Then, $\hat{\bm{\beta}} = \bar{\bm{Y}}_{\rm post} - \hat{\bm{\Lambda}}\hat{\bm{f}} = \bar{\bm{\beta}} + \bar{\bm{\varepsilon}}_{\rm post}+\bm{\Lambda}\bar{\bm{f}}_{\rm post} -\hat{\bm{\Lambda}}\hat{\bm{f}} = \bar{\bm{\beta}} + \bar{\bm{\varepsilon}}_{\rm post} + o_p(1),$ as $N, T_0\to\infty$.
The above proof is valid for arbitrary $T-T_0$.
If $T-T_0\to\infty$, $\bar{\bm{\varepsilon}}_{\rm post} = o_p(1)$ and we have $\hat{\bm{\beta}} = \bar{\bm{\beta}} +o_p(1)$.

\end{proof}

\begin{proof}[Proof of Theorem 2 (ii)]
In this part, $T-T_0\to\infty$ and we assume $T_0/T \to \kappa\in [0,1]$ as $T_*\to\infty$, where recall $T_* = \min\{T_0, T-T_0\}$.
    Consider the first-order equation $\bm{\Phi}(\hat{\bm{f}}) = 0$ and Taylor's expansion:
    \begin{equation}\label{eq: rrtarloyexpan}
    0 = \bm{\Phi}(\hat{\bm{f}}) = 
        \bm{\Phi}\left(\bar{\bm{f}}_{\rm post}\right) + \nabla \bm{\Phi}\left(\bar{\bm{f}}_{\rm post}\right)\left(\hat{\bm{f}}-\bar{\bm{f}}_{\rm post}\right)+o_p\left(\left\|\hat{\bm{f}}-\bar{\bm{f}}_{\rm post}\right\|_2\right).
    \end{equation}
    We have $\hat{\bm{f}} - \bar{\bm{f}}_{\rm post} = -\{\nabla\bm{\Phi}(\bar{\bm{f}}_{\rm post})+o_p(1)\}^{-1}\bm{\Phi}(\bar{\bm{f}}_{\rm post})$ and
    \begin{equation}\label{eq: normalexpan}
        \begin{aligned}
        T_*^{1/2}\left(\hat{\beta}_{i} - \bar{\beta}_i\right) &= T_*^{1/2}\bar{\varepsilon}_{\mathrm{post},i} + T_*^{1/2}\left(\bm{\lambda}_i - \hat{\bm{\lambda}}_i\right)^{\t}\bar{\bm{f}}_{\rm post} -T_*^{1/2}\hat{\bm{\lambda}}_i^{\t}\left(\hat{\bm{f}}-\bar{\bm{f}}_{\rm post}\right)\\
        &= T_*^{1/2}\bar{\varepsilon}_{\mathrm{post},i} + T_*^{1/2}\left(\bm{\lambda}_i - \hat{\bm{\lambda}}_i\right)^{\t}\bar{\bm{f}}_{\rm post} -\hat{\bm{\lambda}}_i^{\t}\left\{\nabla\bm{\Phi}\left(\bar{\bm{f}}_{\rm post}\right)+o_p(1)\right\}^{-1}T_*^{1/2}\bm{\Phi}\left(\bar{\bm{f}}_{\rm post}\right).
    \end{aligned}
    \end{equation}
Below we prove that $T_*^{1/2}\bm{\Phi}(\bar{\bm{f}}_{\rm post}) = o_p(1)$ and $\{\nabla\bm{\Phi}\left(\bar{\bm{f}}_{\rm post}\right)+o_p(1)\}^{-1} = O_p(1)$.
The following is parallel to the proof of \cite{wang2017confounder}.

Consider
\[\begin{aligned}
    \bm{\Phi}\left(\bar{\bm{f}}_{\rm post}\right) &= N^{-1}\sum_{i=1}^N\frac{\hat{\bm{\lambda}}_i}{\hat{\sigma}_i}\varphi\left(\frac{\bar{Y}_{i} -\hat{\bm{\lambda}}_i^{\t}\bar{\bm{f}}_{\rm post}}{\hat{\sigma}_i}\right)\\
    &= N^{-1}\sum_{i=1}^N\frac{\hat{\bm{\lambda}}_i}{\hat{\sigma}_i}\varphi\left(\frac{\bar{\beta}_i +\bm{\lambda}_i^{\t}\bar{\bm{f}}_{\rm post}-\hat{\bm{\lambda}}_i^{\t}\bar{\bm{f}}_{\rm post}+\bar{\varepsilon}_{\mathrm{post},i}}{\hat{\sigma}_i}\right)\\
    &= N^{-1}\sum_{i=1}^N\frac{\hat{\bm{\lambda}}_i}{\hat{\sigma}_i}\varphi\left(\frac{\bar{\beta}_i -\left(T_0^{-1}\sum_{t=1}^{T_0}\bm{f}_t\varepsilon_{it}\right)^{\t}\bar{\bm{f}}_{\rm post}+ \tilde{u}_i +\bar{\varepsilon}_{\mathrm{post},i}}{\sigma_i+\delta_i}\right),
\end{aligned}\]
where $\tilde{u}_i = \bm{u}_i^{\t}\bar{\bm{f}}_{\rm post}$, $\bm{u}_i = \bm{\lambda}_i -\hat{\bm{\lambda}}_i + T_0^{-1}\sum_{t=1}^{T_0}\bm{f}_t\varepsilon_{it}$. 
By Lemma S.\ref{lemma: maxrate} and $\|\bar{\bm{f}}_{\rm post}\|_2 \leq (T-T_0)^{-1}\sum_{t=T_0+1}^T\|\bm{f}_{t}\|_2 \leq C$ given Assumption \ref{asmp: AFM}, we have $\max_{1\leq i\leq N}|\tilde{u}_i|=o_p(T_0^{-1/2})$ and $\max_{1\leq i\leq N}|\delta_i|=o_p(1)$.
Using both $\varphi$ and $\varphi^{\prime}$ are bounded, the results of Lemma S.\ref{lemma: maxrate}, and the sparsity Assumption \ref{asmp: largeNsparse}(b), we have 
\[\bm{\Phi}(\bar{\bm{f}}_{\rm post}) = N^{-1}\sum_{i=1}^N\frac{\hat{\bm{\lambda}}_i}{\hat{\sigma}_i}\varphi\left\{\frac{-\left(T_0^{-1}\sum_{t=1}^{T_0}\bm{f}_t\varepsilon_{it}\right)^{\t}\bar{\bm{f}}_{\rm post}+ \tilde{u}_i +\bar{\varepsilon}_{\mathrm{post},i}}{\sigma_i+\delta_i}\right\} + o_p(T_*^{-1/2}),\]
where we divide the summation as within the index set $\mathcal{S}$ and $\mathcal{S}^c$ and control them separately as in our proof of Theorem \ref{thm: largeNasymptotics}(i).
Let $g_i$ be the expression inside $\varphi$ in the last equation omitting $\tilde{u}_i$ and $\delta_i$, that is,
\[g_i = -\sigma_i^{-1}\left(T_0^{-1}\sum_{t=1}^{T_0}\bm{f}_t\varepsilon_{it}\right)^{\t}\bar{\bm{f}}_{\rm post} +\sigma_i^{-1}\bar{\varepsilon}_{\mathrm{post},i}.\]
The random variable $g_i$ satisfies $E(g_i) = 0$ and $\max_{1\leq i\leq N}E|g_i| = O\{T_0^{-1/2}+(T-T_0)^{-1/2}\}$.
For the latter claim, consider the variance of sums
\[
\begin{aligned}
E\left\|T_0^{-1}\sum_{t=1}^{T_0}\bm{f}_t\,\varepsilon_{it}/\sigma_i\right\|_2^2
&\leq T_0^{-2}\sum_{t=1}^{T_0}\sum_{s=1}^{T_0} \left|\bm{f}_t^{\t} \bm{f}_s\right|\,
\left|E\!\left(\varepsilon_{it}\varepsilon_{is}\right)\right|/\sigma_i^{2}\\
&\leq T_0^{-2}C^4\sum_{t=1}^{T_0}\sum_{s=1}^{T_0}\tau_{ts} \leq C^5T_0^{-1},
\end{aligned}
\]
and 
\[
\begin{aligned}
E(\bar\varepsilon_{{\rm post},i}^2/\sigma_i^2)
&\leq\left(T-T_0\right)^{-2}\sum_{t=T_0+1}^{T}\sum_{s=T_0+1}^{T}
\left|E\left(\varepsilon_{it}\varepsilon_{is}\right)\right|/\sigma_i^{2} \notag\\
&\le C^2\,\left(T-T_0\right)^{-2}\sum_{t=T_0+1}^{T}\sum_{s=T_0+1}^{T}
\tau_{ts} \le C^3\,(T-T_0)^{-1},
\end{aligned}
\]
where we use the correlation restriction in Condition \ref{asmp: AFMdep}(i).
Hence, by $(E|X|)^2\leq E(X^2)$, we have $\max_{1\leq i\leq N}E|T_0^{-1}\sum_{t=1}^{T_0}\bm{f}_t^{\t}\varepsilon_{it}/\sigma_i|=O(T_0^{-1/2})$ and $\max_{1\leq i\leq N}E|\bar{\varepsilon}_{\mathrm{post},i}/\sigma_i| = O\{(T-T_0)^{-1/2}\}$.
And the rate of $\max_{1\leq i\leq N}|g_i|$ follows from the triangle inequality.

Then using $|\varphi^{\prime}|\leq C$, we have
\[\begin{aligned}
&\left\|N^{-1}\sum_{i=1}^{N}\frac{\hat{\bm{\lambda}}_i}{\hat{\sigma}_i}\left\{\varphi\left(g_i + \frac{\tilde{u}_i - \delta_i g_i}{\hat{\sigma}_i}\right) - \varphi\left(g_i\right)\right\}\right\|_2\\
&\leq C \cdot \left\|N^{-1}\sum_{i=1}^{N}\frac{\hat{\bm{\lambda}}_i}{\hat{\sigma}_i^2}\left(\left|\tilde{u}_i\right| + \left|g_i\right|\left|\delta_i\right|\right)\right\|_2 = o_p\left\{T_0^{-1/2}+\left(T-T_0\right)^{-1/2}\right\},
\end{aligned}\]
where the last inequality is because $\max_{1\leq i\leq N}|\tilde{u}_i|=o_p(T_0^{-1/2})$, $\max_{1\leq i\leq N}|\delta_i|=o_p(1)$, $\max_{1\leq i\leq N}|\hat{\bm{\lambda}}_i|=O_p(1)$ by Lemma S.\ref{lemma: maxrate}, and
\[\begin{aligned}
    \left\|N^{-1}\sum_{i=1}^{N} \frac{\hat{\bm{\lambda}}_i}{\hat{\sigma}_i^2}\left|g_i\right| \left|\delta_i\right|\right\|_2&\leq \left(N^{-1}\sum_{i=1}^{N} |g_i|\right)\max_{1\leq i\leq N}\left(\frac{1}{\hat{\sigma}_i^2}\left|\delta_i\right|\left\|\hat{\bm{\lambda}}_i\right\|_2\right)\\
    &= O_p\left(\max_{1\leq i\leq N}E|g_i|\right) \max_{1\leq i\leq N}\left(\frac{1}{\hat{\sigma}_i^2}\left|\delta_i\right|\left\|\hat{\bm{\lambda}}_i\right\|_2\right)\\
    &= O_p\{T_0^{-1/2}+(T-T_0)^{-1/2}\}o_p(1)\\
    &= o_p\left\{T_0^{-1/2}+\left(T-T_0\right)^{-1/2}\right\}.
\end{aligned}\]
Next, we have 
\[\begin{aligned}
\left\|\bm{\Phi}\left(\bar{\bm{f}}_{\rm post}\right)\right\|_2 &= \left\|N^{-1}\sum_{i=1}^{N}\frac{\hat{\bm{\lambda}}_i}{\hat{\sigma}_i}\varphi\left(g_i + \frac{\tilde{u}_i - \delta_i g_i}{\hat{\sigma}_i}\right)\right\|_2 + o_p(T_*^{-1/2})\\
&= \left\|N^{-1}\sum_{i=1}^{N}\frac{\hat{\bm{\lambda}}_i}{\hat{\sigma}_i}\varphi(g_i)\right\|_2 + o_p\left\{T_0^{-1/2}+\left(T-T_0\right)^{-1/2}\right\}\\
&= \left\|N^{-1}\sum_{i=1}^{N}\frac{\bm{\lambda}_i}{\sigma_i}\varphi(g_i)\right\|_2 + o_p\left\{T_0^{-1/2}+\left(T-T_0\right)^{-1/2}\right\}\\
&= \left\|N^{-1}\sum_{i=1}^{N}\frac{\bm{\lambda}_i}{\sigma_i}\varphi'(0)g_i\right\|_2 + o_p\left\{T_0^{-1/2}+\left(T-T_0\right)^{-1/2}\right\}\\
&= o_p\left\{T_0^{-1/2}+\left(T-T_0\right)^{-1/2}\right\}.
\end{aligned}\]
Below, we give proofs for the third, fourth, and fifth equality.
By the triangle inequality, we have
\[
   \left\|N^{-1}\sum_{i=1}^{N}\left(\frac{\hat{\bm{\lambda}}_i}{\hat{\sigma}_i}-\frac{\bm{\lambda}_i}{\sigma_i}\right)\varphi(g_i)\right\|_2\leq \max_{1\leq i\leq N}\left\|\frac{\hat{\bm{\lambda}}_i}{\hat{\sigma}_i}-\frac{\bm{\lambda}_i}{\sigma_i}\right\|_2\left(N^{-1}\sum_{i=1}^N\left|\varphi\left(g_i\right)\right|\right).
\]
The first term of the right-hand is $o_p(1)$ by Lemma S.\ref{lemma: maxrate} and Assumption \ref{asmp: AFM}.
As $\varphi(0)=0$ and $|\varphi^{\prime}|\leq C$, the second term
\[
    N^{-1}\sum_{i=1}^N|\varphi(g_i)| \leq CN^{-1}\sum_{i=1}^N\left|g_i\right| = O_p\left(\max_{1\leq i\leq N}E|g_i|\right) = O_p\left\{T_0^{-1/2}+\left(T-T_0\right)^{-1/2}\right\},
\]
where we use Markov inequality and the previous results about $\max_{1\leq i\leq N}E|g_i|$.
This completes the proof of the third equality.

Next, we use $|\varphi(g_i) - \varphi^{\prime}(0)g_i|\leq Cg_i^2/2$ by $|\varphi^{\prime\prime}|\leq C$
\[N^{-1}\sum_{i=1}^N\left|\varphi(g_i)-\varphi^{\prime}\left(0\right)g_i\right| \leq N^{-1}\sum_{i=1}^Ng_i^2  =  O_p\left\{\max_{1\leq i\leq N}E\left(g_i^2\right)\right\}.\]
Denote $d_t = \bm{f}_t^{\t}\bar{\bm{f}}_{\rm post}$, then $|d_t|\leq \|\bm{f}_t\|_2\cdot \|\bar{\bm{f}}_{\rm post}\|_2\leq C^2$. 
Using the uniformly boundedness of $\|\bm{f}_t\|_2, \|\bar{\bm{f}}_{\rm post}\|_2,\sigma_i^{-1}, d_t$, we have
\[\begin{aligned}
    E\left(g_i^2\right)
&\leq 
\frac{1}{\sigma_i^2}\left\{
T_0^{-2}\sum_{t=1}^{T_0}\sum_{s=1}^{T_0}\left|d_t d_s\right|\tau_{i,ts}
+T_0^{-1}(T-T_0)^{-1}\sum_{t=1}^{T_0}\sum_{s=T_0+1}^{T} \left|d_t\right|\tau_{i,ts}\right.\\
&\qquad\left.
+T_0^{-1}(T-T_0)^{-1}\sum_{t=T_0+1}^{T}\sum_{s=1}^{T_0} \left|d_s\right|\tau_{i,ts}
+(T-T_0)^{-2}\sum_{t=T_0+1}^{T}\sum_{s=T_0+1}^{T} \tau_{i,ts}
\right\}\\
&\leq C^7\left\{T_0^{-1}+T_0^{-1}+\left(T-T_0\right)^{-1}+\left(T-T_0\right)^{-1}\right\}\\
& = 2C^7\left\{T_0^{-1}+\left(T-T_0\right)^{-1}\right\},
\end{aligned}
\]
which implies $\max_{1\leq i\leq N}E(g_i^2) = O_p\{T_0^{-1}+(T-T_0)^{-1}\} = o_p\{T_0^{-1/2}+(T-T_0)^{-1/2}\}$ and completes the proof of the fourth equality.

Finally, let $\bm{a}_i = \bm{\lambda}_i/\sigma_i\varphi^{\prime}(0)$ that $\|\bm{a}_i\|_2\leq C^3$, by Condition \ref{asmp: AFMmoments}, we have
\[
\begin{aligned}
E\left\|N^{-1}\sum_{i=1}^N \bm{a}_i g_i\right\|_2^2
&\le N^{-2}\sum_{i=1}^N\sum_{j=1}^N \left|\bm{a}_i^{\t}\bm{a}_j\right|\left|\operatorname{Cov}\left(g_i,g_j\right)\right| \\
&\le C^6\,N^{-2}\sum_{i=1}^N\sum_{j=1}^N \left|\operatorname{Cov}\left(g_i,g_j\right)\right|\\
&\leq C^6N^{-2}\sum_{i=1}^N\sum_{j=1}^N
\frac{1}{\sigma_i\sigma_j}\left\{
T_0^{-2}\sum_{t=1}^{T_0}\sum_{s=1}^{T_0} \left|d_t d_s\right|\gamma_{ij,ts}
+T_0^{-1}(T-T_0)^{-1}\sum_{t=1}^{T_0}\sum_{s=T_0+1}^{T} \left|d_t\right|\gamma_{ij,ts}\right.\\
&\qquad\left.
+T_0^{-1}(T-T_0)^{-1}\sum_{t=T_0+1}^{T}\sum_{s=1}^{T_0} \left|d_s\right|\gamma_{ij,ts}
+(T-T_0)^{-2}\sum_{t=T_0+1}^{T}\sum_{s=T_0+1}^{T} \gamma_{ij,ts}
\right\}\\
&\leq C^{12}N^{-2}\left\{T_0^{-2}\sum_{i=1}^N\sum_{j=1}^N\sum_{t=1}^{T_0}\sum_{s=1}^{T_0} \gamma_{ij,ts}
+T_0^{-1}(T-T_0)^{-1}\sum_{i=1}^N\sum_{j=1}^N\sum_{t=1}^{T_0}\sum_{s=T_0+1}^{T} \gamma_{ij,ts}\right.\\
&\qquad\left.+T_0^{-1}(T-T_0)^{-1}\sum_{i=1}^N\sum_{j=1}^N\sum_{t=T_0+1}^{T}\sum_{s=1}^{T_0}\gamma_{ij,ts}
+(T-T_0)^{-2}\sum_{i=1}^N\sum_{j=1}^N\sum_{t=T_0+1}^{T}\sum_{s=T_0+1}^{T} \gamma_{ij,ts} \right\}\\
&\leq C^{13}N^{-2}\left\{NT_0^{-1}+N\left(T-T_0\right)^{-1}+NT_0^{-1}+N\left(T-T_0\right)^{-1}\right\}\\
& = O\left(N^{-1}\left\{T_0^{-1}+\left(T-T_0\right)^{-1}\right\}\right).
\end{aligned}
\]
As a result, $E\|N^{-1}\sum_{i=1}^N\bm{a}_ig_i\|_2 = o_p\{T_0^{-1/2}+(T-T_0)^{-1/2}\}$.
By Markov's inequality, $\|N^{-1}\sum_{i=1}^N\bm{a}_ig_i\|_2 = o_p\{T_0^{-1/2}+(T-T_0)^{-1/2}\}$, which completes the proof of the fifth equality.
So far, we have proved that $\bm{\Psi}(\bar{\bm{f}}_{\rm post}) = o_p\{T_0^{-1/2}+(T-T_0)^{-1/2}\}$, so $T_*^{1/2}\bm{\Psi}(\bar{\bm{f}}_{\rm post}) = o_p(1)$.

Now consider
\[\begin{aligned}
\left\{\nabla\Phi(\bar{\bm{f}}_{\rm post})\right\}^{-1} &= \left\{N^{-1}\sum_{i=1}^{N}\varphi'\left(g_i + \frac{\tilde{u}_i - \delta_i g_i}{\hat{\sigma}_i}\right)\hat{\bm{\lambda}}_i\hat{\bm{\lambda}}_i^{\t}/\hat{\sigma}_i^2 + o_p(1)\right\}^{-1}\\
&= \left\{N^{-1}\sum_{i=1}^{N}\varphi'(0)\bm{\lambda}_i\bm{\lambda}_i^{\t}/\sigma_i^2 + o_p(1)\right\}^{-1}\\
&= \left\{\varphi'(0)\lim_{N\to\infty} N^{-1}\sum_{i=1}^{N}\bm{\lambda}_i\bm{\lambda}_i^{\t}/\sigma_i^2\right\}^{-1} +o_p(1).
\end{aligned}\]
By Assumption \ref{asmp: AFM}, $N^{-1}\sum_{i=1}^{N}\bm{\lambda}_j\bm{\lambda}_i^{\t}/\sigma_i^2$ converges to a positive definite matrix in probability, so $\{\nabla\Phi(\bar{\bm{f}}_{\rm post})\}^{-1} = O_p(1)$.

Back to the expansion \eqref{eq: normalexpan}, by Conditions \ref{asmp: AFMCLT} and \ref{asmp: expmixing}, and $\bar{\bm{f}}_{\rm post} = \bm{\alpha}_1+o(1)$, we have 
\[\begin{aligned}
T_*^{1/2}\left(\hat{\beta}_{i} - \bar{\beta}_i\right) &= T_*^{1/2}\bar{\varepsilon}_{\mathrm{post},i} + T_*^{1/2}\left(\bm{\lambda}_i - \hat{\bm{\lambda}}_i\right)^{\t}\bar{\bm{f}}_{\rm post}+o_p(1)\\
    &=T_*^{1/2}\bar{\varepsilon}_{\mathrm{post},i} + T_*^{1/2}\left(\bm{\lambda}_i - \hat{\bm{\lambda}}_i\right)^{\t}\bm{\alpha}_1 +o_p(1)\\
        &=T_*^{1/2}\bar{\varepsilon}_{\mathrm{post},i} + T_*^{1/2}T_0^{-1}\sum_{t=1}^{T_0}\bm{f}_t^{\t}\bm{\alpha}_1\varepsilon_{it} +o_p(1)\\
        &\stackrel{d}{\rightarrow} N\left\{0, \omega_i^2\left(\kappa\right)\right\}.
\end{aligned}\]
Here $\omega_i^2(\kappa) = c_AA+c_B B+2c_A^{1/2}c_B^{1/2} C$, where $c_A = \min\{1,\kappa/(1-\kappa)\}$ for $\kappa \neq 1$ and $c_A = 1$ when $\kappa = 1$, $c_B = \min\{1,(1-\kappa)/\kappa\}$ for $\kappa \neq 0$ and $c_B = 1$ when $\kappa = 0$, and $A = \lim_{T-T_0\to\infty}(T-T_0)^{-1}\sum_{s=T_0+1}^T\sum_{t=T_0+1}^T\tau_{i,st}$, $B = \bm{\alpha}_1^{\t}\left(\lim_{T_0\to\infty}T_0^{-1}\sum_{s=1}^{T_0}\sum_{t=1}^{T_0}
 \bm{f}_t\bm{f}_s^{\t}\varpi_{i,ts}\right)\bm{\alpha}_1$, $C = \lim_{T_0, T-T_0\to\infty}(T-T_0)^{-1/2}T_0^{-1/2}
   \sum_{s=T_0+1}^T\sum_{t=1}^{T_0}
   \bm{f}_t^{\t}\bm{\alpha}_1\varpi_{i,st}.$
   
The asymptotic normality holds for all $\kappa\in[0,1]$ and the asymptotic covariance does not degenerate even on the boundary cases $\kappa = 0,1$.
\end{proof}

\section{Inference with block bootstrap}

We use the simple circular block bootstrap to obtain a variance estimator accounting for time series correlations \citep{politis1992circular}. 
The circular block bootstrap resamples blocks of data and therefore accounts for the inherent dependence structure in the data.
In theory, the block length needs to grow as the sample size grows. In practice, typical choices of block lengths include $T^{1/2},T^{1/3}$, and $T^{1/4}$, where we have assessed the robustness of these choices via simulation in Section \ref{sec:simulation}.

Given the prespecified block size $M$, the bootstrap procedures are summarized as follows.
\begin{enumerate}
    \item \textbf{Pre-intervention resampling.} Let $T_0$ denote the pre-intervention length. Define circular blocks of size $M$ by treating indices $\{1, \ldots, T_0\}$ as wrapped on a circle: for each starting index $k \in \{1, \ldots, T_0\}$, the block is
    \[
    B_k^{(0)} = \bigl( k, k+1, \ldots, k+M-1 \bigr) \mod T_0,
    \]
    where indice $i$ exceeding $T_0$ wraps around to $i-T_0$. To construct the resampled index set $I_0$:
    \begin{enumerate}
        \item[(a)] Draw $\lceil T_0 / M \rceil$ block starting points $k_1, k_2, \ldots, k_{\lceil T_0/M \rceil}$ independently and uniformly from $\{1, \ldots, T_0\}$ with replacement.
        \item[(b)] Concatenate the corresponding blocks to form a sequence of indices: \\
        $I_0' = (B_{k_1}^{(0)}, B_{k_2}^{(0)}, \ldots, B_{k_{\lceil T_0/M \rceil}}^{(0)})$.
        \item[(c)] Truncate $I_0'$ to retain exactly the first $T_0$ indices, yielding $I_0$.
    \end{enumerate}
    
    \item \textbf{Post-intervention resampling.} Define circular blocks of size $M$ on the post-intervention indices $\{T_0+1, \ldots, T\}$ analogously: for each starting index $k \in \{1, \ldots, T-T_0\}$, the block is
    \[
    B_k^{(1)} = \bigl( T_0 + k, T_0 + k + 1, \ldots, T_0 + k + M - 1 \bigr) \mod T-T_0,
    \]
    where indice $i$ exceeding $T$ wrap around to $i - (T-T_0)$. Construct the resampled index set $I_1$ following the same steps (a)--(c) with $T-T_0$ in place of $T_0$.
    
    \item \textbf{Data concatenation.} Resample the data using $I_0$ and $I_1$ to obtain the bootstrap sample:
    \[
    (Z_t^*, \bm{Y}_t^*)_{t=1}^{T} = \left\{ (Z_t, \bm{Y}_t)_{t \in I_0}, \, (Z_t, \bm{Y}_t)_{t \in I_1} \right\}.
    \]
    
    \item \textbf{Bootstrap estimation.} Compute the bootstrap estimator $\hat{\bm{\beta}}^*$ from the resampled data $(Z_t^*, \bm{Y}_t^*)_{t=1}^{T}$.
    
    \item \textbf{Replication and inference.} Repeat Steps 1--4 independently $B$ times to obtain bootstrap replicates $\{\hat{\bm{\beta}}^{*(b)}\}_{b=1}^B$. Confidence intervals are constructed using either the empirical quantiles of $\{\hat{\bm{\beta}}^{*(b)}\}_{b=1}^B$ entrywise or the bootstrap variance estimate 
    \[
    \hat{\omega}_i^{*2} = \frac{1}{B} \sum_{b=1}^B \left(\hat{\beta}_i^{*(b)} - \bar{\beta}^*_i\right)^2, \quad \text{where } \bar{\beta}^*_i = \frac{1}{B} \sum_{b=1}^B \hat{\beta}_i^{*(b)},
    \]
    to form Wald-type confidence intervals.
\end{enumerate}

\section{Additional simulations} 
\subsection{Simulation under perturbed post-intervention loadings}\label{sec:suppsimupert}

In this section, we investigate a different DGP beyond the model specification in the main paper. 
We consider the case where there exists a mean-zero perturbation to the factor loadings in the post-intervention period, that is, $Y_{it}(0) = (\bm{\lambda}_i + \bm{\xi}_{it})^{\t}\bm{f}_t + \varepsilon_{it}$ in the post-treatment periods, where the perturbation terms $\bm{\xi}_{it}$ are i.i.d. from $N(0,\bm{I}_2)$.  
Apart from this additional perturbation, the remaining data-generating process and the causal effects are identical to those in Section 5 for both fixed-$N$ and large-$N$ settings.

\subsubsection{Fixed-$N$ setting}

\begin{figure}[H]
    \captionsetup[subfloat]{captionskip=-1pt}
    \subfloat[$T_0=100, T=200$]{
  		 	\includegraphics[width=0.93\textwidth, height=0.31\textwidth]{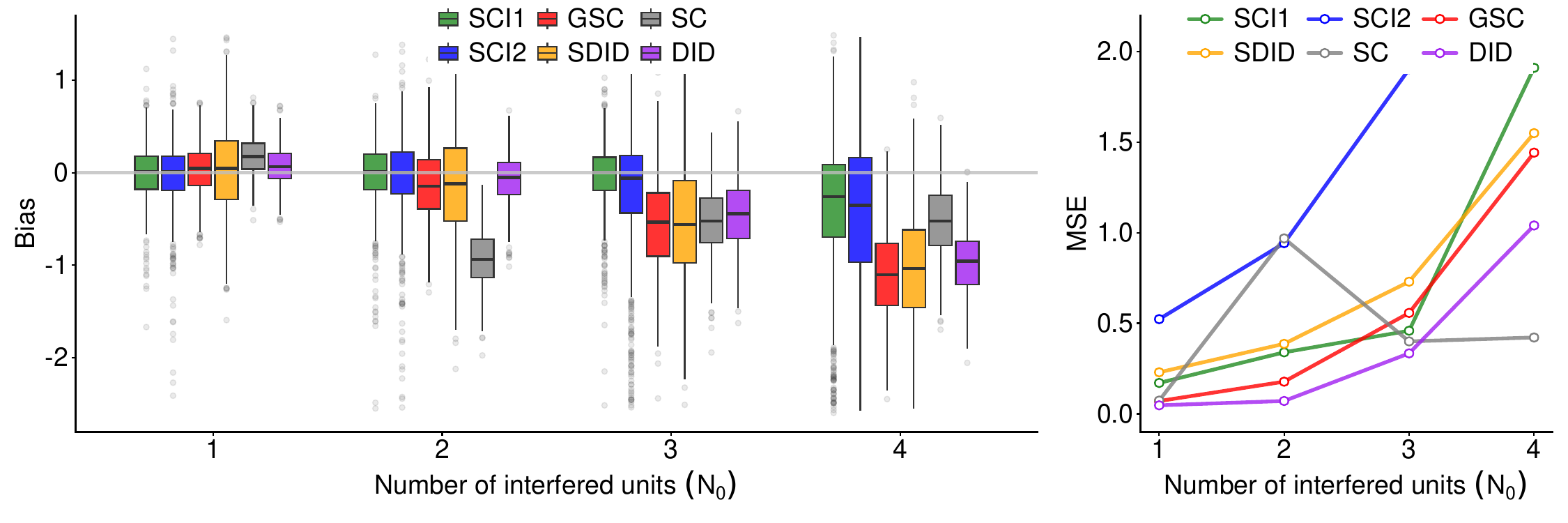} 
    	}\\
     \subfloat[$T_0=200, T=400$]{
  		 	\includegraphics[width=0.93\textwidth, height=0.31\textwidth]{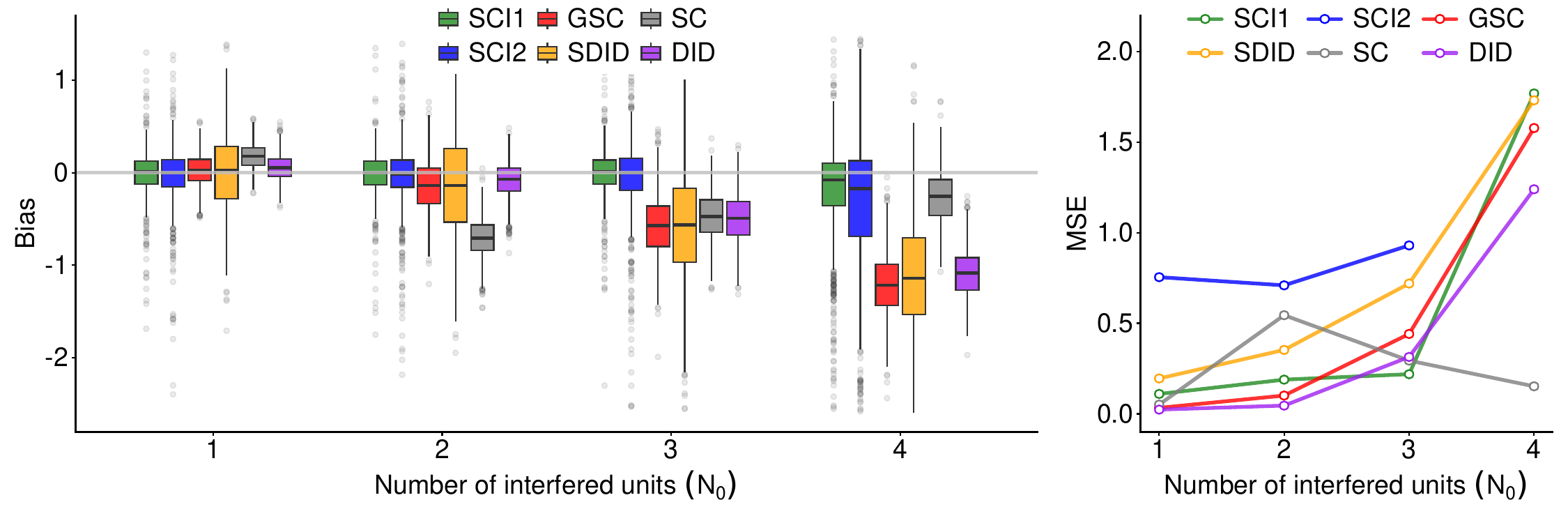}
    	}
	\caption{Boxplots and MSE of different estimators for $\bar{\beta}_1$ under fixed-$N$ setting with perturbed loadings.} 
 \label{fig:interfereS1}
\end{figure}

Figure \ref{fig:interfereS1} presents the bias and mean squared error (MSE) of different estimators under two time dimensions, $T=200$ and $T=400$, with varying numbers of interfered units $N_0$.

The proposed SCI estimators display overall favorable performance in terms of both bias and MSE. In particular, SCI$_1$ exhibits negligible bias and the smallest MSE when the number of interfered units is small, except when $N_0=4$, where the assumption is violated.
This indicates that while the proposed method is robust to loading perturbations, its performance is more sensitive to violations of the majority valid control assumption than in the constant-loading setting.
This highlights the importance of correctly identifying the interference structure in practice.
In contrast, the competing methods, including GSC, SDID, SC, and DID, suffer from non-negligible biases in the presence of interference. Both bias and MSE tend to increase as the number of interfered units $N_0$ grows, indicating that misspecification of control units can substantially deteriorate estimation accuracy.

Table \ref{tab:cvpS1} reports the Monte Carlo coverage probabilities of 95\% confidence intervals based on the circular block bootstrap. The empirical coverage rates are slightly above the nominal level across different block lengths and interference intensities, suggesting that the proposed inference procedure provides reliable uncertainty quantification in finite samples. This may be due to the instability of factor analysis under the fixed-$N$ setting, leading to more outliers and larger variances for bootstrap samples than in the correctly specified setting.

\begin{table}[H]
\renewcommand{\arraystretch}{0.5} 
\caption{Monte Carlo coverage probabilities (\%) of the 95\% confidence intervals using the circular block bootstrap with different block length for $\bar{\beta}_{1}$ under fixed-$N$ setting with perturbed loadings}
\centering
\begin{tabular}[t]{ccccccccc}
\toprule
\multirow{2}{*}{Block length}& \multicolumn{4}{c}{$T_0 = T/2=100$} & \multicolumn{4}{c}{$T_0 =T/2=200$} \\
\cmidrule(lr){2-5} \cmidrule(lr){6-9}
 & $N_0=1$ & 2 & 3 & 4 & 1 & 2 & 3 & 4 \\
\midrule
$T^{1/4}$ & 97.0 & 98.8 & 99.4 & 99.2 & 97.5 & 98.5 & 99.2 & 99.3 \\
$T^{1/3}$ & 97.0 & 98.8 & 99.4 & 99.1 & 97.5 & 98.6 & 99.2 & 99.7 \\
$T^{1/2}$ & 96.4 & 98.5 & 99.2 & 98.9 & 97.0 & 98.1 & 99.0 & 99.6 \\
\bottomrule
\end{tabular}
\label{tab:cvpS1}
\end{table}

\subsubsection{Large-$N$ setting}

For the large-$N$ setting, the performance of the proposed estimator and other estimators is similar to the results in Section 5.2. Again, the proposed method has the smallest bias and MSE, showing that it is relatively robust to the perturbed factor-loading model.

\begin{figure}[H]
    \captionsetup[subfloat]{captionskip=-1pt}
    \subfloat[$T_0=100, T=200$]{
  		 	\includegraphics[width=0.93\textwidth, height=0.31\textwidth]{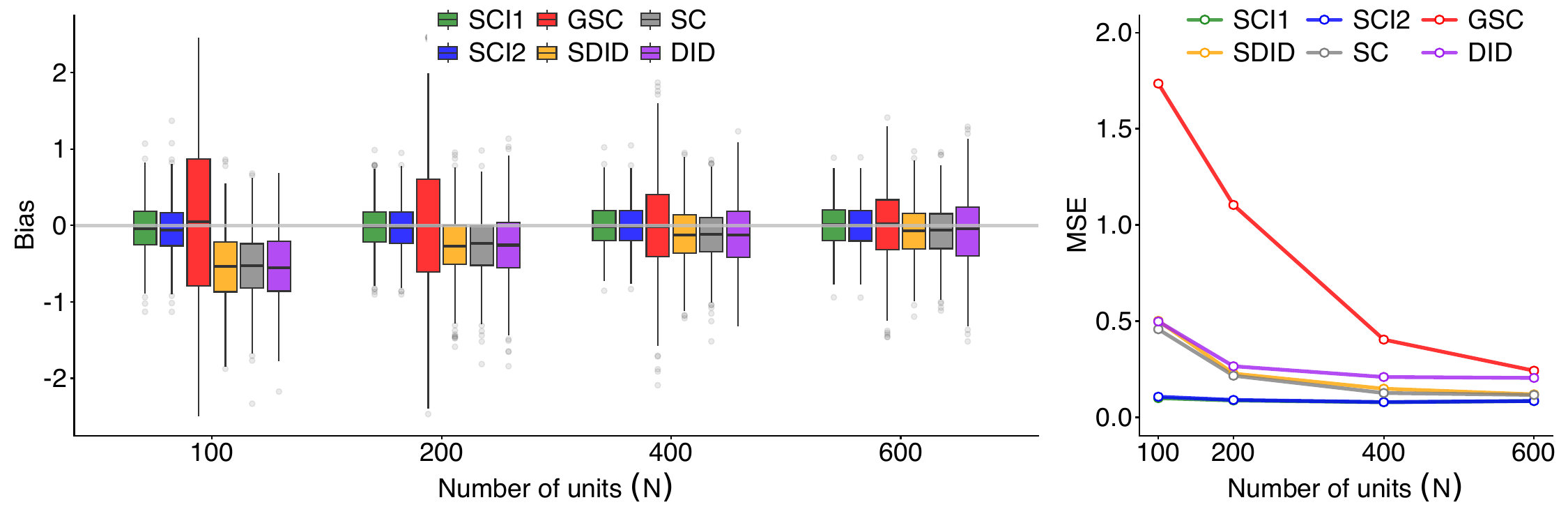} 
    	}\\
     \subfloat[$T_0=200, T=400$]{
  		 	\includegraphics[width=0.93\textwidth, height=0.31\textwidth]{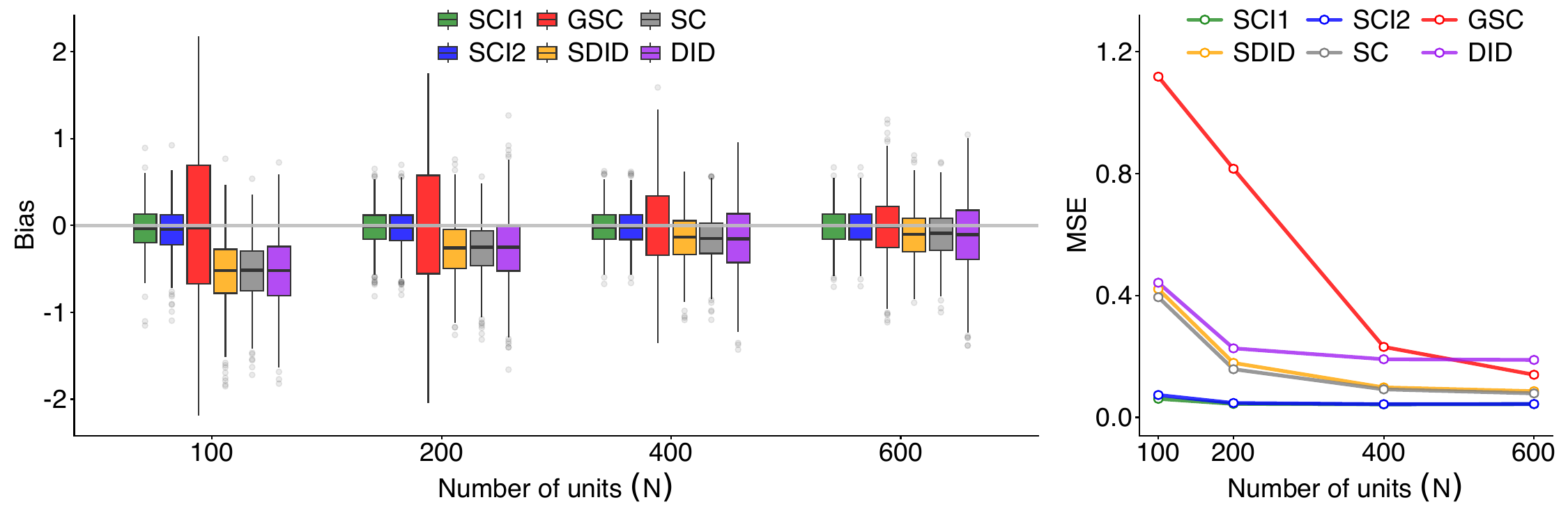}
    	}
	\caption{Boxplots and MSE of different estimators for $\bar{\beta}_1$ under large-$N$ setting with perturbed loadings.} 
 \label{fig:interfereS2}
\end{figure}

\begin{table}[H]
\renewcommand{\arraystretch}{0.5} 
\caption{Monte Carlo coverage probabilities (\%) of the 95\% confidence intervals using the circular block bootstrap with different block lengths for $\bar{\beta}_{1}$ under large-$N$ setting with perturbed loadings}
\centering
\begin{tabular}[t]{ccccccccc}
\toprule
\multirow{2}{*}{Block length}& \multicolumn{4}{c}{$T_0 = T/2=100$} & \multicolumn{4}{c}{$T_0 =T/2=200$} \\
\cmidrule(lr){2-5} \cmidrule(lr){6-9}
 & $N=100$ & 200 & 300 & 400 & 100 & 200 & 300 & 400 \\
\midrule
$T^{1/4}$ & 94.0 & 93.7 & 94.7 & 92.7 & 93.1 & 95.0 & 95.2 & 95.1 \\
$T^{1/3}$ & 94.5 & 94.0 & 95.5 & 94.2 & 93.7 & 95.7 & 95.5 & 96.1 \\
$T^{1/2}$ & 92.8 & 92.3 & 94.0 & 92.3 & 89.3 & 92.2 & 91.6 & 92.3 \\
\bottomrule
\end{tabular}
\label{tab:cvpS2}
\end{table}

\subsection{Simulation results for the average interference effect $\bar{\beta}_2$}

\begin{figure}[H]
    \captionsetup[subfloat]{captionskip=-1pt}
    \subfloat[$T_0=100, T=200$]{
  		 	\includegraphics[width=0.93\textwidth, height=0.31\textwidth]{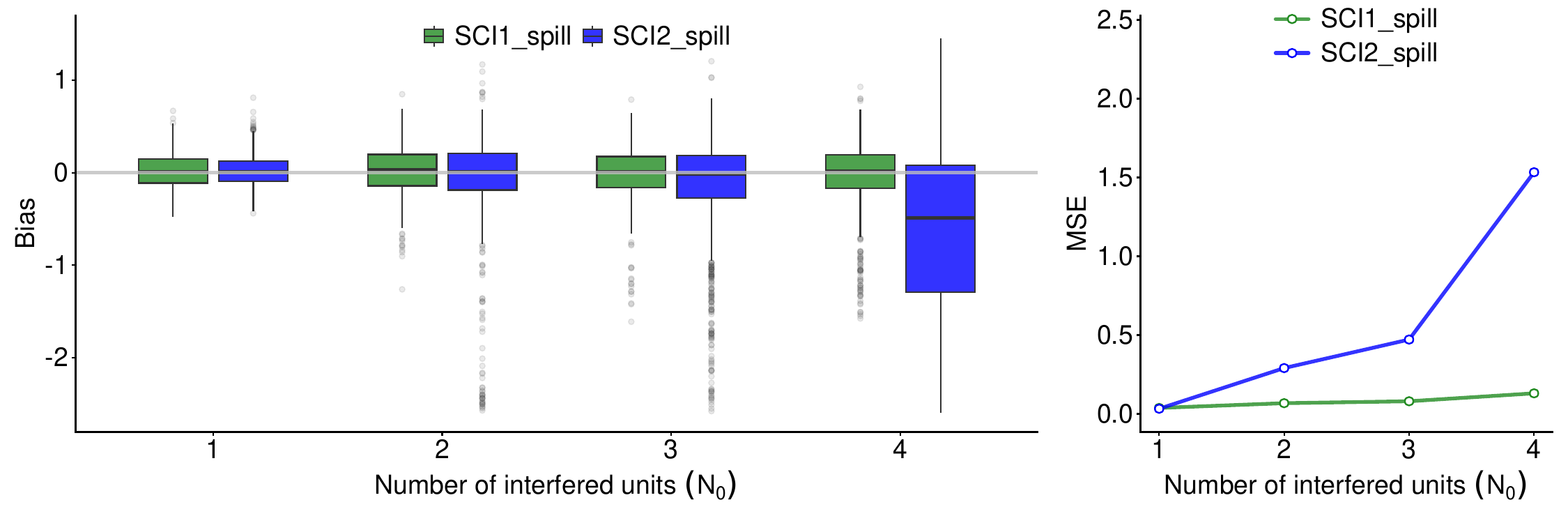} 
    	}\\
     \subfloat[$T_0=200, T=400$]{
  		 	\includegraphics[width=0.93\textwidth, height=0.31\textwidth]{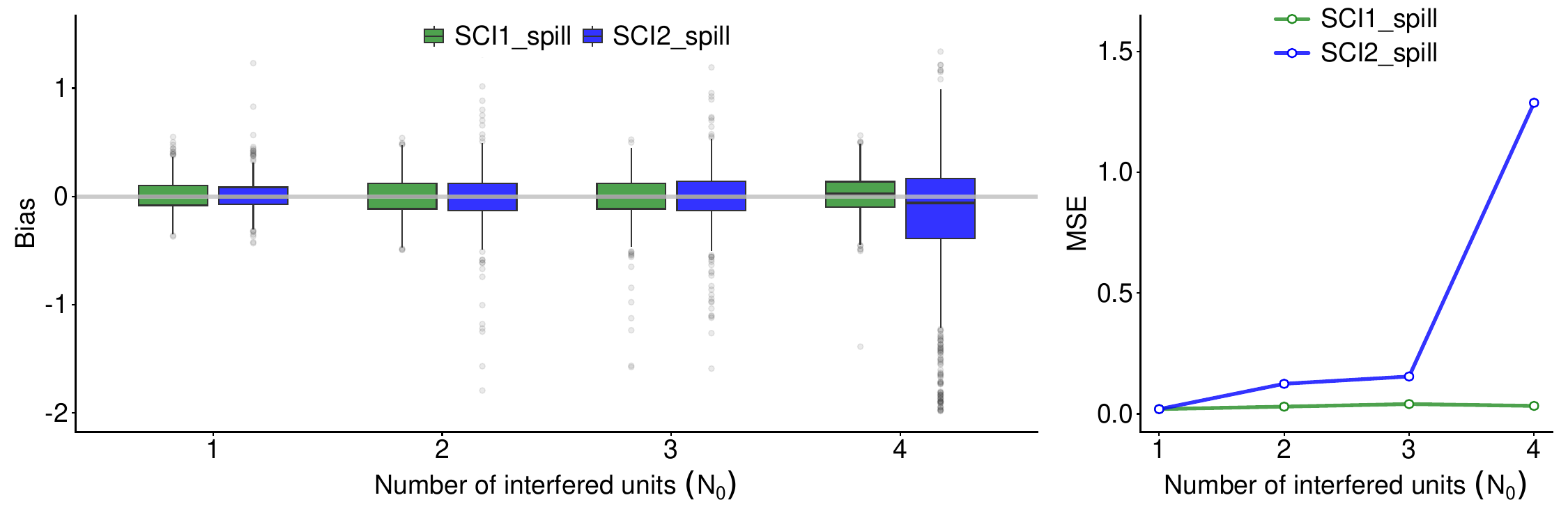}
    	}\\
         \captionsetup{skip=-5pt}
    \caption{Finite-sample performance of proposed method for $\bar{\beta}_2$ under fixed-$N$ setting in Section 5. }
 \label{fig:b2f}
\end{figure}

\begin{figure}[H]
    \captionsetup[subfloat]{captionskip=-1pt}
    \subfloat[$T_0=100, T=200$]{
  		 	\includegraphics[width=0.93\textwidth, height=0.31\textwidth]{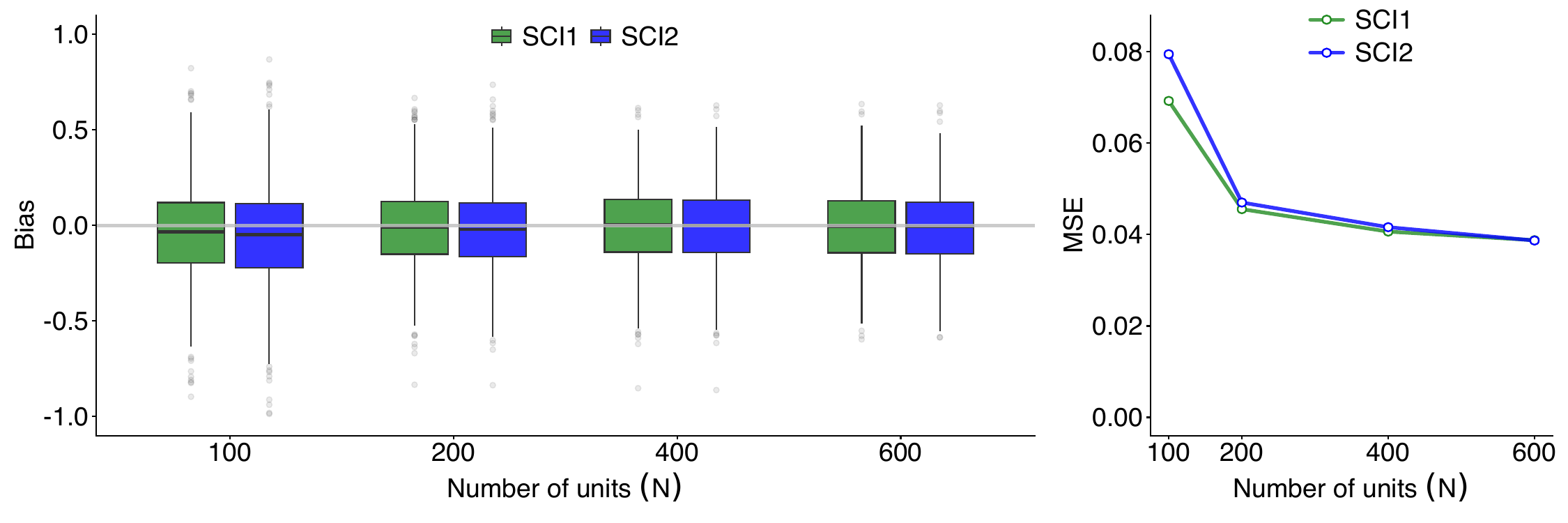} 
    	}\\
     \subfloat[$T_0=200, T=400$]{
  		 	\includegraphics[width=0.93\textwidth, height=0.31\textwidth]{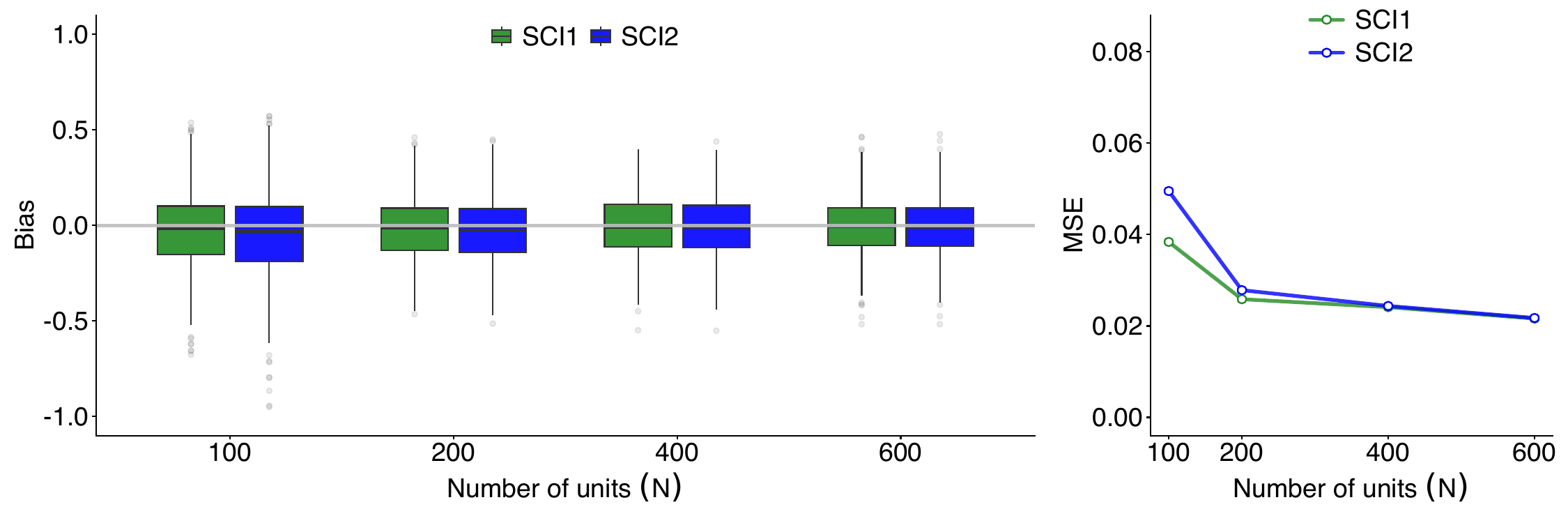}
    	}\\
         \captionsetup{skip=-5pt}
    \caption{Finite-sample performance of proposed method for $\bar{\beta}_2$ under large-$N$ setting in Section 5. }
 \label{fig:b2l}
\end{figure}

In this subsection, we report simulations illustrating performance for estimating average interference effects. Since results are qualitatively similar across $\bar{\beta}_i$ and closely mirror those for the average direct effect $\bar{\beta}_1$, we present $\bar{\beta}_2$ as a representative case. 
Besides, competing methods cannot estimate interference effects, so we focus on the performance of the proposed methods without comparing them to other methods.
The data-generating process matches Section 5 of the main text. Results for the fixed-$N$ setting appear in Figure~\ref{fig:b2f} and Table~\ref{tab:b2f}, and for the large-$N$ setting in Figure~\ref{fig:b2l} and Table~\ref{tab:b2l}.
Overall, the conclusions are very similar to the average direct effect. Correctly specified proposed fixed-$N$ and large-$N$ methods (SCI$_1$) consistently yield very little bias across all scenarios.
The proposed fixed-$N$ method with a misspecified number of factors has larger bias and MSE as the interference level increases, while the proposed large-$N$ method shows very little influence from such misspecification, especially for large panels.
The circular block bootstrap confidence intervals yield coverage rates close to the nominal level.
The performance is not sensitive to block length in the fixed-$N$ setting, but $T^{1/2}$ leads to undercoverage in the large-$N$ setting.
\subsubsection{Fixed-$N$ setting}

\begin{table}[H]
\renewcommand{\arraystretch}{0.5} 
\caption{Monte Carlo coverage probabilities (\%) of the 95\% confidence intervals using the circular block bootstrap with different block lengths for $\bar{\beta}_{2}$ under fixed-$N$ setting in Section \ref{sec:simulation}}
\centering
\begin{tabular}[t]{ccccccccc}
\toprule
\multirow{2}{*}{Block length}& \multicolumn{4}{c}{$T_0 = T/2=100$} & \multicolumn{4}{c}{$T_0 =T/2=200$} \\
\cmidrule(lr){2-5} \cmidrule(lr){6-9}
 & $N_0=1$ & 2 & 3 & 4 & 1 & 2 & 3 & 4 \\
\midrule
$T^{1/4}$ & 93.3 & 94.6 & 96.1 & 97.1 & 93.5 & 94.0 & 94.9 & 95.3 \\
$T^{1/3}$ & 93.9 & 95.1 & 96.5 & 97.2 & 93.9 & 94.4 & 95.2 & 95.5 \\
$T^{1/2}$ & 92.5 & 94.3 & 95.6 & 96.2 & 92.8 & 93.3 & 94.5 & 94.5 \\
\bottomrule
\end{tabular}
\label{tab:b2f}
\end{table}

\subsubsection{Large-$N$ setting}

\begin{table}[H]
\renewcommand{\arraystretch}{0.5} 
\caption{Monte Carlo coverage probabilities (\%) of the 95\% confidence intervals using the circular block bootstrap with different block lengths for $\bar{\beta}_{2}$ under large-$N$ setting in Section \ref{sec:simulation}}
\centering
\begin{tabular}[t]{ccccccccc}
\toprule
\multirow{2}{*}{Block length}& \multicolumn{4}{c}{$T_0 = T/2=100$} & \multicolumn{4}{c}{$T_0 =T/2=200$} \\
\cmidrule(lr){2-5} \cmidrule(lr){6-9}
 & $N=100$ & 200 & 400 & 600 & 100 & 200 & 400 & 600 \\
\midrule
$T^{1/4}$ & 89.70 & 93.10 & 92.70 & 94.70 & 90.10 & 93.20 & 93.50 & 95.50 \\
$T^{1/3}$ & 90.00 & 93.30 & 94.40 & 95.50 & 90.40 & 93.50 & 94.70 & 95.40 \\
$T^{1/2}$ & 89.10 & 92.10 & 93.10 & 94.60 & 85.20 & 89.40 & 90.70 & 92.30 \\
\bottomrule
\end{tabular}
\label{tab:b2l}
\end{table}

\subsection{Simulation under data-generating process from \citet{cao2019estimation}}

In addition to the data-generating processes in the main text and the previous part, we consider the $\mathcal{I}(1)$ data-generating process from \cite{cao2019estimation}, which also studies synthetic control with interference. 
As this data setting involves heavy non-stationarity, we only implement the large-$N$ method.
For the large-$N$ setting, the outcomes are generated from a factor model with a mixture of integrated and stationary latent factors. Specifically, the three-dimensional factor process $\boldsymbol{f}_t = (f_{1t}, f_{2t}, f_{3t})^{\t}$ is generated as
\[
f_{1t} = f_{1,t-1} + 0.5 u_{1t}, \quad
f_{2t} = f_{2,t-1} + 0.5 u_{2t}, \quad
f_{3t} = 0.5 f_{3,t-1} + u_{3t},
\]
where $\{u_{kt}\}_{k=1}^3$ are i.i.d. $N(0,1)$. Thus, the first two factors follow random walks, while the third factor is stationary.
The first four units have fixed loadings $\bm{\lambda}_1=(1,0,0)^{\t}$, $\bm{\lambda}_2=(0,1,0)^{\t}$, $\bm{\lambda}_3=(1,0,0)^{\t}$, and $\bm{\lambda}_4=(0,1,0)^{\t}$. For $j \ge 5$, the loadings are generated as $\bm{\lambda}_j = (x_{j1},x_{j2},x_{j3})^{\t} / \sum_{k=1}^3 x_{jk}$ with $x_{jk} \sim \text{Unif}(0,1)$ independently, ensuring positive loadings that sum to one.
The time-varying treatment effects are periodic. Define $\text{base}_t = 5 + \sin(\pi t / 12), t > T_0.$
For the treated units, we set
\[
\beta_{1t} = \,\text{base}_t \mathbbm{1}(t > T_0), \quad
\beta_{it} = 0.75\text{base}_t \mathbbm{1}(t > T_0), \quad 1 < i \le N_0.
\]

The idiosyncratic errors are generated as $\varepsilon_{it} \sim N(0,1)$ independently across $i$ and $t$, and the observed outcome is $Y_{it} = \beta_{it} + \bm{\lambda}_i^{\t}\bm{f}_t + \varepsilon_{it}.$
We consider two choices of time periods, $T_0 =T/2\in \{100, 200\}$, and for each $T$, we vary the cross-sectional dimension $N \in \{100, 200, 400, 600\}$. Each design is replicated 1,000 times.

\begin{figure}[H]
    \captionsetup[subfloat]{captionskip=-1pt}
    \subfloat[$T_0=100, T=200$]{
  		 	\includegraphics[width=0.93\textwidth, height=0.31\textwidth]{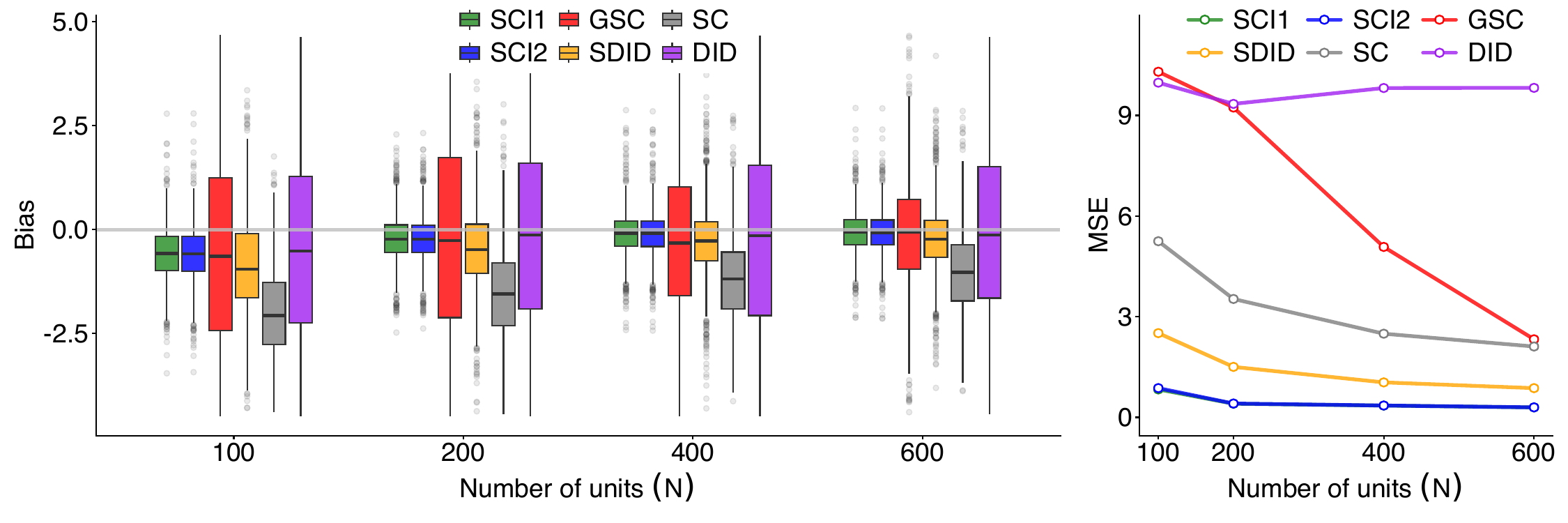} 
    	}\\
     \subfloat[$T_0=200, T=400$]{
  		 	\includegraphics[width=0.93\textwidth, height=0.31\textwidth]{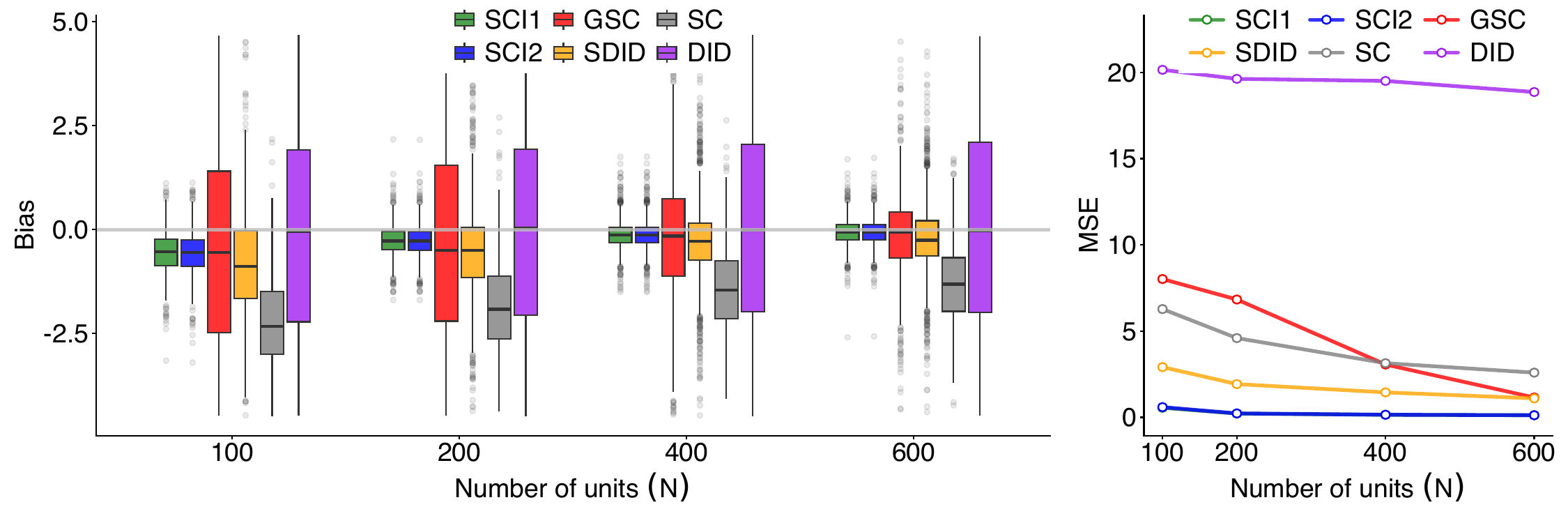}
    	}
	\caption{Boxplots and MSE of different estimators for $\bar{\beta}_1$ under DGP of \cite{cao2019estimation}} 
 \label{fig:interfereS3}
\end{figure}

Figure~\ref{fig:interfereS3} shows that the proposed method has the smallest MSE in this setting, and the bias vanishes as $N$ increases. Meanwhile, the other competing methods exhibit larger MSE due to either large bias or large variances, or both.

Table~\ref{tab:cvpS3} reports the empirical coverage probabilities of the 95\% confidence intervals for SCI$_1$ based on the circular block bootstrap with block lengths $T^{1/4}$, $T^{1/3}$, and $T^{1/2}$ and 500 bootstrap replications. 
The intervals exhibit undercoverage for the panel sizes considered in the main paper, likely reflecting finite-sample bias and nonstationarity. 
To further assess this finite-sample behavior, we additionally consider larger panels with $(T_0,N)=(400,800)$ and $(800,800)$. 
Coverage generally improves as the panel dimensions increase, reaching approximately 90--94\% for these larger designs and suggesting improved bootstrap accuracy in larger panels.

\begin{table}[H]
\centering
\caption{Monte Carlo coverage probabilities (\%) of the 95\% confidence intervals using the circular block bootstrap with different block lengths for $\bar{\beta}_{1}$ under the DGP of \cite{cao2019estimation}}
\label{tab:cvp-largeN-cao}

\begin{tabular}{@{}l*{10}{r}@{}}
\toprule
\multirow{2}{*}{Block length} & \multicolumn{4}{c}{$T_0=T/2=100$} & \multicolumn{4}{c}{$T_0=T/2=200$} & \multicolumn{1}{c}{$T_0=400$} & \multicolumn{1}{c}{$T_0=800$} \\
\cmidrule(lr){2-5} \cmidrule(lr){6-9} \cmidrule(lr){10-10} \cmidrule(lr){11-11}
 & $N=100$ & $200$ & $400$ & $600$ & $100$ & $200$ & $400$ & $600$ & $800$ & $800$ \\
\midrule
$T^{1/4}$ & 62.8 & 69.3 & 72.6 & 75.1 & 56.0 & 75.4 & 87.0 & 89.7 & 93.0 & 93.7 \\
$T^{1/3}$ & 64.3 & 71.4 & 74.7 & 76.1 & 57.7 & 76.3 & 86.5 & 90.4 & 93.0 & 93.3 \\
$T^{1/2}$ & 67.9 & 72.4 & 74.5 & 77.9 & 57.2 & 73.4 & 84.2 & 86.6 & 90.2 & 90.3 \\
\bottomrule
\end{tabular}
\label{tab:cvpS3}
\end{table}

\subsection{Simulation under stronger serial dependence}
We examine the impact of serial dependence by considering a stronger serial dependence generating model: $
\bm{f}_t = {\bm{\alpha}}_1Z_t + {\bm{\alpha}}_0(1-Z_t) + \bm{\zeta}_t,\  \bm{Y}_{t} =  \bm{\beta}_{t}Z_t + \bm{\Lambda} \bm{f}_t + \bm{\ee}_{t}, \ \bm{\ee_t} = 0.5\bm{\ee}_{t-1} + 0.2\bm{\ee}_{t-2} + \bm{\nu}_t, 
$
where $\bm{\zeta}_t \sim N(\bm{0}, \mathbf{I}_2), \bm{\nu}_t \thicksim  N(\bm{0}, \mathbf{I}_N), {\bm{\alpha}}_0 = (0,0)^{\t}, {\bm{\alpha}}_1 = (1,1)^{\t},$ the others are the same as in Section 5.
Compared to the base data-generating process in Section 5, we use 
$\bm{\ee_t} = 0.5\bm{\ee}_{t-1} + 0.2\bm{\ee}_{t-2} + \bm{\nu}_t$ in replace of $\bm{\ee_t} = 0.2\bm{\ee}_{t-1} + 0.1\bm{\ee}_{t-2} + \bm{\nu}_t$, substantively making the serial dependence stronger.

\subsubsection{Fixed-$N$ setting}
\begin{figure}[H]
    \captionsetup[subfloat]{captionskip=-1pt}
    \subfloat[$T_0=100, T=200$]{
  		 	\includegraphics[width=0.93\textwidth, height=0.31\textwidth]{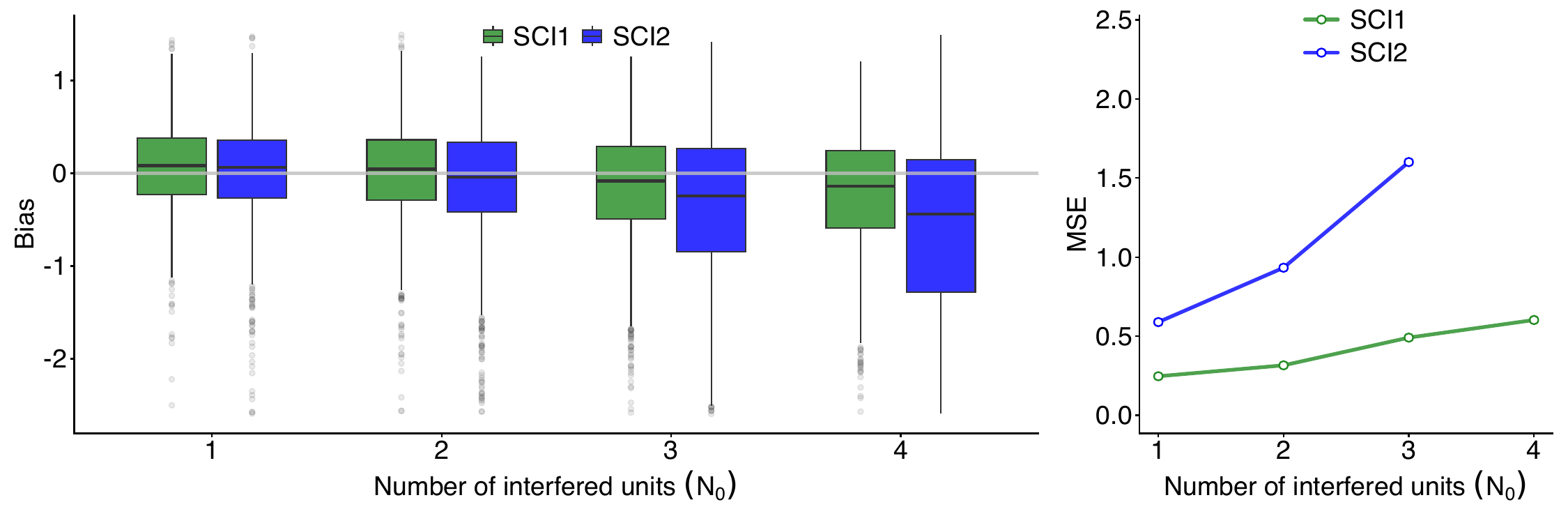} 
    	}\\
     \subfloat[$T_0=200, T=400$]{
  		 	\includegraphics[width=0.93\textwidth, height=0.31\textwidth]{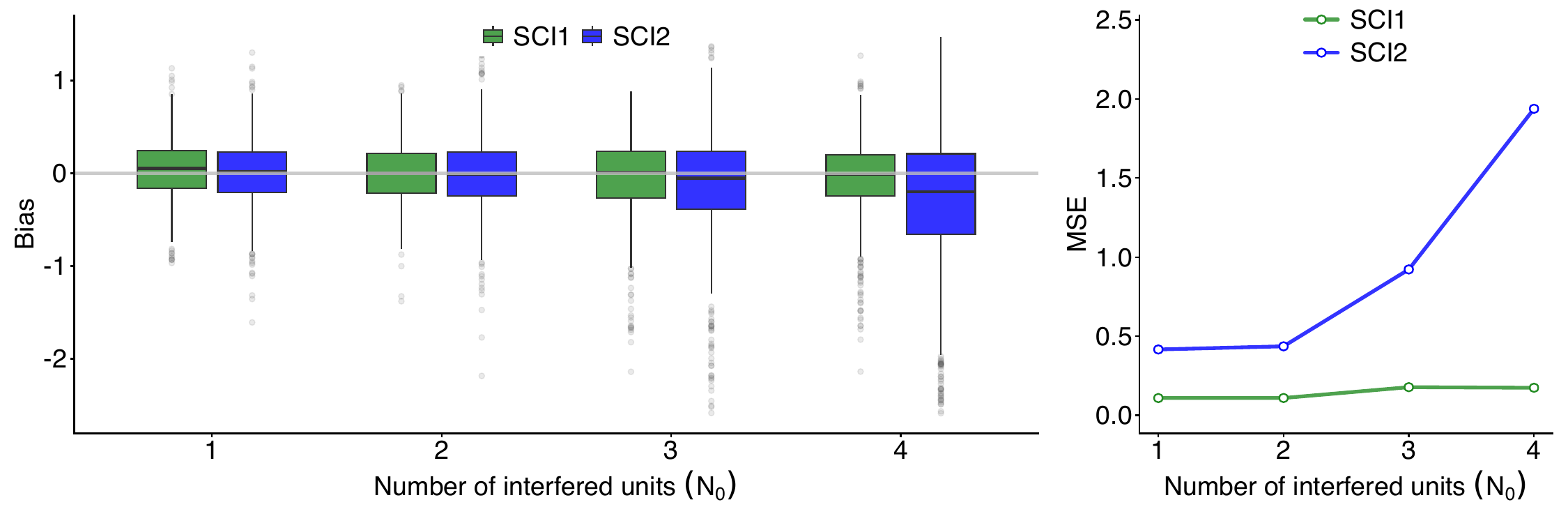}
    	}\\
         \captionsetup{skip=-5pt}
    \caption{Finite-sample performance of proposed method for $\bar{\beta}_1$ under fixed-$N$ setting with stronger serial dependence} 
 \label{fig:supp_dep}
\end{figure}

Figure~\ref{fig:supp_dep} displays the bias and MSE of the proposed methods under different interference levels. With correctly specified factors, SCI$_1$ exhibits minimal bias and low MSE across all scenarios. Slight bias appears when the post-intervention period is relatively short but vanishes as $T-T_0$ doubles. 
The misspecified SCI$_2$ shows little bias when interference is low, but both bias and MSE increase with $N_0$. 
When both the factor model and majority control assumption are violated, SCI$_2$ exhibits non-negligible bias and large variance.

Table \ref{tab:supp_cvp_dep} presents the coverage probabilities of 95\%  confidence intervals for SCI$_1$ based on circular block bootstrap with block lengths $T^{1/4}, T^{1/3}$ and $T^{1/2}$ and 500 bootstrap replicates under the strong serial dependence. 
Coverage probabilities are slightly lower in some scenarios due to the strong serial dependence, but remain close to the nominal level in most cases.

\begin{table}[H]
\renewcommand{\arraystretch}{0.5} 
\caption{Monte Carlo coverage probabilities (\%) of the 95\% confidence intervals using the circular block bootstrap with different block lengths for $\bar{\beta}_{1}$ under fixed-$N$ setting with stronger serial dependence}
\centering
\begin{tabular}[t]{ccccccccc}
\toprule
\multirow{2}{*}{Block length}& \multicolumn{4}{c}{$T_0 = T/2=100$} & \multicolumn{4}{c}{$T_0 =T/2=200$} \\
\cmidrule(lr){2-5} \cmidrule(lr){6-9}
 & $N_0=1$ & 2 & 3 & 4 & 1 & 2 & 3 & 4 \\
\midrule
$T^{1/4}$ & 87.4 & 91.5 & 93.1 & 94.1 & 88.3 & 89.2 & 92.0 & 92.9 \\
$T^{1/3}$ & 89.9 & 93.4 & 94.7 & 95.5 & 90.9 & 91.9 & 94.2 & 94.8 \\
$T^{1/2}$ & 91.6 & 94.9 & 95.4 & 95.7 & 92.6 & 93.4 & 95.5 & 95.8 \\
\bottomrule
\end{tabular}
\label{tab:supp_cvp_dep}
\end{table}

\subsubsection{Large-$N$ setting}
\begin{figure}[H]
    \captionsetup[subfloat]{captionskip=-1pt}
    \subfloat[$T_0=100, T=200$]{
  		 	\includegraphics[width=0.93\textwidth, height=0.31\textwidth]{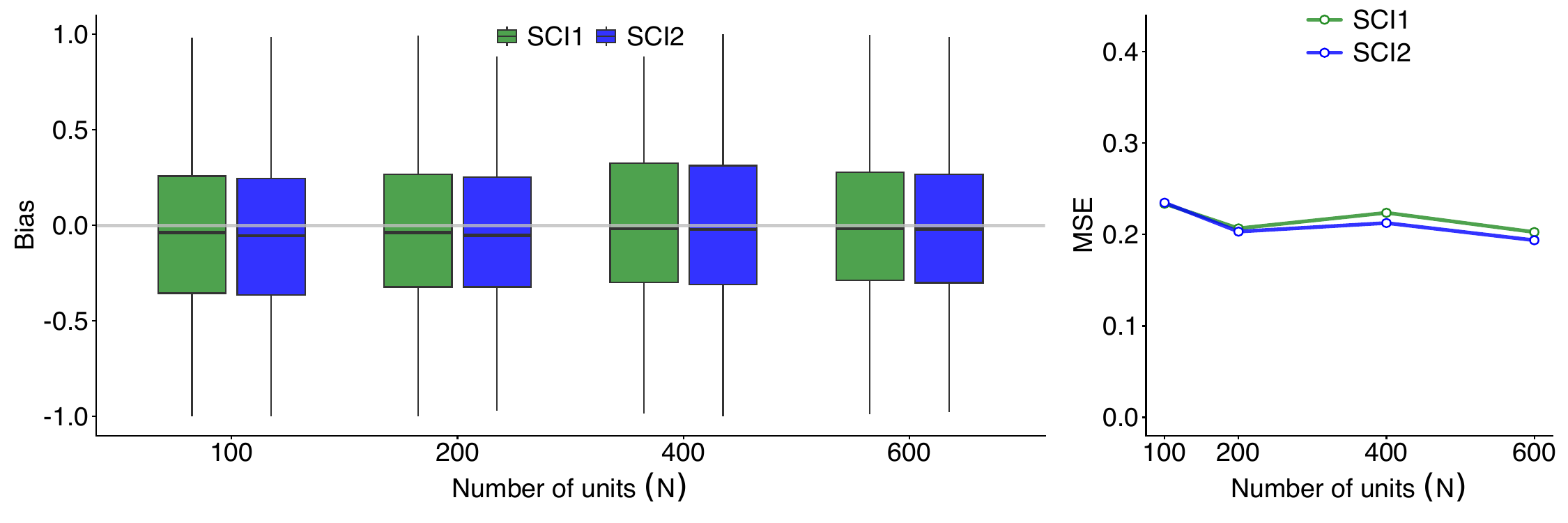} 
    	}\\
     \subfloat[$T_0=200, T=400$]{
  		 	\includegraphics[width=0.93\textwidth, height=0.31\textwidth]{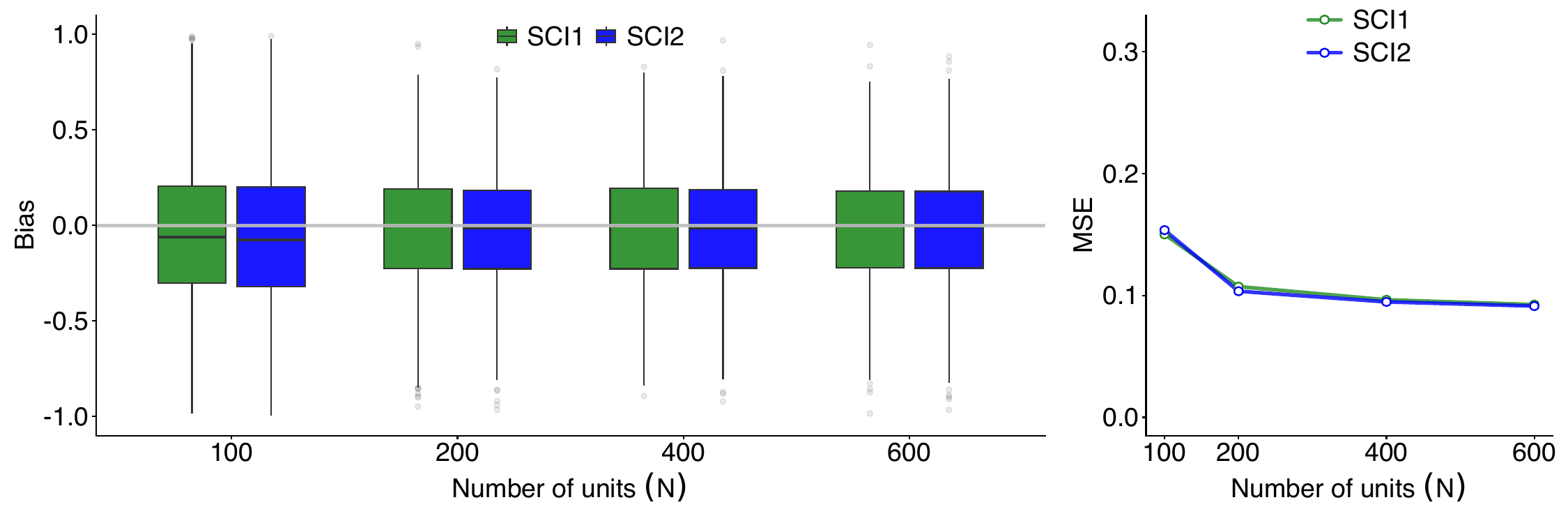}
    	}\\
         \captionsetup{skip=-5pt}
    \caption{Finite-sample performance of proposed method for $\bar{\beta}_1$ under large-$N$ setting with stronger serial dependence} 
 \label{fig:supp_dep_largeN}
\end{figure}

Figure~\ref{fig:supp_dep_largeN} displays the bias and MSE of the proposed methods under different interference levels. 
We found that the proposed method consistently yields very little bias and small MSE across all scenarios, even when the number of factors is misspecified.

Table~\ref{tab:supp_cvp_dep_largeN} reports the empirical coverage probabilities of the 95\% confidence intervals for SCI$_1$ under stronger serial dependence. 
The intervals exhibit noticeable undercoverage for the smaller panel sizes, with no clear monotone improvement as $N$ increases. 
To further assess this finite-sample behavior, we additionally consider larger panels with $(T_0,N)=(400,800)$ and $(800,800)$. 
Coverage improves for these larger designs, reaching approximately 88--94\%, suggesting that the bootstrap approximation becomes more accurate as the panel dimensions increase, although some undercoverage remains under stronger serial dependence.

\begin{table}[H]
\centering
\caption{Monte Carlo coverage probabilities (\%) of the 95\% confidence intervals using the circular block bootstrap with different block lengths for $\bar{\beta}_{1}$ under large-$N$ setting with stronger serial dependence}

\begin{tabular}{@{}l*{10}{r}@{}}
\toprule
\multirow{2}{*}{Block length} & \multicolumn{4}{c}{$T_0=T/2=100$} & \multicolumn{4}{c}{$T_0=T/2=200$} & \multicolumn{1}{c}{$T_0=400$} & \multicolumn{1}{c}{$T_0=800$} \\
\cmidrule(lr){2-5} \cmidrule(lr){6-9} \cmidrule(lr){10-10} \cmidrule(lr){11-11}
 & $N=100$ & $200$ & $400$ & $600$ & $100$ & $200$ & $400$ & $600$ & $800$ & $800$ \\
\midrule
$T^{1/4}$ & 76.6 & 73.9 & 70.6 & 74.8 & 78.1 & 79.5 & 78.1 & 79.1 & 88.9 & 91.9 \\
$T^{1/3}$ & 79.7 & 78.3 & 75.1 & 78.9 & 80.7 & 81.5 & 80.2 & 79.9 & 90.3 & 93.4 \\
$T^{1/2}$ & 80.9 & 80.5 & 77.7 & 79.7 & 78.7 & 79.4 & 78.2 & 78.4 & 88.1 & 93.5 \\
\bottomrule
\end{tabular}
\label{tab:supp_cvp_dep_largeN}
\end{table}

\section{Supporting information on the applications}
In the US embassy relocation example, Syria is excluded from the analysis due to a lack of records before January 2017.
Four countries, Kuwait, Oman, Qatar, and the United Arab Emirates, are excluded from the set of control units due to their extremely low numbers of conflicts.
Conflict counts in these relatively peaceful countries are unlikely to be driven by the same latent factors as those in more turbulent Middle Eastern countries.
Table \ref{tab:sumfour} presents the outcome frequencies of the four countries.
As noted, the recorded conflicts among the four countries are extremely low: the maximum number is 2 over the study period and the majority of values are zero. 
It suggests that these countries are peaceful in terms of the level of conflict, making them inappropriate for comparison with more turbulent countries and thus providing justification for their exclusion. Furthermore, the presence of almost all-zero outcomes renders these unit negligible for predicting counterfactual outcomes through synthetic control, which supports the comparability of our results to those of \cite{muhlbach2021tree}.
 	\begin{table}[H]
    \renewcommand{\arraystretch}{0.5} 
   		\centering
 		\caption{Summary of the number of conflicts in Kuwait, Oman, Qatar, and the United Arab Emirates}
            \label{tab:sumfour}
 		\begin{tabular}{lcccc}
 			\toprule
 			&\multicolumn{4}{c}{Frequency} \\
 			Country&0&1&2&{$>$2} \\
 			\midrule
 			Kuwait&129&18&2&0 \\
 			Oman&142&6&1&0 \\
 			Qatar&147&1&1&0 \\
    United Arab Emirates&147&2&0&0\\
    \bottomrule
 		\end{tabular}
 	\end{table}

\putbib
\end{bibunit}

\end{document}